\documentclass[preprint]{aastex631}
\renewcommand{\added}[1]{{#1}}
\renewcommand{\deleted}[1]{}
\renewcommand{\replaced}[2]{{#2}}
\newcommand \versnum {6}
\newcommand{\fwidth}{10.0cm}   
\newcommand{\fwidthb}{7.0cm}   
\newcommand{\fwidthc}{8.0cm}   
\usepackage{color}
\usepackage{natbib}
\usepackage{amsmath}
\usepackage{nicefrac}
\tightenlines

\newcommand \bahat      {\boldsymbol{\hat{a}}}
\newcommand \bB         {\boldsymbol{B}}

\newcommand \bE         {\boldsymbol{E}}

\newcommand \bkhat      {\boldsymbol{\hat{k}}}

\newcommand \br         {\boldsymbol{r}}

\newcommand \beq        {\begin{equation}}
\newcommand \beqa	{\begin{eqnarray}}

\newcommand \cm         {\,{\rm cm}}

\newcommand \eeq	{\end{equation}}
\newcommand \eeqa	{\end{eqnarray}}

\newcommand \gm         {\,{\rm g}}

\newcommand \gtsim	{\gtrsim}		 

\newcommand \HH	        {{\rm H}_2}

\newcommand \keV	{\,{\rm keV}}

\newcommand \K  	{\,{\rm K}}
\newcommand \kms	{\,{\rm km~s}^{-1}}

\newcommand \lambdap    {\lambda_{\rm p}}

\newcommand \ltsim	{\lesssim}		 

\newcommand \nH         {n_{\rm H}}

\newcommand \pfirmax    {[p_{\rm em}({\rm FIR})]_{\rm max}}
\newcommand \pmax       {p_{\rm max}}
\newcommand \poro       {{\cal P}}
\newcommand \poromacro  {{\cal P}_{\rm macro}}
\newcommand \poromicro  {{\cal P}_{\rm micro}}

\newcommand \Rone       {{\cal A}^\prime}
\newcommand \Asymm       {{\cal A}}
\newcommand \Stretch       {{\cal S}}
\newcommand \s	        {\,{\rm s}}
\newcommand \sigmap     {\sigma_{\rm p}}

\newcommand \xtimes     {{\!\,\times\!\,}}

\newcommand \yr  	{\,{\rm yr}}

\newcommand \mm         {\,{\rm mm}}

\newcommand \aeff       {a_{\rm eff}}
\newcommand \aeffp      {a_{\rm eff,p}}

\newcommand \Cabs       {C_{\rm abs}}

\newcommand \CpolPSA    {C_{\rm pol,PSA}}

\newcommand \falign     {f_{\rm align}}
\newcommand \PhiPSA     {\Phi_{\rm PSA}}
\newcommand \Qabs       {Q_{\rm abs}}
\newcommand \Qext       {Q_{\rm ext}}
\newcommand \Qextran    {Q_{\rm ext,ran}}

\newcommand \QpolPSA    {Q_{\rm pol,PSA}}
\newcommand \Qran       {Q_{\rm ran}}

\newcommand \Vsolid     {V_{\rm solid}}
\newcommand \rhosolid   {\rho_{\rm solid}}

\newcommand \website    {\url{http://www.astro.princeton.edu/~draine/Draine_2024b_suppmat.html}}

\countdef\decade=200
\advance\decade by \year
\countdef\hours=201
\advance\hours by \time
\divide\hours by 60
\countdef\mins=202
\advance\mins by \hours
\multiply\mins by 60
\multiply\hours by 100
\countdef\miltime=203
\advance\miltime by \hours
\advance\miltime by \time
\advance\miltime by -\mins
\renewcommand\today{\number\decade.\number\month.\number\day.\number\miltime}
\begin{document}

\title{%
        \vspace*{-2.0em}
        {\normalsize\rm {\it Astrophysical Journal}, accepted}\\ 
        \vspace*{1.0em}
        {\bf Sensitivity of Polarization to Grain Shape: II. Aggregates}
	}

\author[0000-0002-0846-936X]{B.~T.~Draine}
\affiliation{Dept.\ of Astrophysical Sciences,
  Princeton University, Princeton, NJ 08544, USA}
\affiliation{Institute for Advanced Study,
  Princeton, NJ 08540, USA}

\email{draine@astro.princeton.edu}

\begin{abstract}
A previous study (Paper I) investigated the polarization properties of
a variety of simple convex grain shapes, some of which were found to
be consistent with the observed polarization properties of
interstellar dust from far-ultraviolet to far-infrared.  Here we study
the optical properties of 45 non-convex shapes, all aggregates of $N$
equal-sized spheres.  We consider $N=2$, $N=3$, and $N=256$ random
aggregates obtained from 3 different aggregation schemes.  We also
consider ``trimmed'' $N=256$ aggregates obtained by systematically
trimming initially random aggregates to increase either flattening or
elongation.  The ``macroporosities'' of the studied aggregates range
from $\poromacro=0.18$ (for the $N=2$ bisphere) to $\poromacro\approx
0.85$ (for the $N=256$ ``BA'' aggregates).  The only aggregates
consistent with observations of starlight polarization and polarized
thermal emission are shapes that have been trimmed to increase their
asymmetry.  If interstellar grains are high-porosity aggregates, there
must be processes causing extreme elongation or flattening; if not,
interstellar grains must be dominated by fairly compact structures,
with at most moderate porosities.  The ratio of polarization in the
10$\micron$ silicate feature to starlight polarization in the optical
is shown to be insensitive to porosity and shape.  X-ray scattering
may be the best tool to determine the porosity of interstellar grains.
We propose that modest porosities of interstellar grains could be the
result of ``photolytic densification''.  \added{High polarization
  fractions observed in some Class-0 cores require processes to reduce
  porosities and/or increase asymmetries of aggregates in dense
  regions.}
\end{abstract}


\section{Introduction
         \label{sec:intro}}

Interstellar dust obscures and reddens stars, emission nebulae, and
active galactic nuclei, and adds ``foreground'' emission to the cosmic
microwave background.  The need to correct for these effects has
provided motivation for continuing efforts to improve our
understanding of interstellar dust.

In addition, it is now recognized that interstellar dust directly
affects the dynamics of the interstellar medium, and thereby the
structure and evolution of star-forming galaxies.  The shielding
effects of dust permit molecules to survive; catalysis of molecular
hydrogen on grain surfaces initiates most interstellar chemistry;
photoelectrons from dust grains heat the interstellar medium;
recombination of ions and electrons on grain surfaces lowers the
degree of ionization in predominantly neutral regions; charged dust
grains help to couple neutral gas to the magnetic field; and radiation
pressure acting on dust can be dynamically important for interstellar
gas.

Dust is also valuable as a \emph{diagnostic} of the interstellar
medium -- the spectrum of infrared emission from the dust informs us
of the intensity of starlight heating the dust, the polarization of
starlight and the polarized thermal emission from dust reveals the
ordering of the interstellar magnetic field.  Finally, the abundance,
composition, and size distribution of dust grains have much to tell us
of the history of the interstellar medium within which grains are
grown, shattered, and sputtered.  Every grain is itself a historical
record, if only we knew how to read it.

Despite the astrophysical importance of interstellar dust, there
remain substantial uncertainties regarding the chemical composition
and morphology of the dust grains.

\begin{figure}
\begin{center}
\includegraphics[angle=0,height=6.0cm,
                 clip=true,trim=0.0cm 0.0cm 0.0cm 0.0cm]
{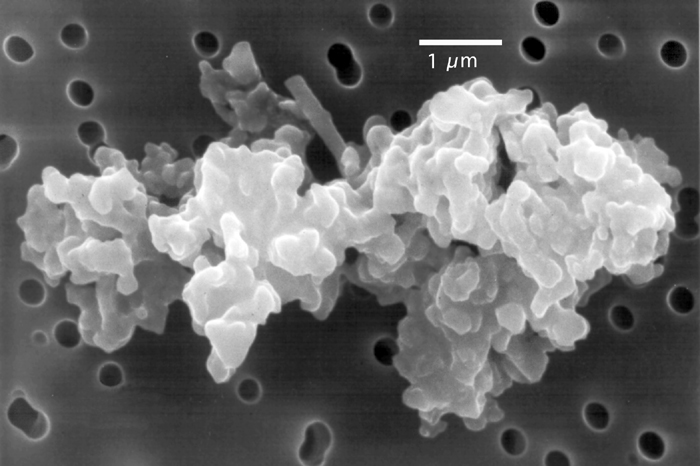}
\includegraphics[angle=0,height=6.0cm,
                 clip=true,trim=0.0cm 0.0cm 0.0cm 0.0cm]
{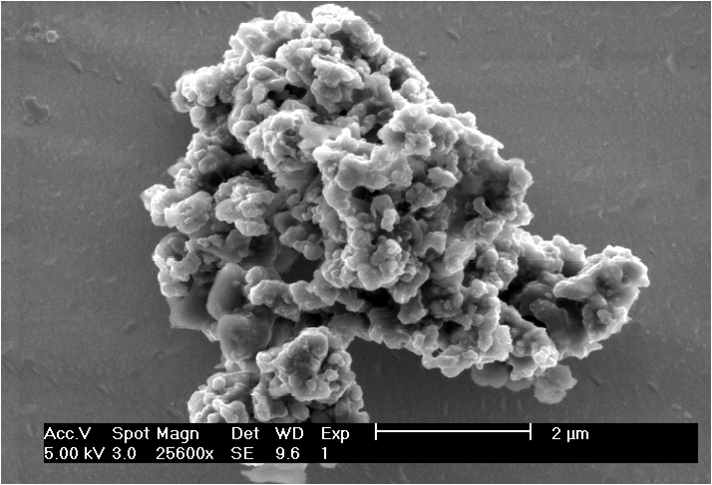}
\caption{\label{fig:IDP}\footnotesize Two porous-chondrite
  interplanetary dust particles.  Left: Credit Don Brownlee
  (U.\ Washington).  Right: Credit Nicole Spring (U.\ Manchester) and
  Henner Busemann (ETH Zurich).}
\end{center}
\end{figure}

It is often suggested that interstellar grains are high-porosity
structures -- sometimes characterized as ``fluffy''-- formed by
aggregation of smaller particles.  Interplanetary dust particles
(IDPs) -- thought to originate in comets, and collected in the
stratosphere -- often have this geometry \citep{Bradley_2003b}.  The
most common type of IDPs linked to short-period comets are the
``anhydrous chondritic-porous IDPs'' (see Figure \ref{fig:IDP}).
These have a substantial silicate component, a carbonaceous matrix,
and a porous structure \citep{Keller+Messenger_2005}.  The IDPs in
Figure \ref{fig:IDP} are much larger than the typical submicron
interstellar dust particle, but demonstrate that high-porosity
aggregate structures are present in some astrophysical settings.

High porosity (vacuum fraction) has also been invoked to explain
polarized scattered light from particles in the AU Mic debris disk;
these particles may resemble the IDPs in our solar system.
\citet{Graham+Kalas+Matthews_2007} argued for porosities as high as
$\poro=0.91-0.94$; \citet{Shen+Draine+Johnson_2009} later showed that
light scattering by the AU Mic debris disk is consistent with more
modest grain porosities, $\poro\approx 0.6$.

The emission spectra of protoplanetary disks have been interpreted as
indicative of grain growth to mm or even cm sizes
\citep{Beckwith+Sargent_1991,Testi+Natta+Shepherd+Wilner_2003,Draine_2006a}.
Polarized emission from the HL Tau protoplanetary disk was modeled by
\citet{Zhang+Zhu+Ueda+etal_2023} using mm-sized particles with high
porosities, $\poro \approx 0.70-0.97$.

High porosities have also been proposed for submicron-sized
interstellar grains.  \citet{Mathis_1996} suggested that porosity
could alleviate perceived stress between the observed extinction of
starlight, and the available abundances of dust-forming elements such
as C and Si, the idea being that the increased surface area of fluffy
grains might result in increased exinction cross section per unit
mass.  However, \citet{Dwek_1997} and \citet{Li_2005} later concluded
that increasing porosity doesn't actually reduce the mass in grains
required to account for the observed extinction.

High porosities have been invoked to help explain the relatively
high far-infrared (FIR) and submm opacities of interstellar dust.
\citet{Mathis+Whiffen_1989} argued that interstellar grains with sizes
$\gtsim 0.01\micron$ are composites, with porosity $\poro \approx
0.8$.  \citet{Voshchinnikov+Ilin+Henning+Dubkova_2006} proposed models
with porosities ranging from $\poro=0.90$ to $0.98$.
\citet{Hirashita+Ilin+Pagani+Lefevre_2021} argued that interstellar
particles larger than $\sim$$0.1\micron$ should have porosities
$\poro>0.7$.

Despite the numerous papers arguing that interstellar grains have high
porosity, it is not clear that such high-porosity grains are
consistent with the observed properties of interstellar dust.
\citet{Heng+Draine_2009} argued that the X-ray scattering halo
observed around GX13+1 required \emph{low} porosities $\poro\ltsim
0.55$ to match the angular profile of the halo.

In the present work we ask whether high-porosity grains are consistent
with the observed polarizing properties of interstellar dust.  The
polarizing efficiency of interstellar dust is observed to be quite
high; this strongly constrains the shapes of interstellar dust grains
\citep{Rogers+Martin_1979,
       Kim+Martin_1995b,
       Draine+Fraisse_2009,
       Fanciullo+Guillet+Boulanger+Jones_2017,
       Draine+Hensley_2021c,
       Hensley+Draine_2023,
       Ysard+Jones+Guillet+etal_2024}.
Paper I \citep{Draine_2024a} considered a variety of simple convex
shapes.  Some of the considered shapes were found to be consistent
with the observed polarization of starlight and polarized FIR
emission, whereas certain other shapes could be ruled out.

The present work extends this study to a variety of non-convex grain
shapes formed by aggregation of spheres, resulting in irregular
structures with substantial ``macroporosity''.  Using the discrete
dipole approximation (DDA), we calculate scattering and absorption
cross sections for complex grain geometries over a wavelength range
running from the far-ultraviolet (FUV) to the FIR.  The polarizing
properties of aligned grains with these shapes are compared to the
observed polarization of starlight and polarized FIR emission from
aligned dust grains.  We find that many of the considered porous
geometries can be ruled out because they are incapable of reproducing
the polarizations observed on some interstellar sightlines \emph{even
if the $a\gtsim0.1\micron$ grains are in ``perfect spinning
alignment'' (PSA)}.  We conclude that the 0.1-0.3$\micron$ grains
primarily responsible for the observed polarization of starlight are
not produced by random aggregation alone.  However, we show that
porous aggregates are allowed, provided that they are ``trimmed'' to
have extreme flattenings or elongations.  Examples of such aggregates
are presented.

Section \ref{sec:porosity} discusses quantitative measures of porosity
and overall asymmetry of individual grains, and section
\ref{sec:targets} presents the considered grain geometries.  Section
\ref{sec:DDA} describes the methodology for calculating optical cross
sections; calculated cross sections are presented in Section
\ref{sec:Q}.  Section \ref{sec:lambdap} discusses the effective
wavelength $\lambdap$ for polarization as a function of grain size and
shape, as well as the dimensionless width $\sigmap$ of the
polarization profile. For each considered shape, the starlight
polarization efficiency integral $\PhiPSA$ is calculated in Section
\ref{sec:poleffint}.  Section \ref{sec:infrared} presents the
polarizing properties of the grains near the $10\micron$ silicate
feature, and in the FIR.

The results are discussed in Section \ref{sec:discuss},
and summarized in Section \ref{sec:summary}.  

\medskip


\section{\label{sec:porosity}
         Asymmetry and Porosity}

The optical properties of a grain depend on composition, size, and --
particularly for polarized extinction and emission -- on the grain
shape.  The size of an irregular particle can be characterized by the
``mass-equivalent'' effective radius
\beq \label{eq:aeffdef}
\aeff \equiv \left(\frac{3 M}{4\pi\rhosolid}\right)^{1/3}
~~~,
\eeq
where $M$ is the mass, and $\rhosolid$ is 
the mass density of the solid material.

The grain has moments of inertia $I_1\geq I_2\geq I_3$ for rotation
around each of the principal axes, $\bahat_1,\bahat_2,\bahat_3$.  Let
\beq
\alpha_j\equiv\frac{I_j}{0.4 M \aeff^2}
~~~,~j=1,2,3
~~~.
\eeq
A sphere has $\alpha_1=\alpha_2=\alpha_3=1$.  

Let $\br=x\bahat_1+y\bahat_2+z\bahat_3$, with the center of mass at
$\br=0$.  For any given shape, the dimensionless $\alpha_j$ can be
used to define an asymmetry parameter
\beq \label{eq:Anew}
\Asymm \equiv 
\left(
\frac{\langle y^2\rangle+\langle z^2\rangle}{2\langle x^2\rangle}
\right)^{1/2} =
\left(
\frac{\alpha_1}{\alpha_2+\alpha_3-\alpha_1}
\right)^{1/2}
~~.
\eeq
A sphere has $\Asymm=1$; elongated or flattened shapes have
$\Asymm>1$.\footnote{
Note that $\Asymm$ defined by Eq.\ (\ref{eq:Anew}) differs from the
parameter $\Rone=\alpha_1/(\alpha_2\alpha_3)^{1/2}$ defined in Paper I
\citep[][Eq.\ 2]{Draine_2024a}.  We find that $\Asymm$ defined by
(\ref{eq:Anew}) is a better predictor of polarizing properties of
irregular grains than $\Rone$.}

We distinguish between ``microporosity'' $\poromicro$ -- due to
voids and defects on atomic scales\footnote{
Normal polycrystalline materials have some level of microporosity.  If
the ideal mineral has density $\rho_c$, but a solid sample with the
same chemical composition has density $\rhosolid$, we define the
microporosity to be $\poromicro=1-\rhosolid/\rho_c$.  For example, in
both graphite and glassy carbon $\sim$100\% of the C atoms are thought to be
in sites with $sp^2$ bonding \citep{Robertson_1986}; however,
crystalline graphite has $\rho_c=2.26\gm\cm^{-3}$, whereas ``glassy
carbon'' samples have $\rhosolid=1.3-1.55\gm\cm^{-3}$, corresponding
to $\poromicro=0.31 - 0.43$.
``Pyrolitic graphite'' has $\rhosolid=1.22-2.22\gm\cm^{-3}$
for production temperatures $1900-2400\K$ 
\citep{Yajima+Satow+Hirai_1965}, corresponding to $\poromicro=0.02-0.46$.
}
 -- and ``macroporosity''
$\poromacro$, characterizing the presence of voids on scales much
larger than the interatomic separation.

A number of different approaches have been taken to quantify the
macroporosity of irregular grains.
\citet{Mukai+Ishimoto+Kozasa+etal_1992} set ${\cal P}_{\rm
  M92}=1-\Vsolid/V_{\rm M92}$, where $V_{\rm M92}$ is the volume of a
sphere with the same radius of gyration.  Note that with this
definition a solid spheroid or ellipsoid, by virtue of being
nonspherical, would have ${\cal P}_{\rm M92}>0$, even though the
actual macroporosity is zero.  \citet{Ossenkopf_1993} used the
angle-averaged projected area to define a ``fluffiness parameter''.

In the present work we employ the macroporosity defined by
\citet{Shen+Draine+Johnson_2008}:
\beq \label{eq:poromacro}
\poromacro \equiv 1-\frac{\Vsolid}{V_{\rm S08}}
\equiv 1- \frac{1}
{\left[
 (\alpha_2+\alpha_3-\alpha_1)
 (\alpha_3+\alpha_1-\alpha_2)
 (\alpha_1+\alpha_2-\alpha_3)
\right]^{1/2}}
~~~,
\eeq
where $V_{\rm S08}$ is the volume of an ellipsoid with the same
$I_j/M$ as the grain.  With this definition, solid spheres, spheroids,
and ellipsoids have $\poromacro=0$, as desired.\footnote{
  Note, however, that non-ellipsoidal solids have $\poromacro>0$; for
  example, cylinders have $\poromacro=0.070$, and rectangular prisms
  have $\poromacro=0.112$.}

Ellipsoidal aggregates formed from $N$ close-packed single-size
spheres have $\poromacro=1-\pi/(3\sqrt{2}) = 0.2595...$ for
$N\rightarrow\infty$ \citep{Gauss_1831w}.  Depending on the growth
process, random aggregates can have values of $\poromacro$ approaching
unity [e.g., random aggregates formed by ``diffusion limited
  aggregation''\citep{Witten+Cates_1986}].

If the grain material has ``microporosity'' $\poromicro$ and the
shape has macroporosity $\poromacro$, the total porosity is
\beq 
\poro = 1-\left(1-\poromicro\right)\left(1-\poromacro\right)
~~~.
\eeq

\medskip

\section{\label{sec:targets}
         Aggregates}

\added{Aggregation of dust grains is expected to occur in the
  interstellar medium, and to play an important role in the evolution
  of distribution of grain sizes.  Aggregation is also expected to result
  in porous grains.

Grain models that reproduce the observed interstellar extinction curve
typically have most of the grain mass in grains with $\aeff\gtsim
0.05\micron$,\footnote{
  E.g., the MRN size distribution \citep{Mathis+Rumpl+Nordsieck_1977}
  has 64\% of the mass in grains with $0.05\micron < a <
  0.25\micron$.}
but with smaller grains still accounting for an appreciable fraction
of the mass, and greatly outnumbering the larger grains.  The larger
aggregates likely contain subunits (``monomers'') spanning a range of
sizes.

The objective of the present work is to study the polarization
properties of porous aggregates.  We do not consider that the details
of the aggregate geometry are realistic -- only that they span an
interesting range of porosities.  To this end, we investigate
aggregates of $N$ equal-sized spherical monomers.  We limit study to
aggregates of equal-size spheres for simplicity, and for computational
feasibility.  We study the cases of $N=2$ and $N=3$, but most of the
study concerns $N=256$ aggregates resulting from different aggregation
schemes, with macroporosities $\poromacro$ ranging from $\sim$0.53 to
$\sim$0.85 .

Aggregates incoporating a wide range of monomer sizes would be more
realistic than assuming equal-size monomers: the smaller particles may
tend to fill in the spaces between the larger particles, lowering the
porosity, although this will depend on the assumed dynamics of
aggregation.  For the results to be accurate, DDA calculations of
scattering and absorption by aggregates require that each monomer be
represented by a sufficient number of dipoles that the shape of its
surface is adequately emulated.  If the smallest monomers are allowed
to be very small (by volume) compared to the overall aggregate, the
overall number of dipoles required quickly becomes computationally
prohibitive.

While recognizing the greater realism of aggregates of polydisperse
monomers, we defer their investigation to future work, here limiting
study to aggregates of equal-size spheres.  }

\subsection{$N=2$ and $N=3$ Aggregates}


\newcommand{\figheight}{3.9cm}
\newcommand{\triml}{0.7cm}
\newcommand{\trimr}{1.5cm}
\newcommand{\trimb}{8.0cm}
\newcommand{\trimt}{2.5cm}
\newcommand{\negspace}{-0.3cm}
\newcommand{\midspace}{0.3cm}
\begin{figure}
\begin{center}

\includegraphics[angle=0,height=\figheight,
                 clip=true,trim={\triml} {\trimb} {\trimr} \trimt]
{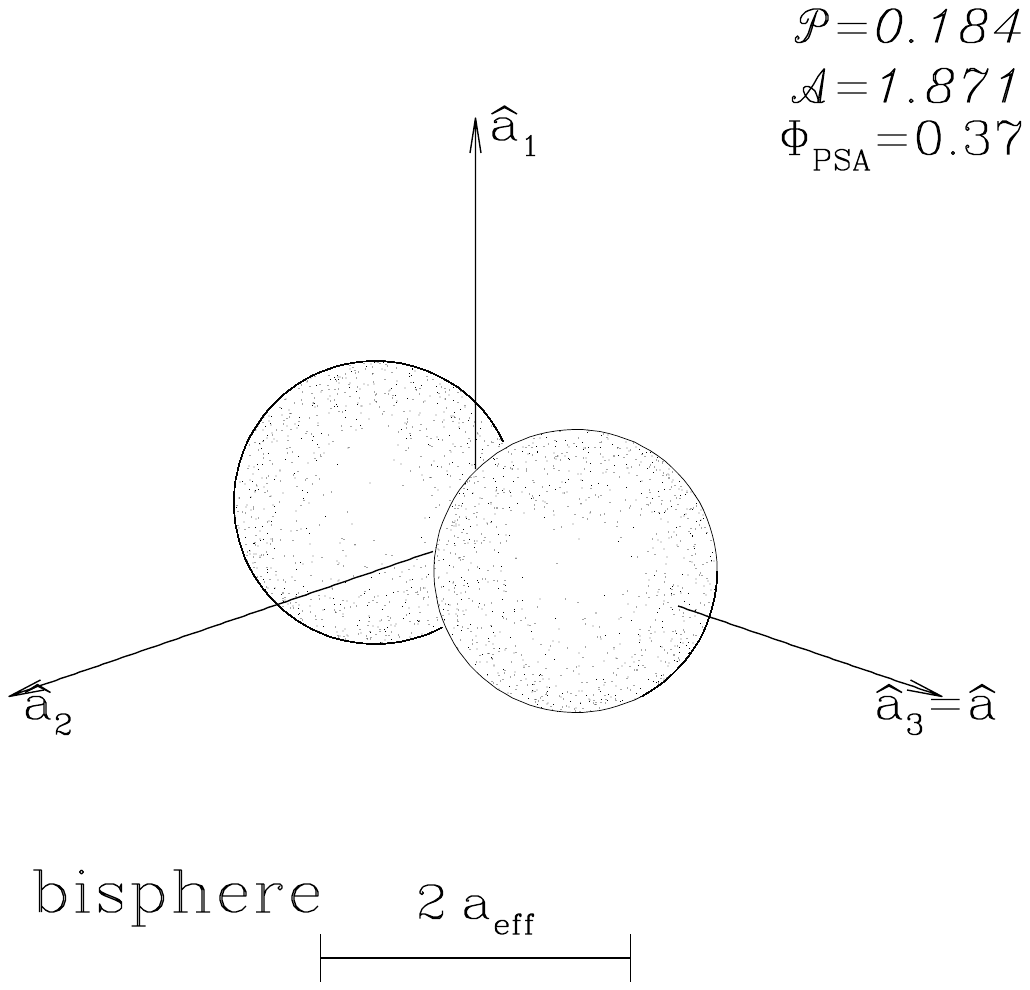}
\includegraphics[angle=0,height=\figheight,
                 clip=true,trim={\triml} {\trimb} {\trimr} \trimt]
{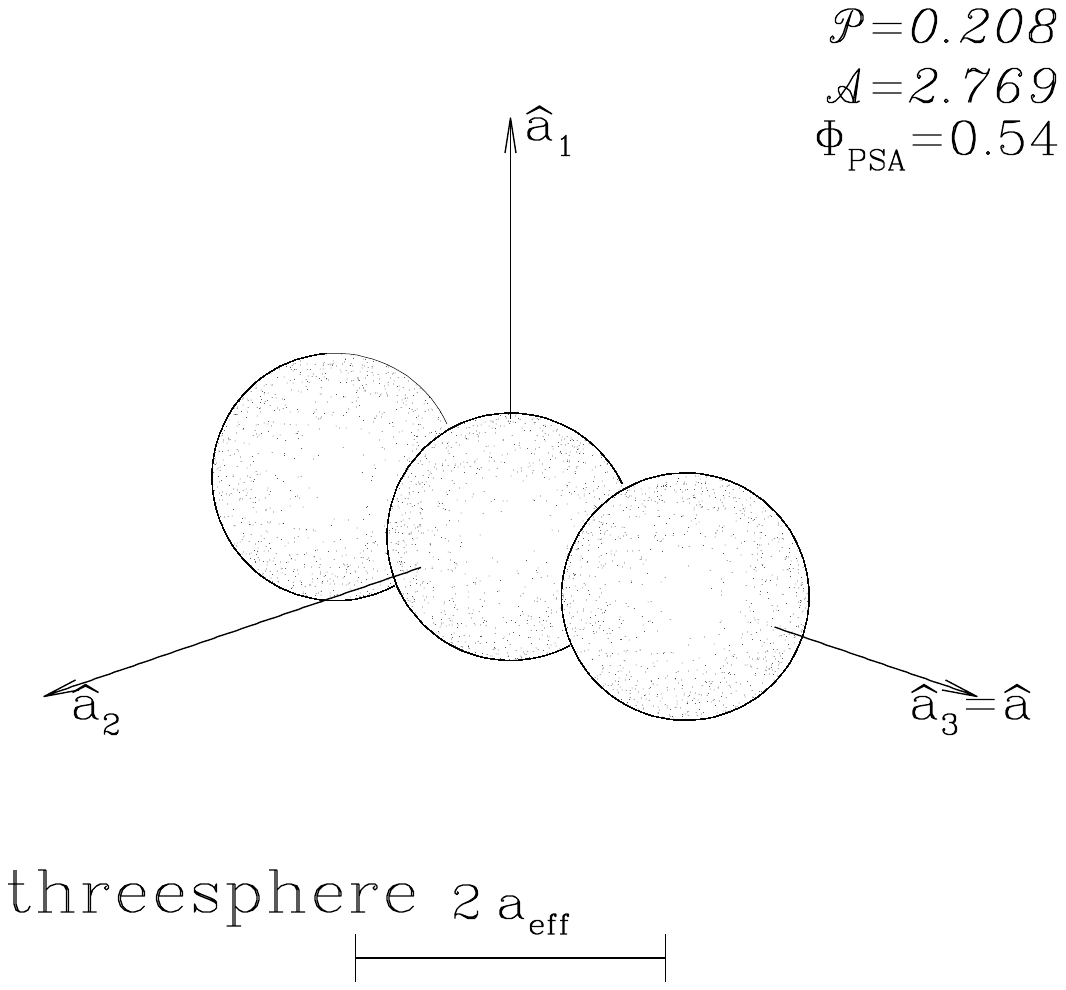}
\includegraphics[angle=0,height=\figheight,
                 clip=true,trim={\triml} {\trimb} {\trimr} \trimt]
{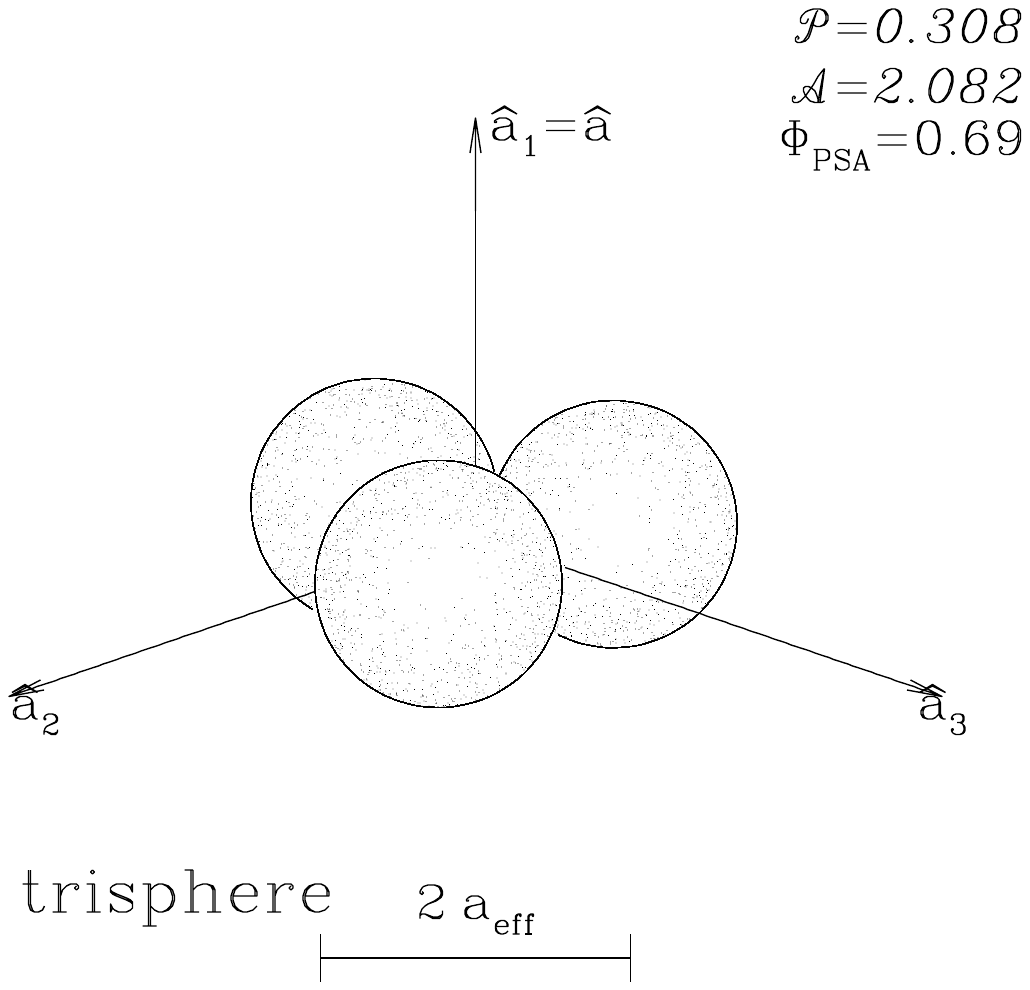}
\caption{\label{fig:shapes_23}\footnotesize Bisphere, threesphere, and
  trisphere geometries.  The scale bars show $2\aeff$, the diameter of
  an equal-volume sphere.  $\bahat_1$ is the principal axis of largest
  moment of inertia; grains are asssumed to spin around this
  axis.}
\end{center}
\end{figure}

There is only one structure consisting of two equal-sized touching
spheres:
\begin{itemize}
\item {\bf ``bisphere''}: 2 touching spheres, with 
$\poromacro=0.184$, and asymmetry factor
$\Asymm=1.871$.
\end{itemize}
We consider 2 examples of structures made up of 3 equal-size spheres:
\begin{itemize}
\item {\bf ``threesphere''}: 3 collinear touching spheres,
with $\poromacro=0.208$, and $\Asymm=2.769$.
\item {\bf ``trisphere''}: 3 touching spheres in close-packed
  geometry, with $\poromacro=0.308$, and $\Asymm=2.082$.
\end{itemize}
These simple shapes are shown in Figure \ref{fig:shapes_23}.
Geometric parameters are given in Table \ref{tab:geom}.

\subsection{$N=256$ Random Aggregates}

We also consider irregular structures comprised of $N=256$ equal-size
spheres, created following three different growth rules described by
\citet{Shen+Draine+Johnson_2008}:
A library of BA, BAM1, and BAM2 aggregates, including the examples
studied here, can be found at
\url{www.astro.princeton.edu/~draine/agglom.html}.

\begin{figure}
\begin{center}
\includegraphics[angle=0,height=\figheight,
                 clip=true,trim={\triml} {\trimb} {\trimr} \trimt]
{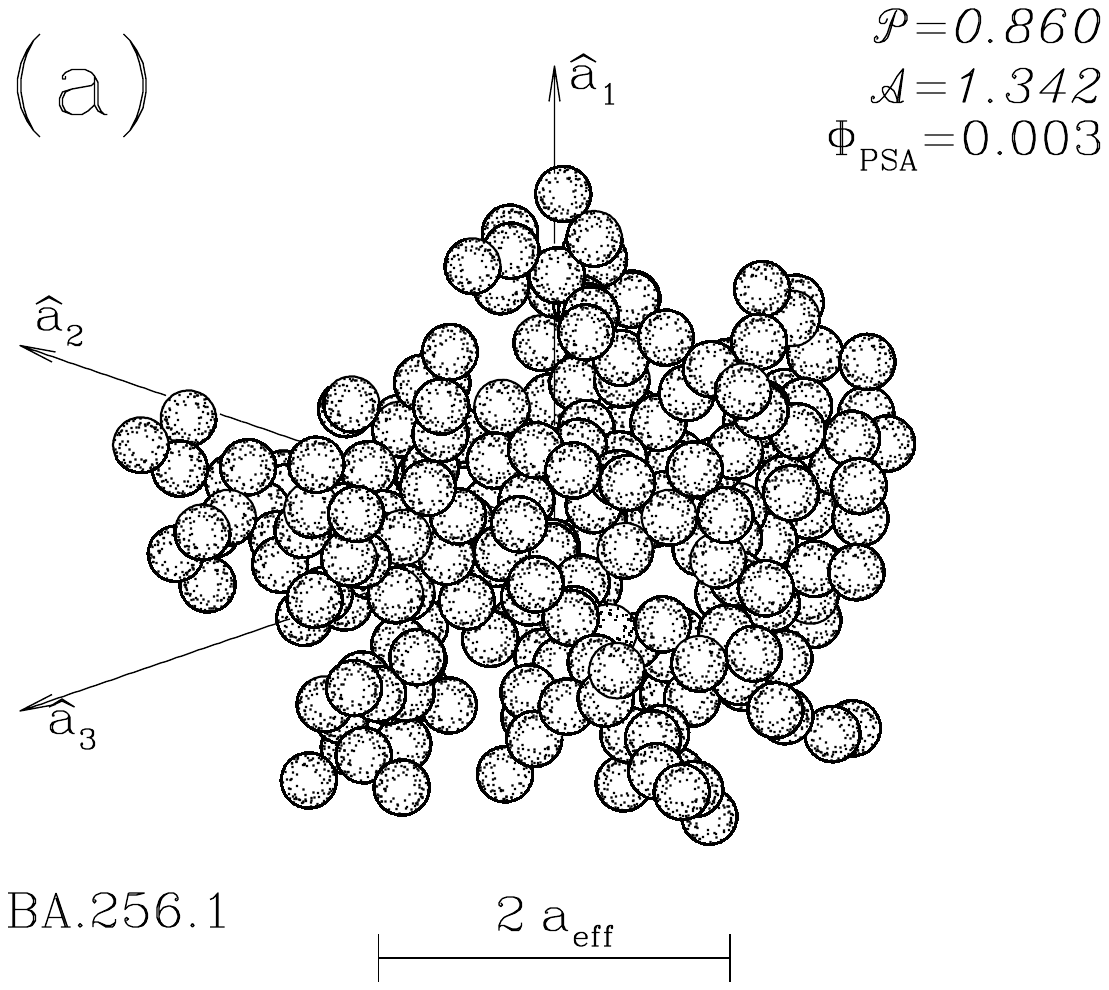}
\hspace*{\negspace}
\includegraphics[angle=0,height=\figheight,
                 clip=true,trim={\triml} {\trimb} {\trimr} \trimt]
{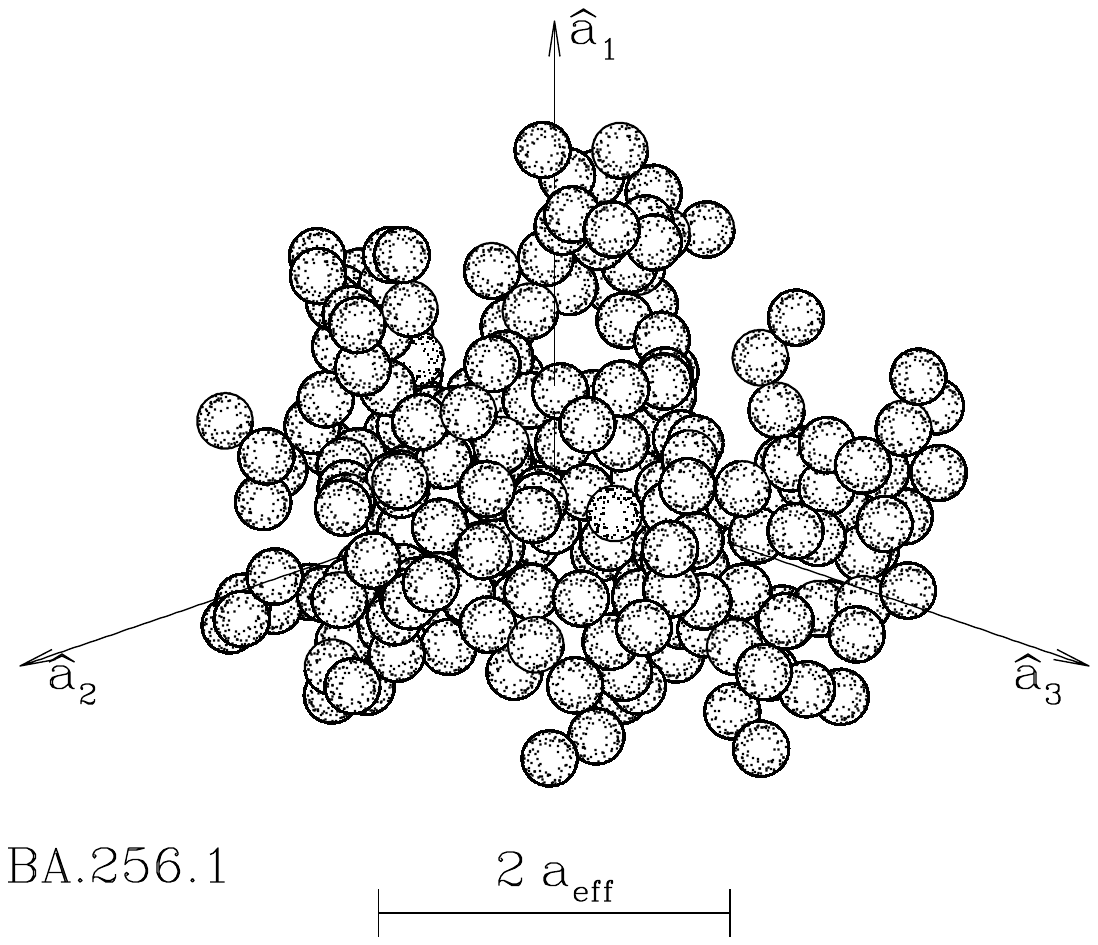}
\hspace*{\negspace}
\includegraphics[angle=0,height=\figheight,
                 clip=true,trim={\triml} {\trimb} {\trimr} \trimt]
{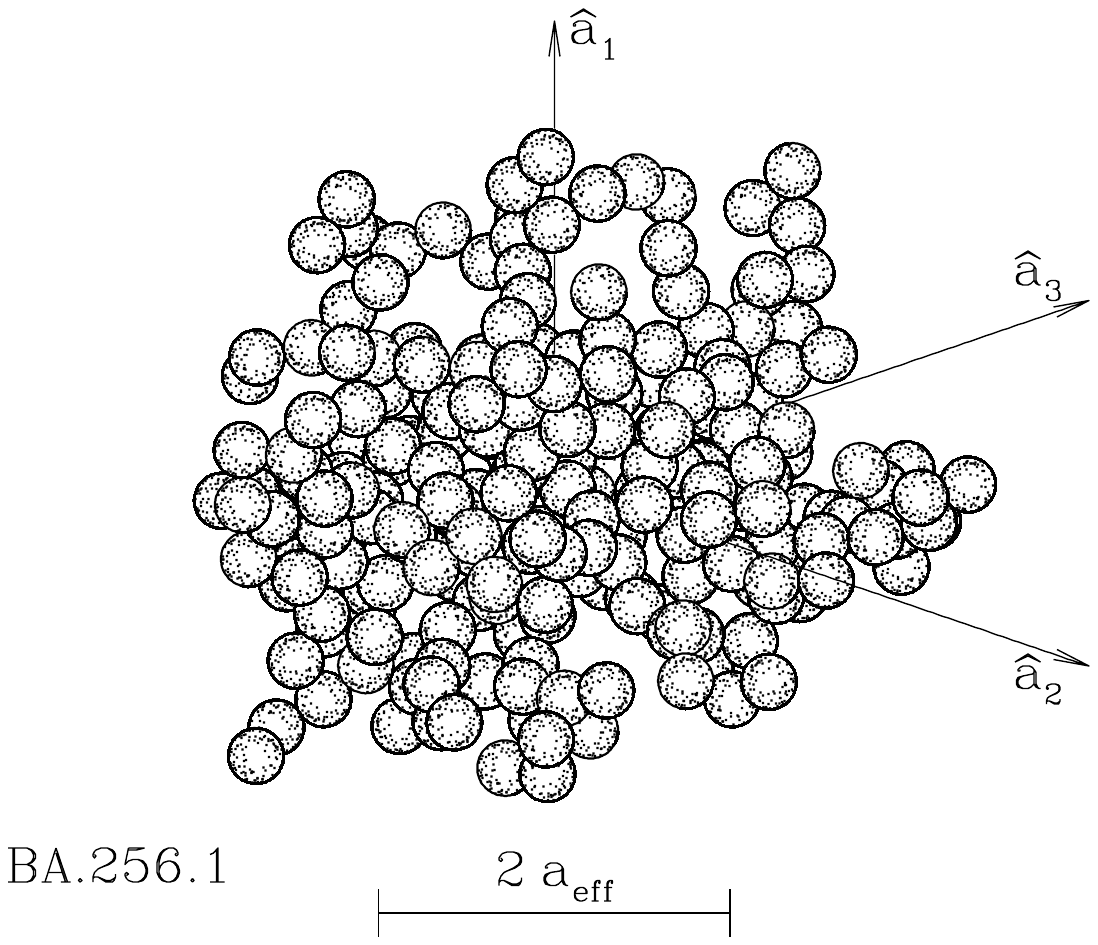}

\includegraphics[angle=0,height=\figheight,
                 clip=true,trim={\triml} {\trimb} {\trimr} \trimt]
{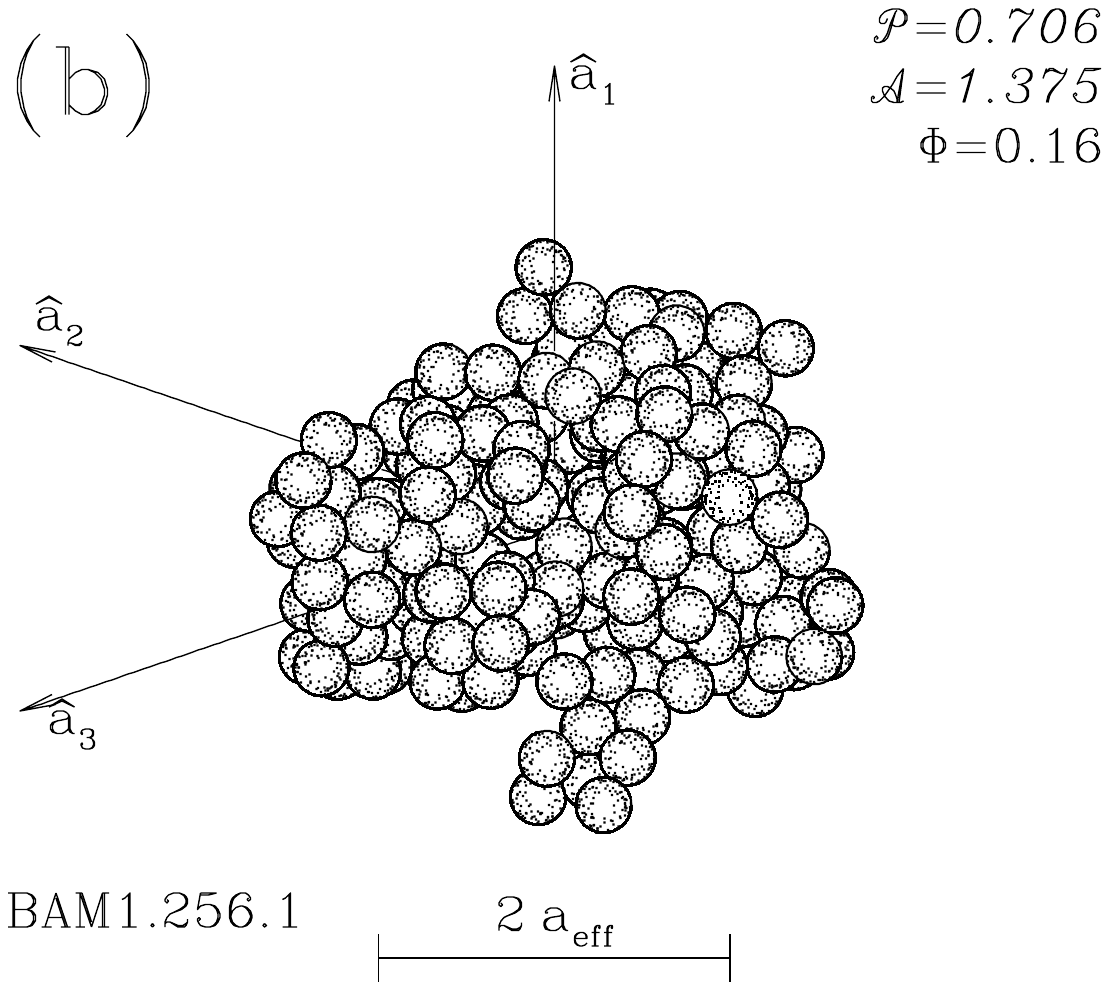}
\hspace*{\negspace}
\includegraphics[angle=0,height=\figheight,
                 clip=true,trim={\triml} {\trimb} {\trimr} \trimt]
{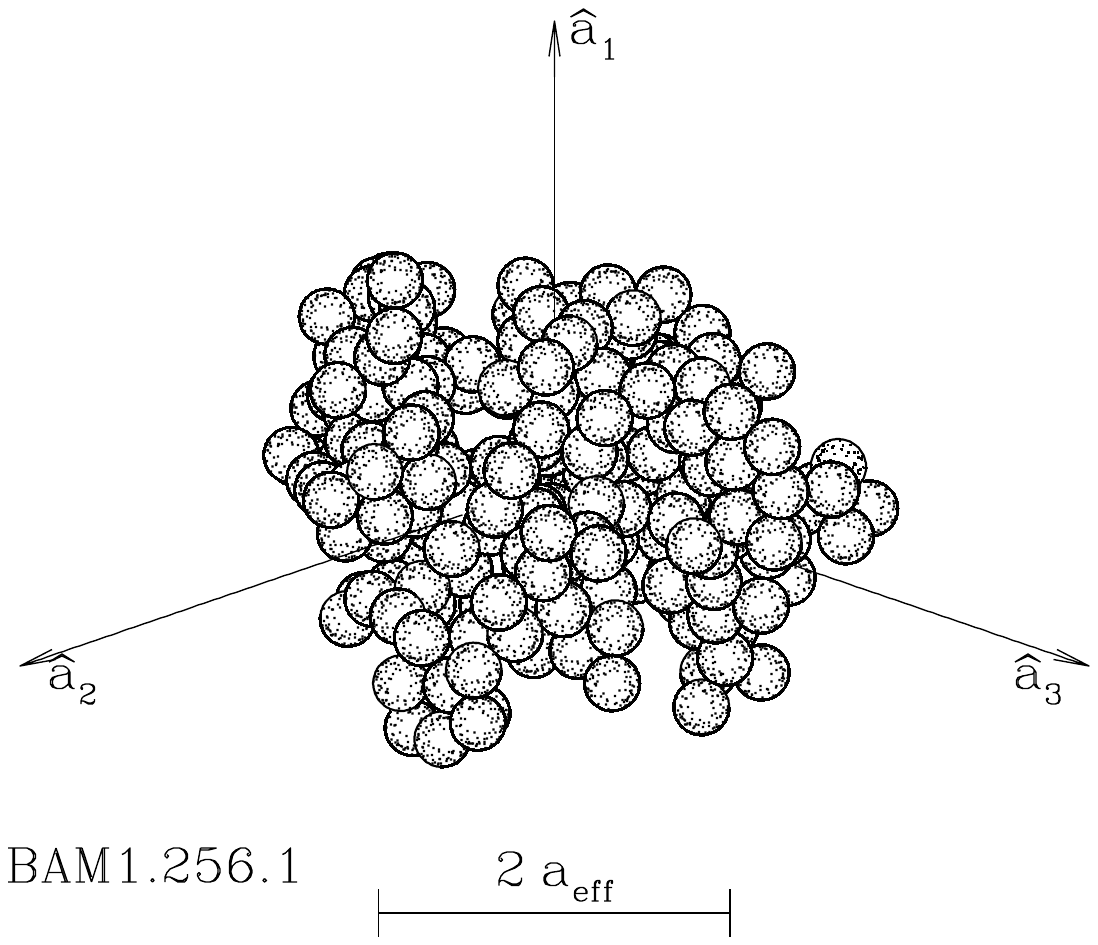}
\hspace*{\negspace}
\includegraphics[angle=0,height=\figheight,
                 clip=true,trim={\triml} {\trimb} {\trimr} \trimt]
{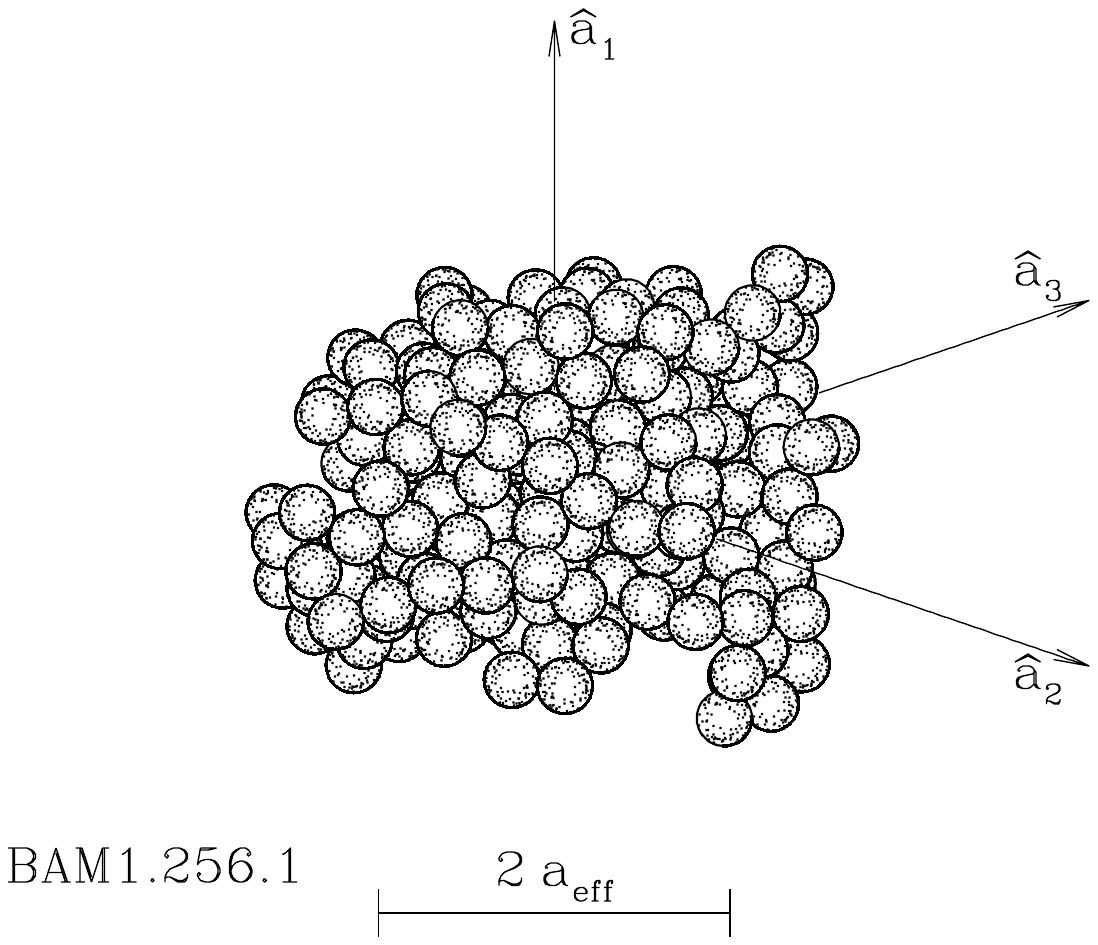}\\

\includegraphics[angle=0,height=\figheight,
                 clip=true,trim={\triml} {\trimb} {\trimr} \trimt]
{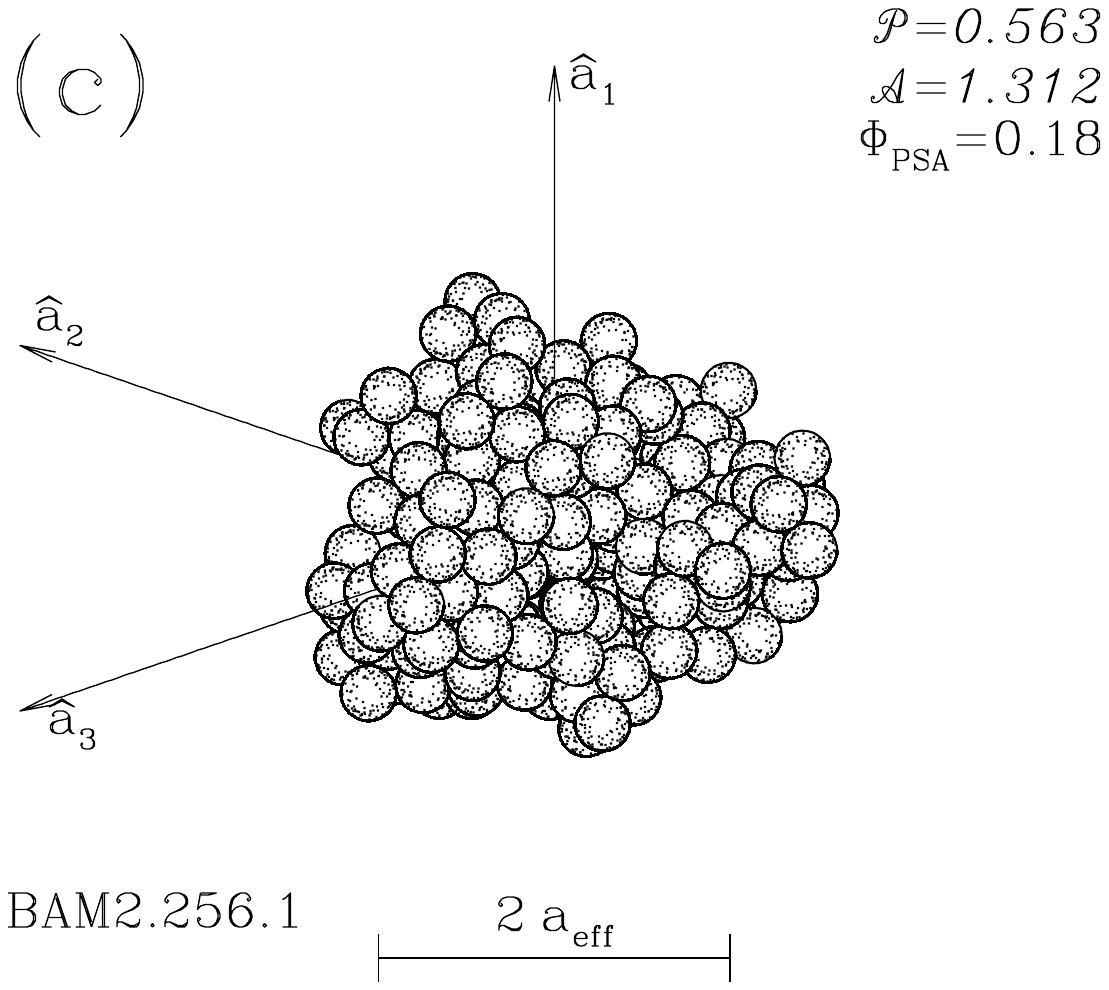}
\hspace*{\negspace}
\includegraphics[angle=0,height=\figheight,
                 clip=true,trim={\triml} {\trimb} {\trimr} \trimt]
{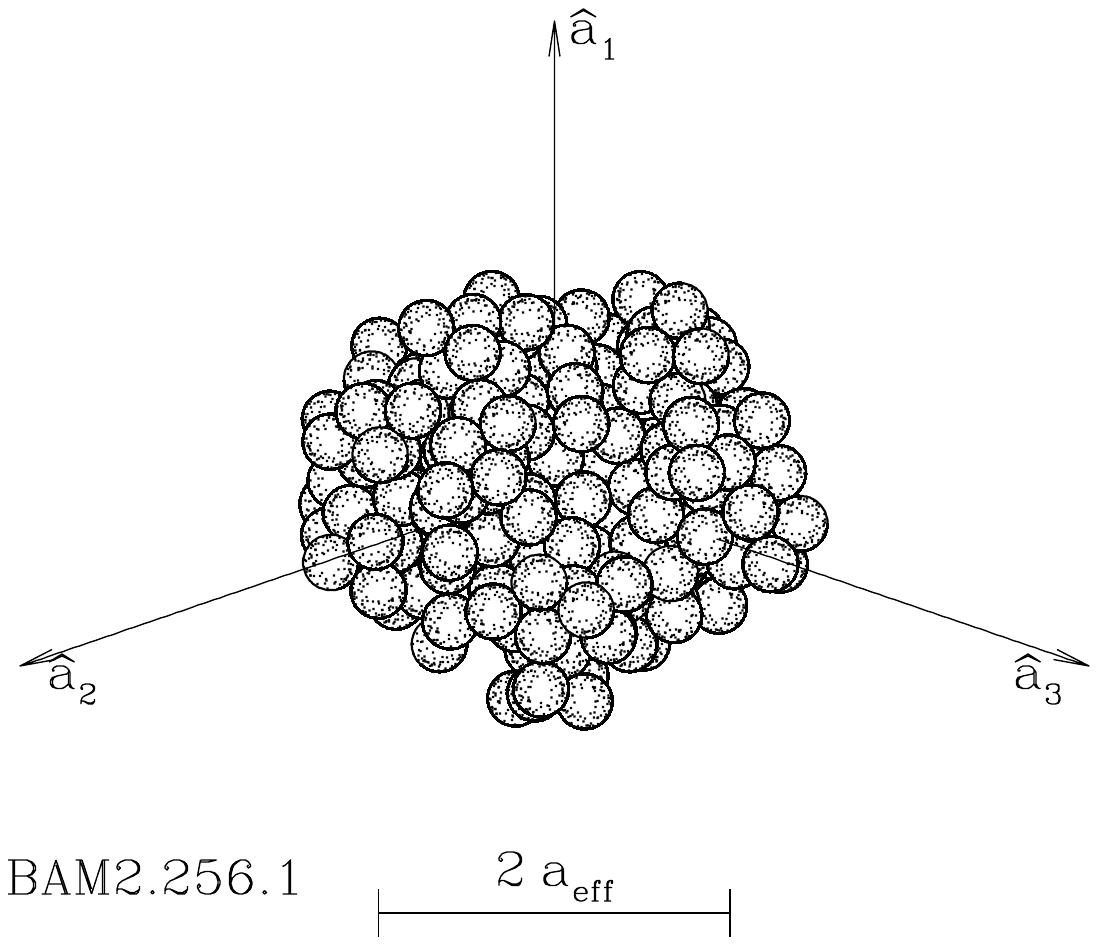}
\hspace*{\negspace}
\includegraphics[angle=0,height=\figheight,
                 clip=true,trim={\triml} {\trimb} {\trimr} \trimt]
{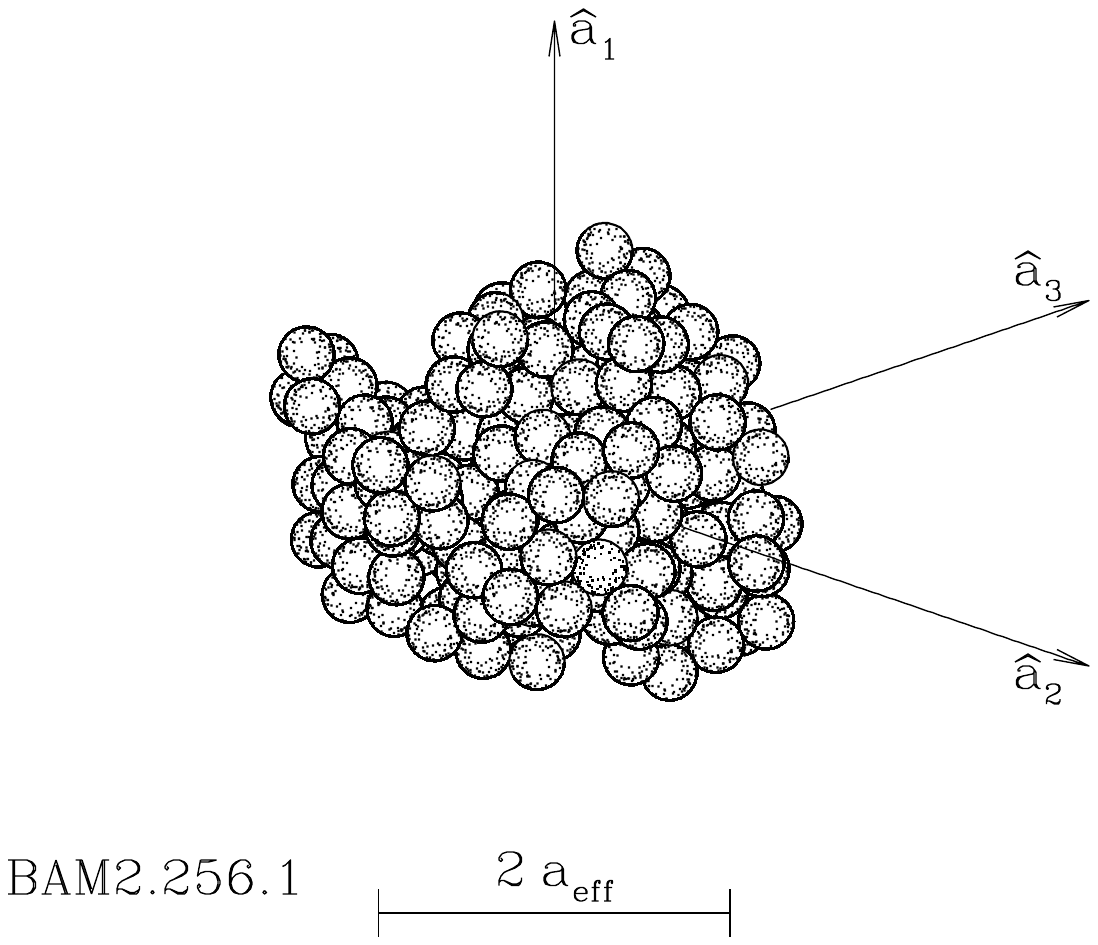}
\caption{\label{fig:shapes_babam1bam2}\footnotesize Random aggregates
  (a) BA.256.1, (b) BAM1.256.1, and (c) BAM2.256.1, each from 3
  viewing angles.$^{\ref{fn:website}}$ $\bahat_1$ is the principal
  axis of largest moment of inertia; grains are assumed to spin around
  this axis.  Macroporosity $\poromacro$ and asymmetry factor $\Asymm$
  are given.  Scale bars show $2\aeff$, the diameter of an
  equal-solid-volume sphere.}
\end{center}
\end{figure}

\begin{itemize}

\item {\bf BA} (``Ballistic Aggregation'', also known as
  ``Particle-Cluster Aggregation''): Spherical monomers arrive on
  random trajectories, and stick wherever they first touch.  This
  growth scheme creates high-porosity structures.  For $N=256$, BA
  aggregates have $\poromacro\approx0.847\pm0.013$ \citep[][the
    ``uncertainty'' is the standard deviation of the distribution of
    $\poromacro$]{Shen+Draine+Johnson_2008}.

\item {\bf BAM1} (Ballistic Aggregation with one Migration): Spherical
  monomers arrive on random trajectories.  For $N>2$: after first
  contact, the new arrival rolls to the nearest point where it can
  be in contact with a second sphere, and ``sticks'' there.  For
  $N=256$, BAM1 aggregates have $\poromacro\approx 0.739\pm0.018$
  \citep{Shen+Draine+Johnson_2008}.

\item {\bf BAM2} (Ballistic Aggregation with two Migrations):
  Spherical ``monomers'' arrive on random trajectories.  For $N>2$:
  after first contact, the new arrival rolls to the nearest point
  where it can be in contact with two other spheres.  If $N>3$, it
  then rolls to the nearest point where it can make contact with a
  third sphere, and ``sticks'' there.  For $N=256$, BAM2 aggregates
  have $\poromacro\approx 0.579\pm0.026$
  \citep{Shen+Draine+Johnson_2008}.

\end{itemize}
Figure \ref{fig:shapes_babam1bam2} shows one random realization of
each of these aggregate types.\footnote{
See \website\ for additional examples of each of these random
aggregate types, as well as extinction and polarization cross sections
for every example in Table \ref{tab:geom}.
\label{fn:website}}

\begin{figure}
\begin{center}
\includegraphics[angle=0,height=\figheight,
                 clip=true,trim={\triml} {\trimb} {\trimr} \trimt]
{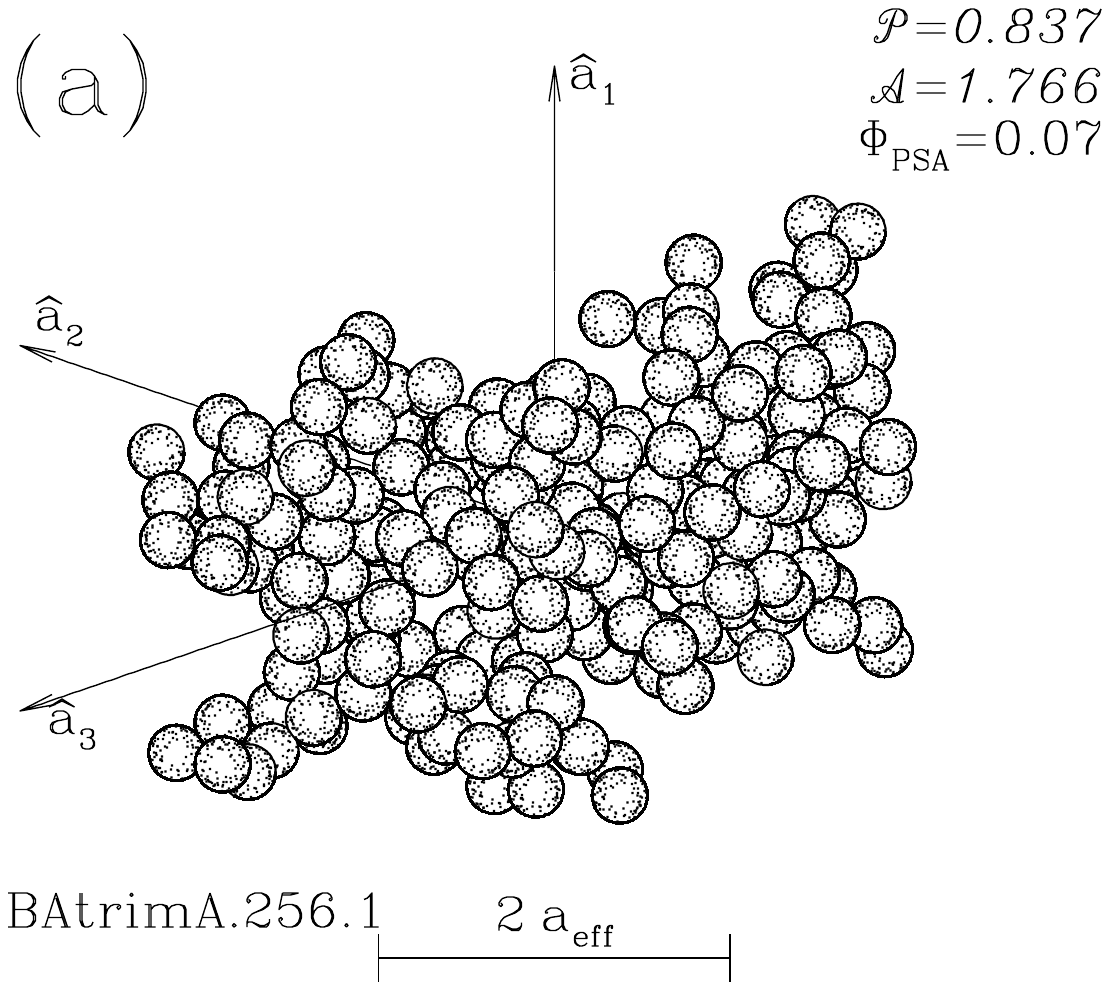}
\hspace*{\negspace}
\includegraphics[angle=0,height=\figheight,
                 clip=true,trim={\triml} {\trimb} {\trimr} \trimt]
{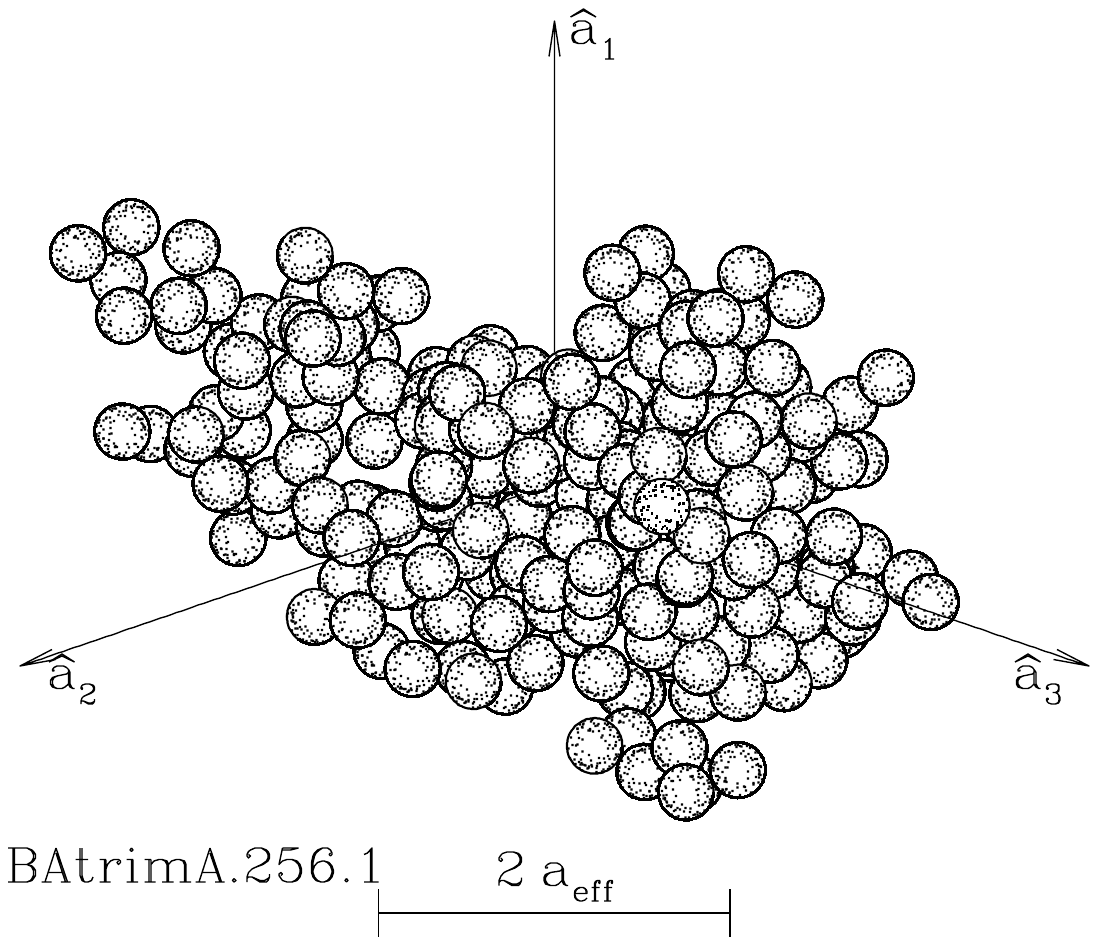}
\hspace*{\negspace}
\includegraphics[angle=0,height=\figheight,
                 clip=true,trim={\triml} {\trimb} {\trimr} \trimt]
{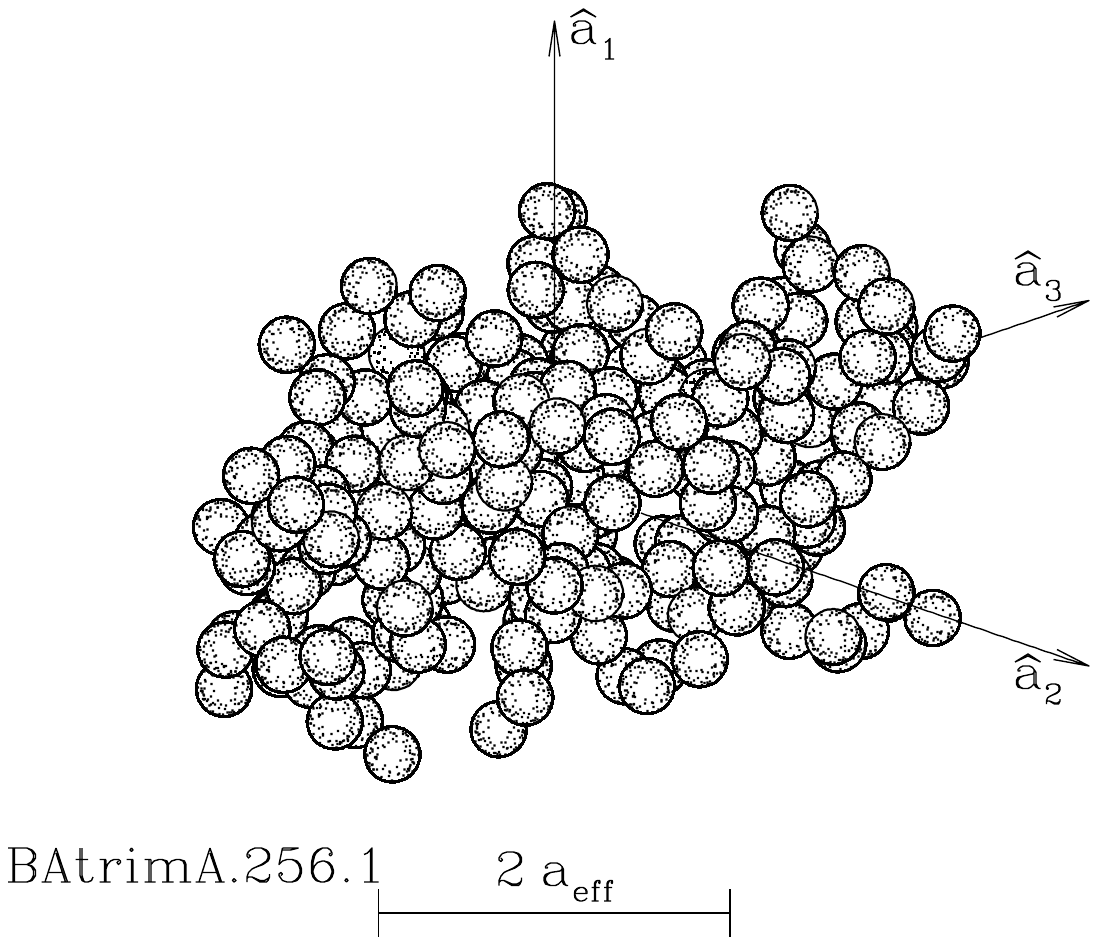}\\

\includegraphics[angle=0,height=\figheight,
                 clip=true,trim={\triml} {\trimb} {\trimr} \trimt]
{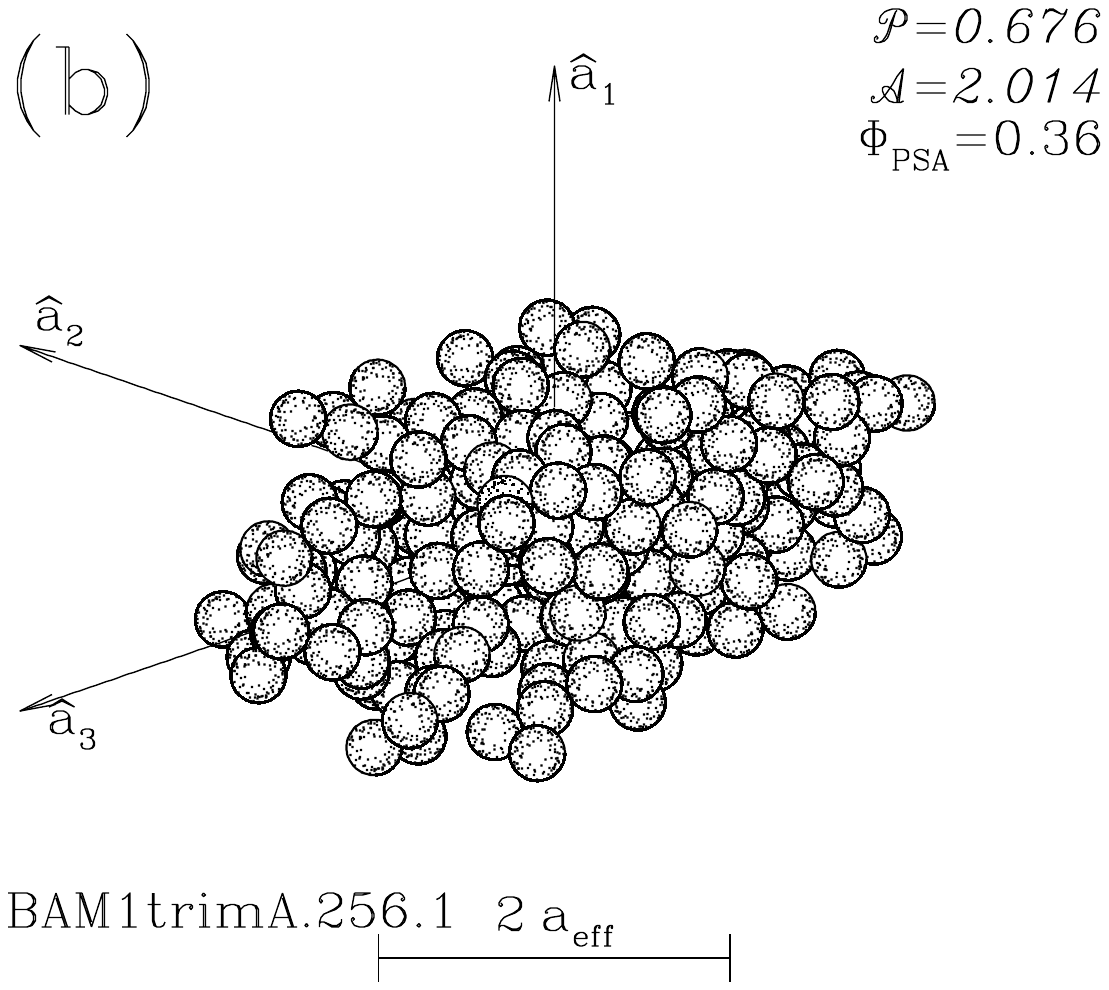}
\hspace*{\negspace}\includegraphics[angle=0,height=\figheight,
                 clip=true,trim={\triml} {\trimb} {\trimr} \trimt]
{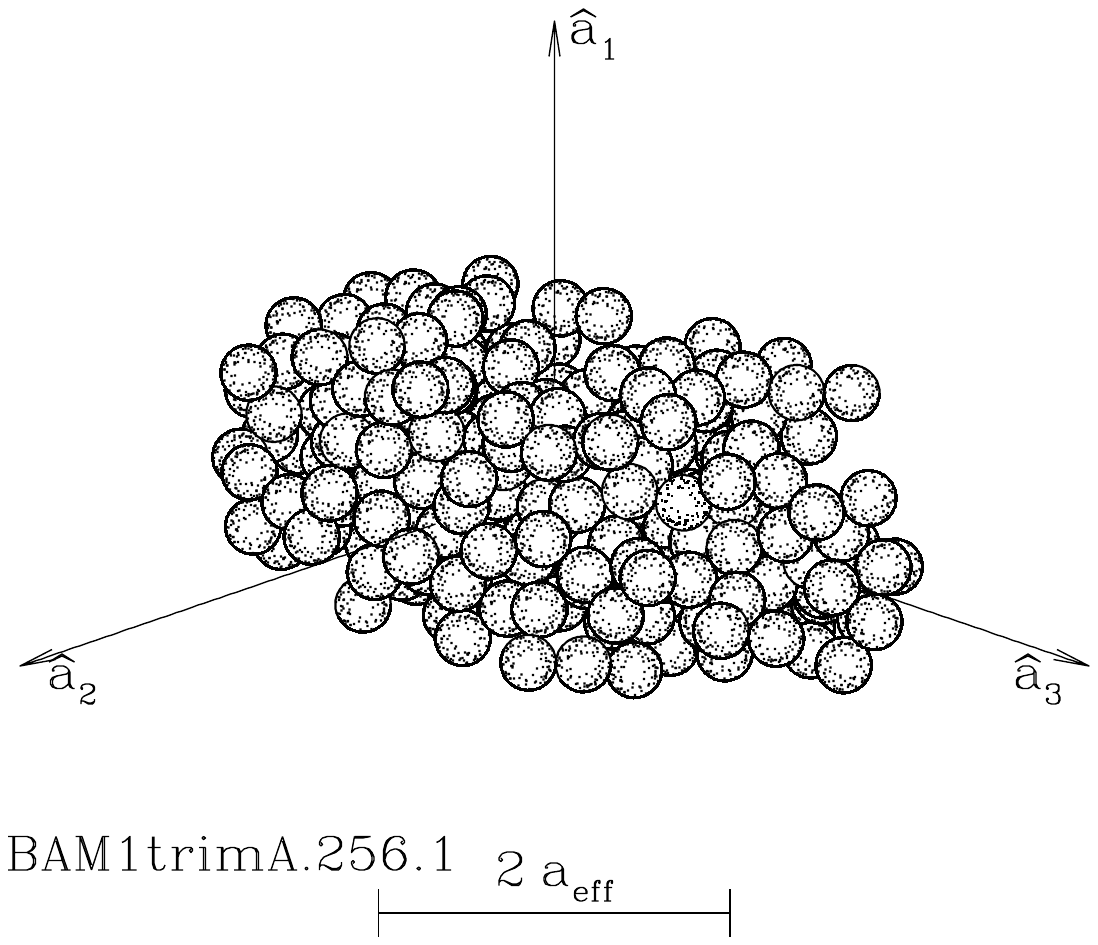}
\hspace*{\negspace}\includegraphics[angle=0,height=\figheight,
                 clip=true,trim={\triml} {\trimb} {\trimr} \trimt]
{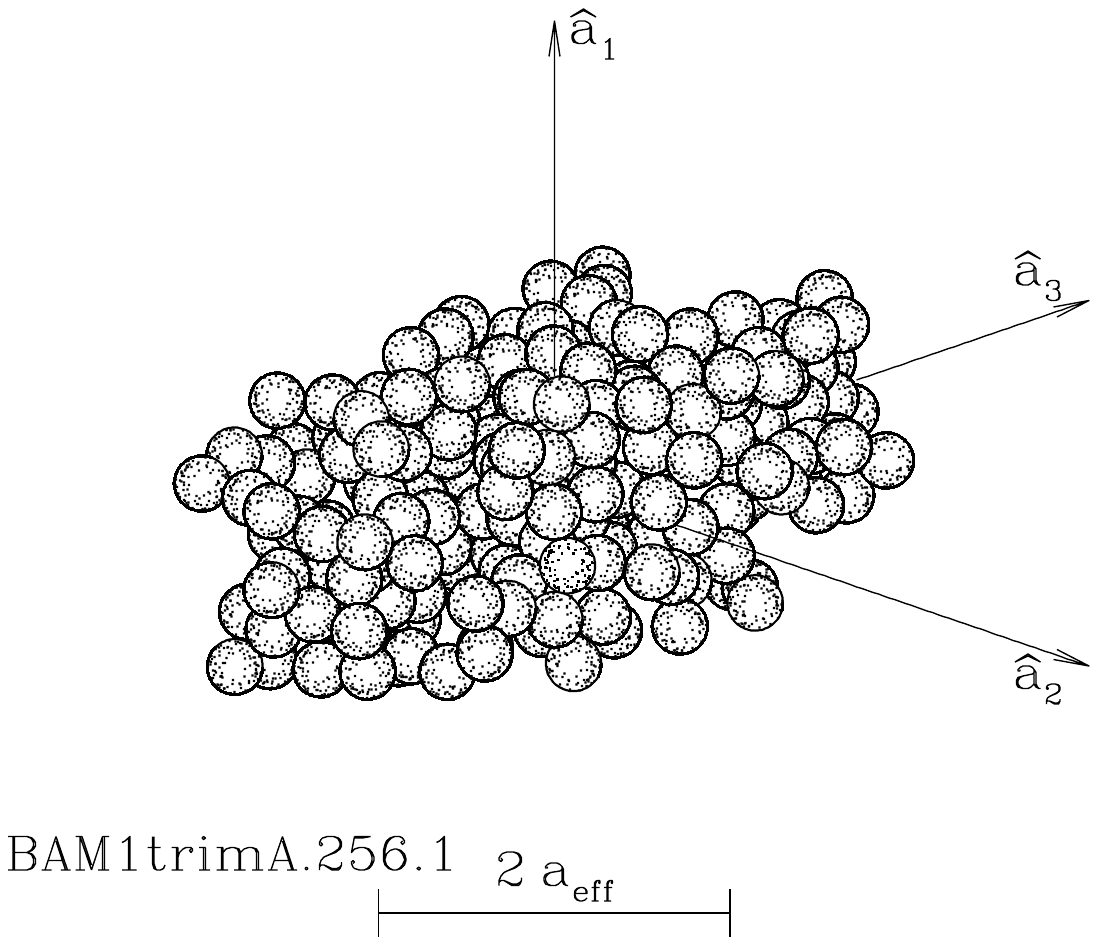}
\\

\includegraphics[angle=0,height=\figheight,
                 clip=true,trim={\triml} {\trimb} {\trimr} \trimt]
{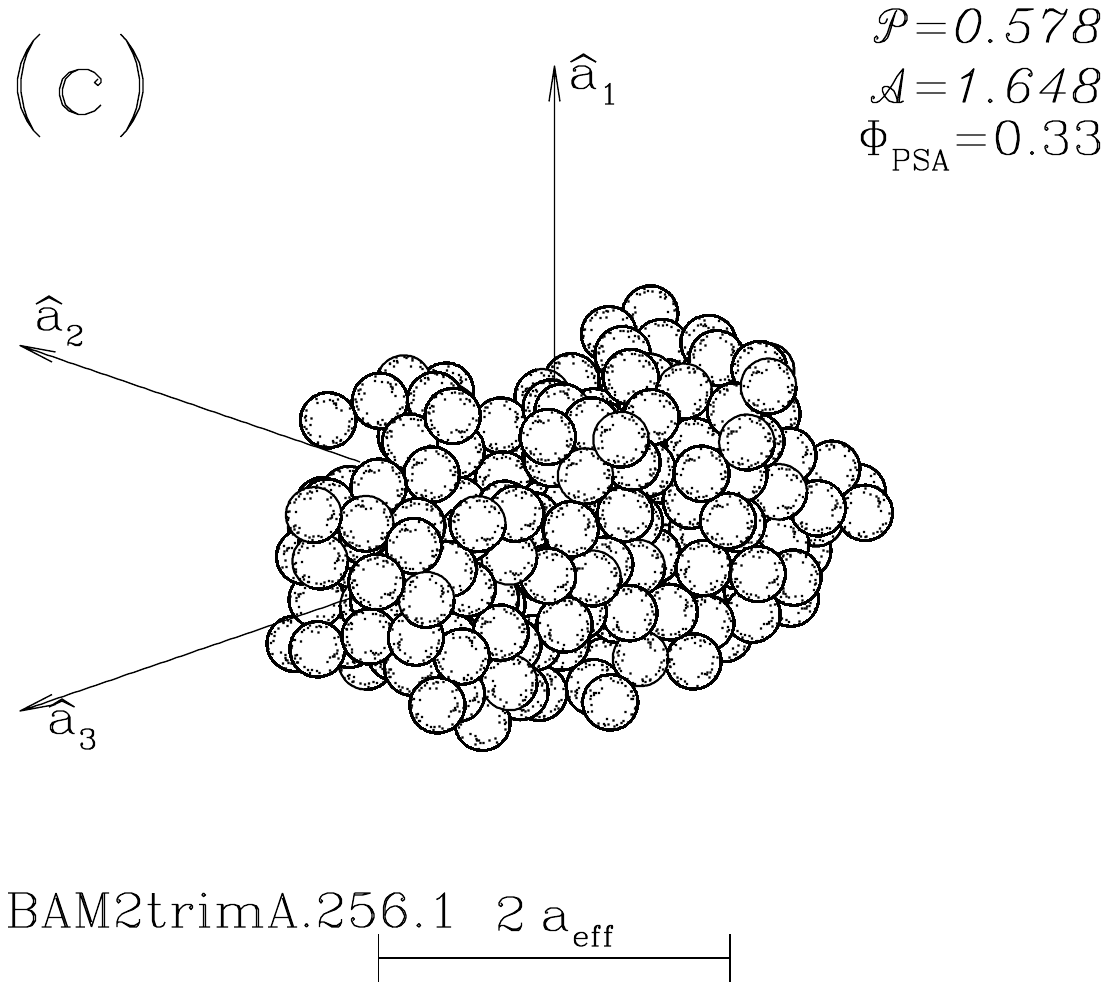}
\hspace*{\negspace}
\includegraphics[angle=0,height=\figheight,
                 clip=true,trim={\triml} {\trimb} {\trimr} \trimt]
{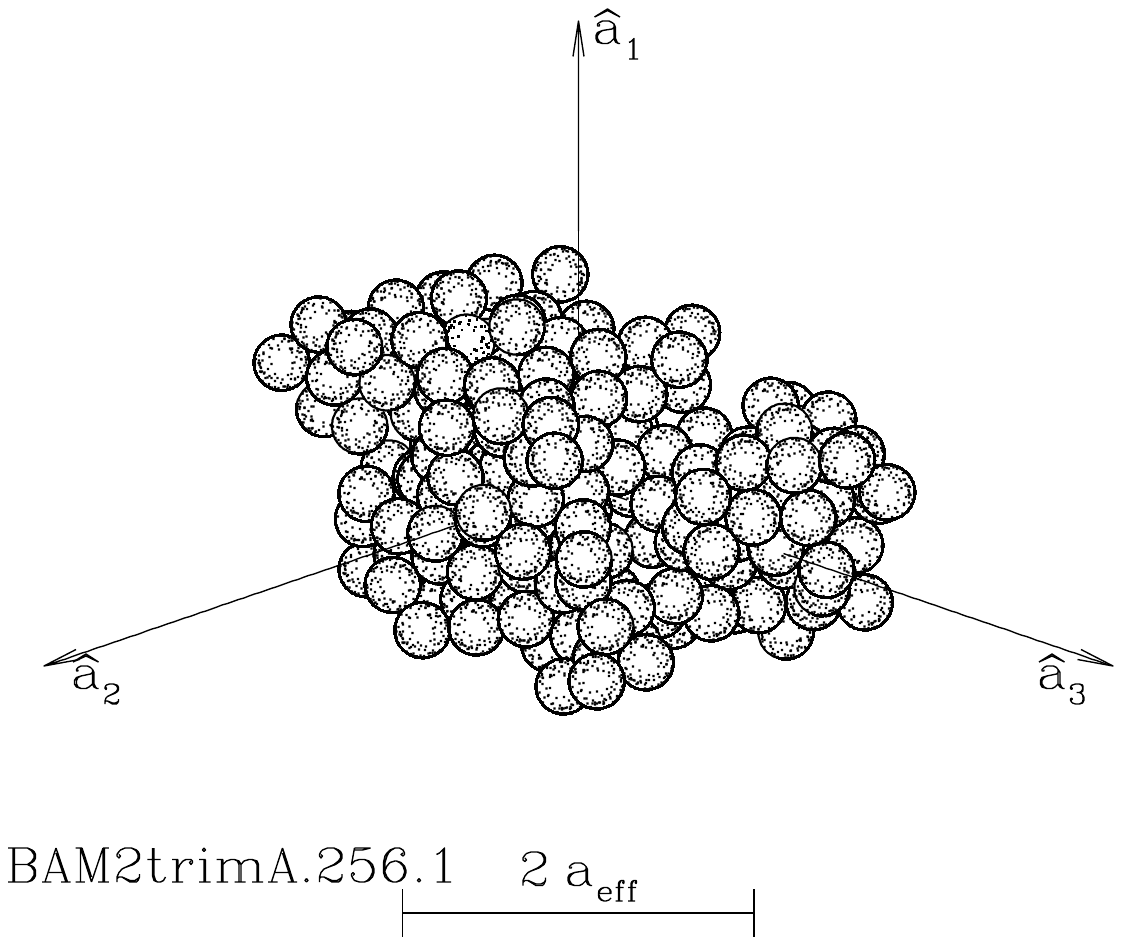}
\hspace*{\negspace}
\includegraphics[angle=0,height=\figheight,
                 clip=true,trim={\triml} {\trimb} {\trimr} \trimt]
{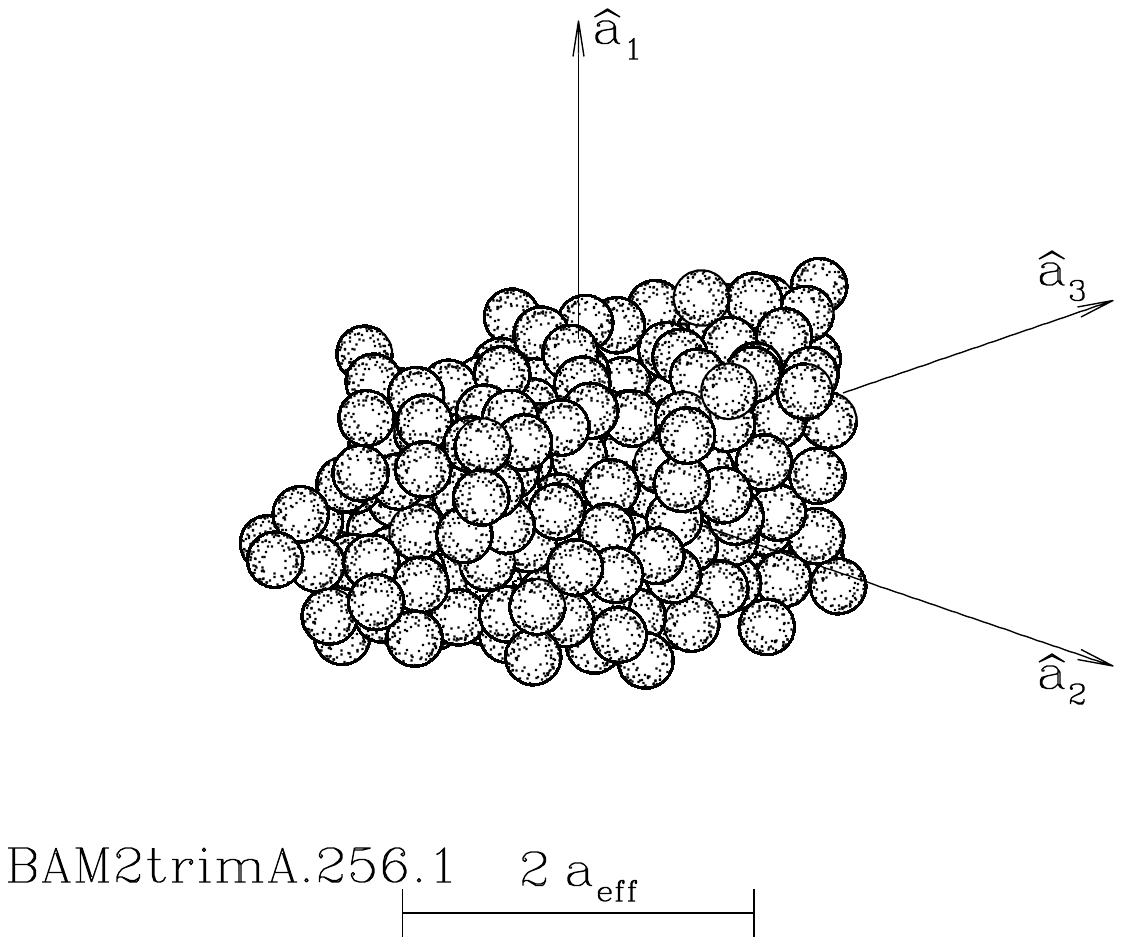}

\caption{\label{fig:shapes_trimA}\footnotesize Same as Figure
  \ref{fig:shapes_babam1bam2}, but for BA, BAM1, and BAM2 clusters
  modified by the \emph{trimA} procedure.$^{\ref{fn:website}}$}
\end{center}
\end{figure}
\begin{figure}
\begin{center}
\includegraphics[angle=0,height=\figheight,
                 clip=true,trim={\triml} {\trimb} {\trimr} \trimt]
{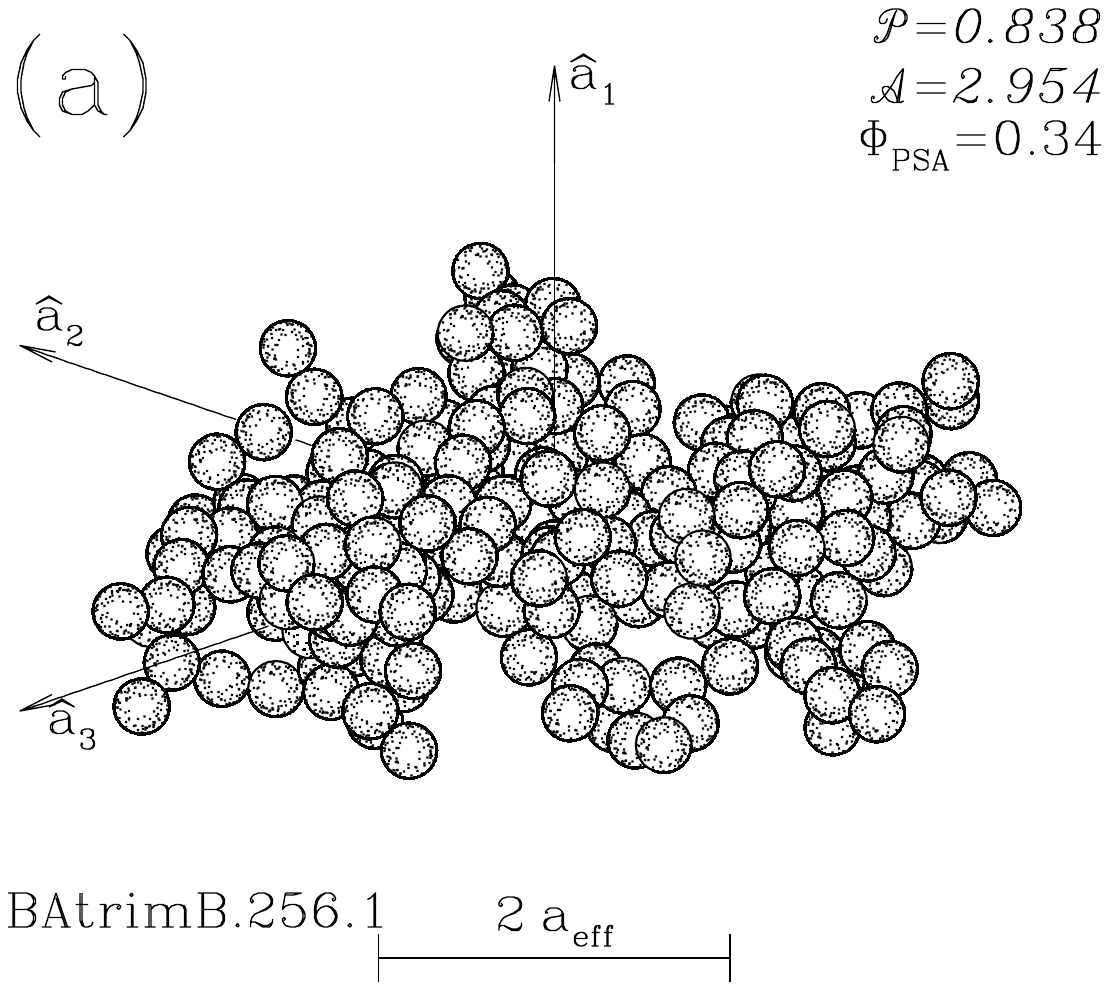}
\hspace*{\negspace}
\includegraphics[angle=0,height=\figheight,
                 clip=true,trim={\triml} {\trimb} {\trimr} \trimt]
{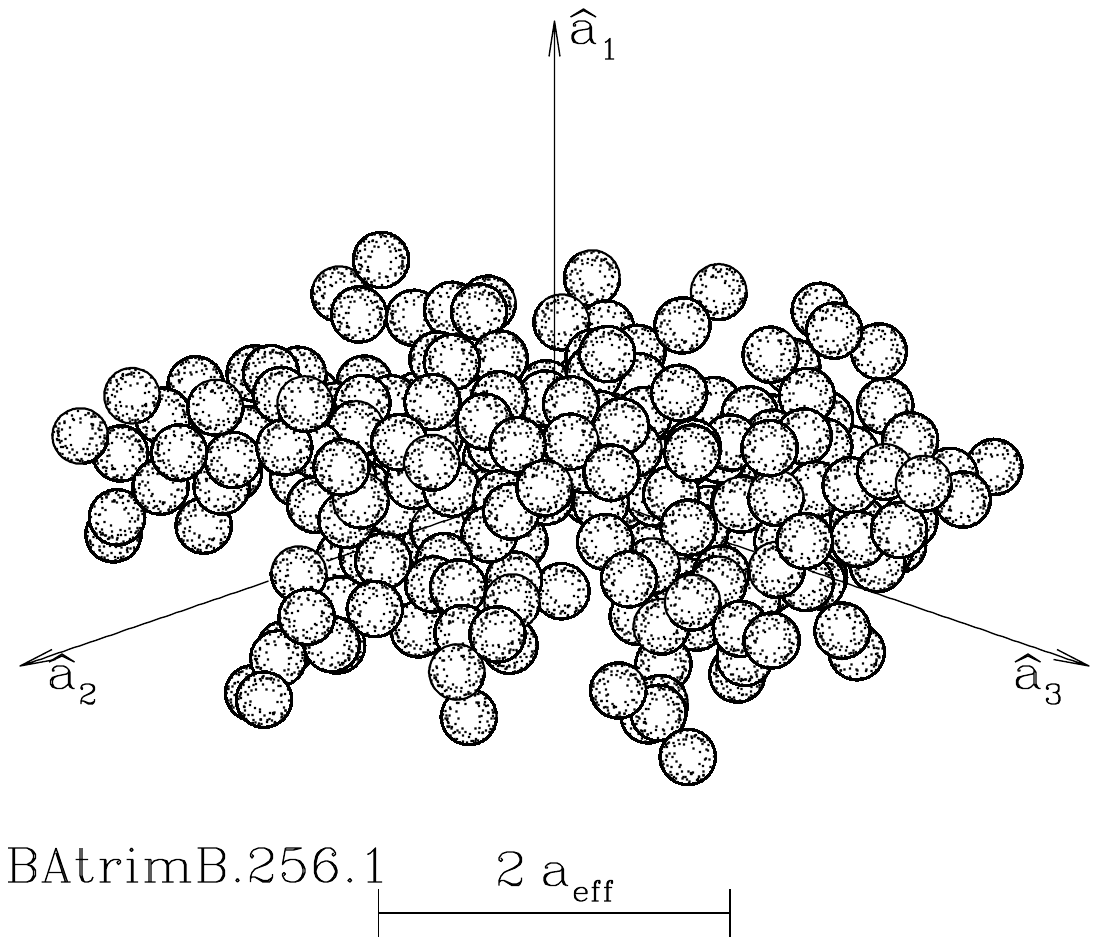}
\hspace*{\negspace}
\includegraphics[angle=0,height=\figheight,
                 clip=true,trim={\triml} {\trimb} {\trimr} \trimt]
{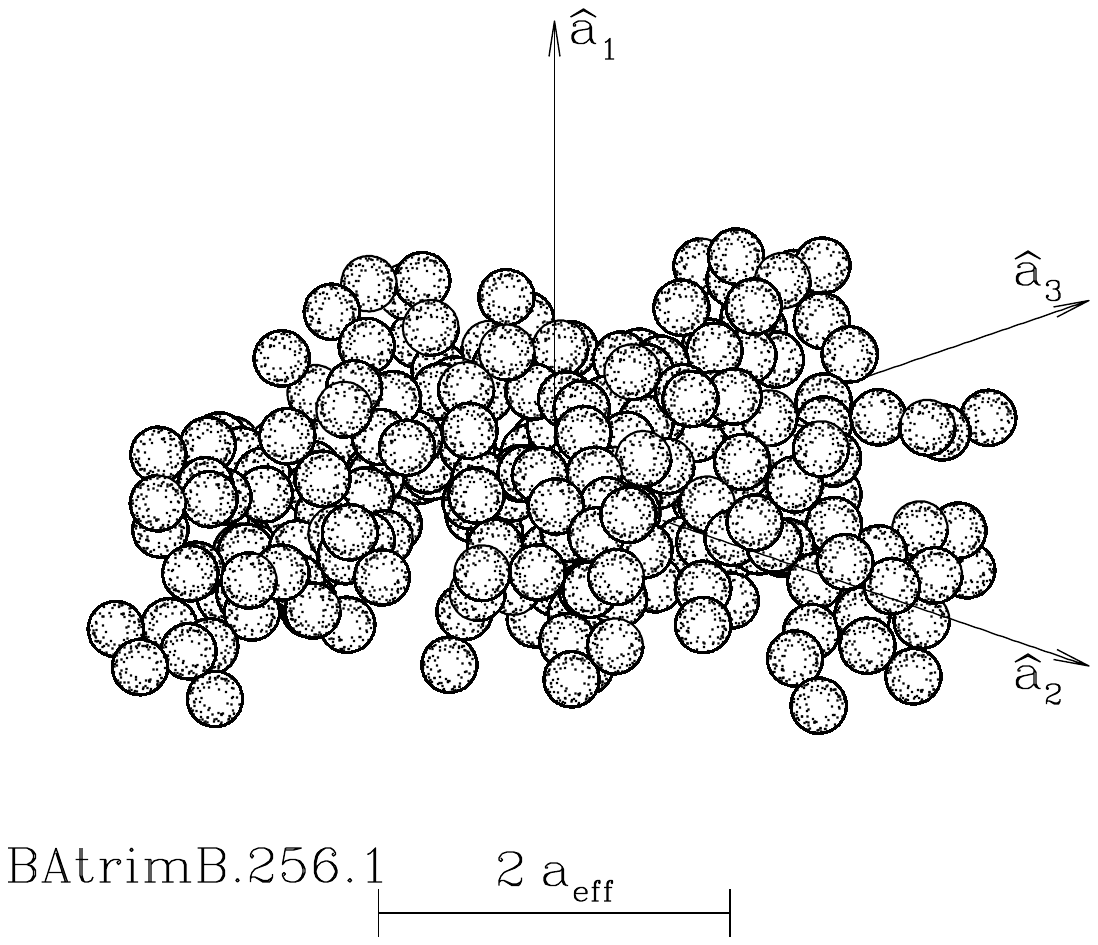}\\

\includegraphics[angle=0,height=\figheight,
                 clip=true,trim={\triml} {\trimb} {\trimr} \trimt]
{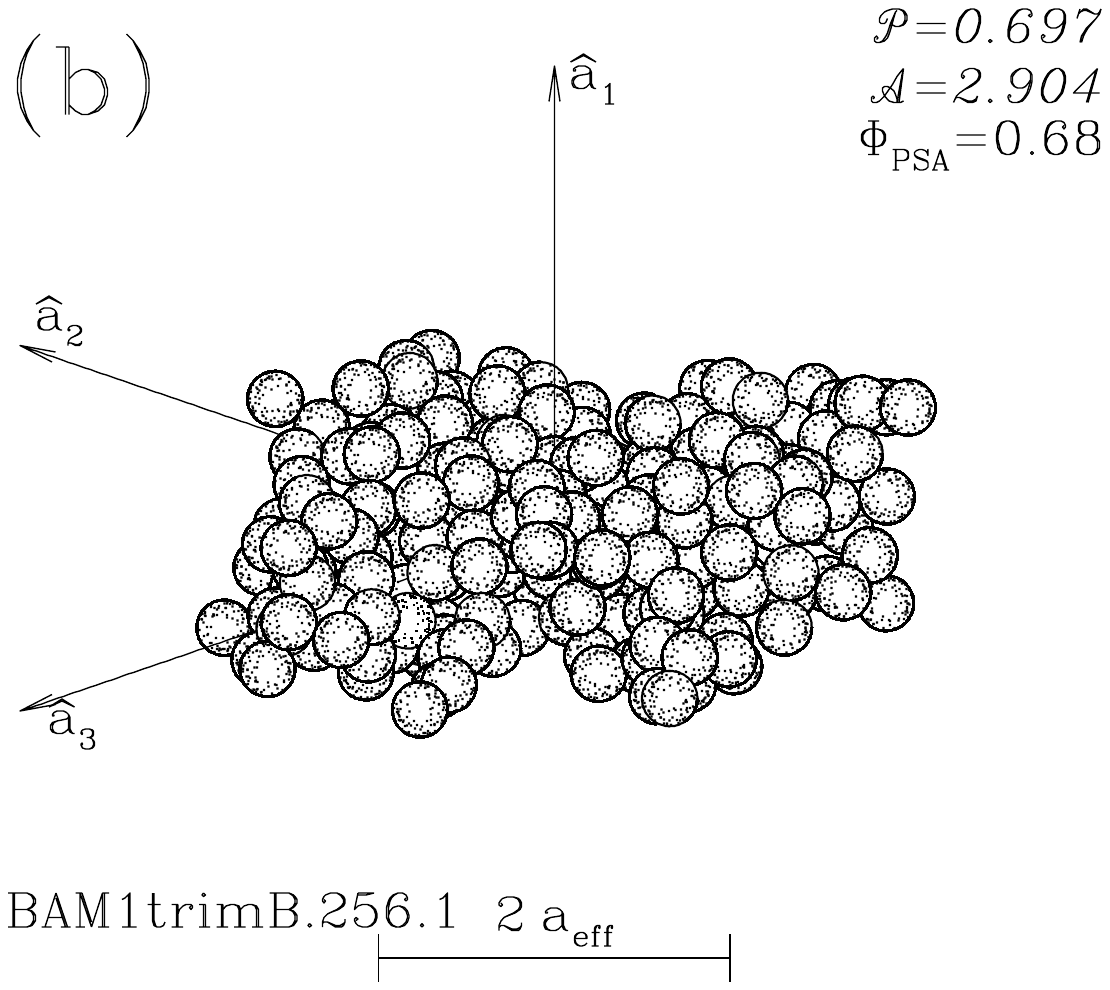}
\hspace*{\negspace}
\includegraphics[angle=0,height=\figheight,
                 clip=true,trim={\triml} {\trimb} {\trimr} \trimt]
{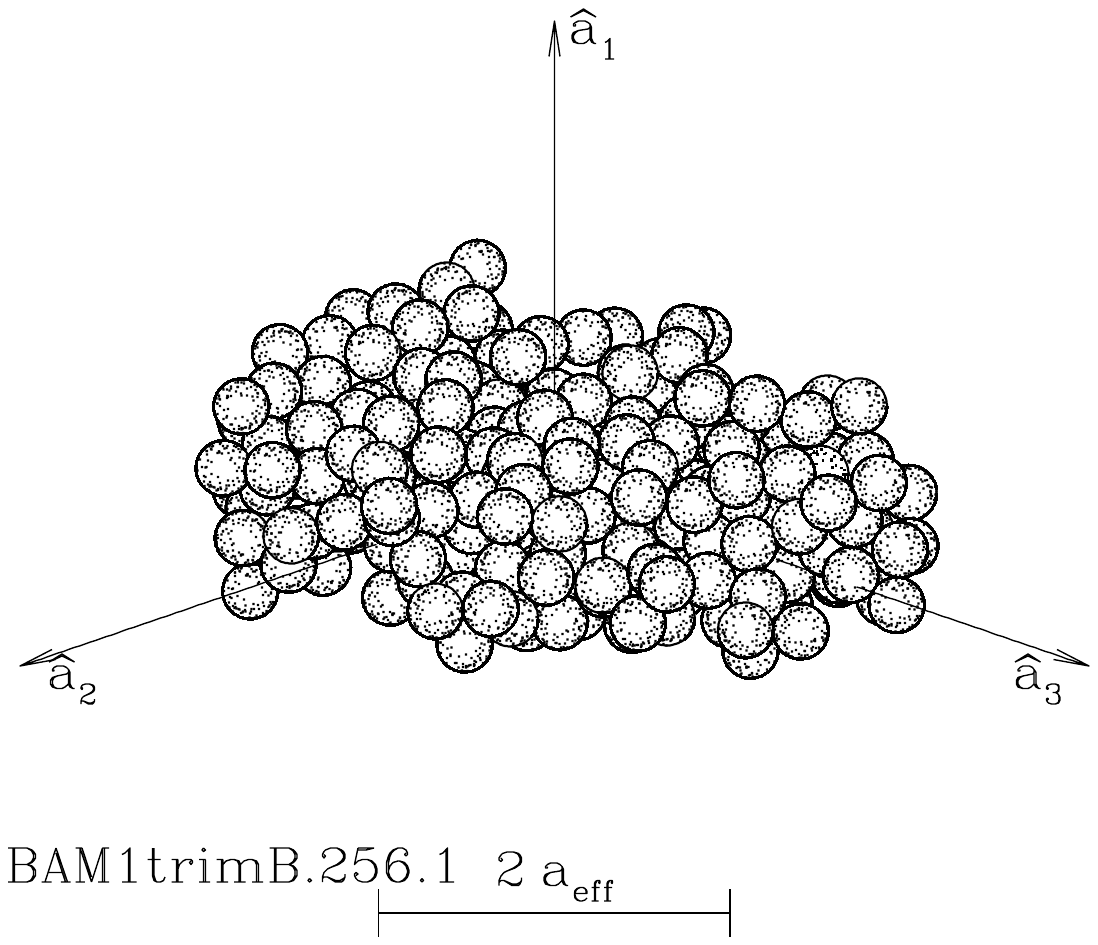}
\hspace*{\negspace}
\includegraphics[angle=0,height=\figheight,
                 clip=true,trim={\triml} {\trimb} {\trimr} \trimt]
{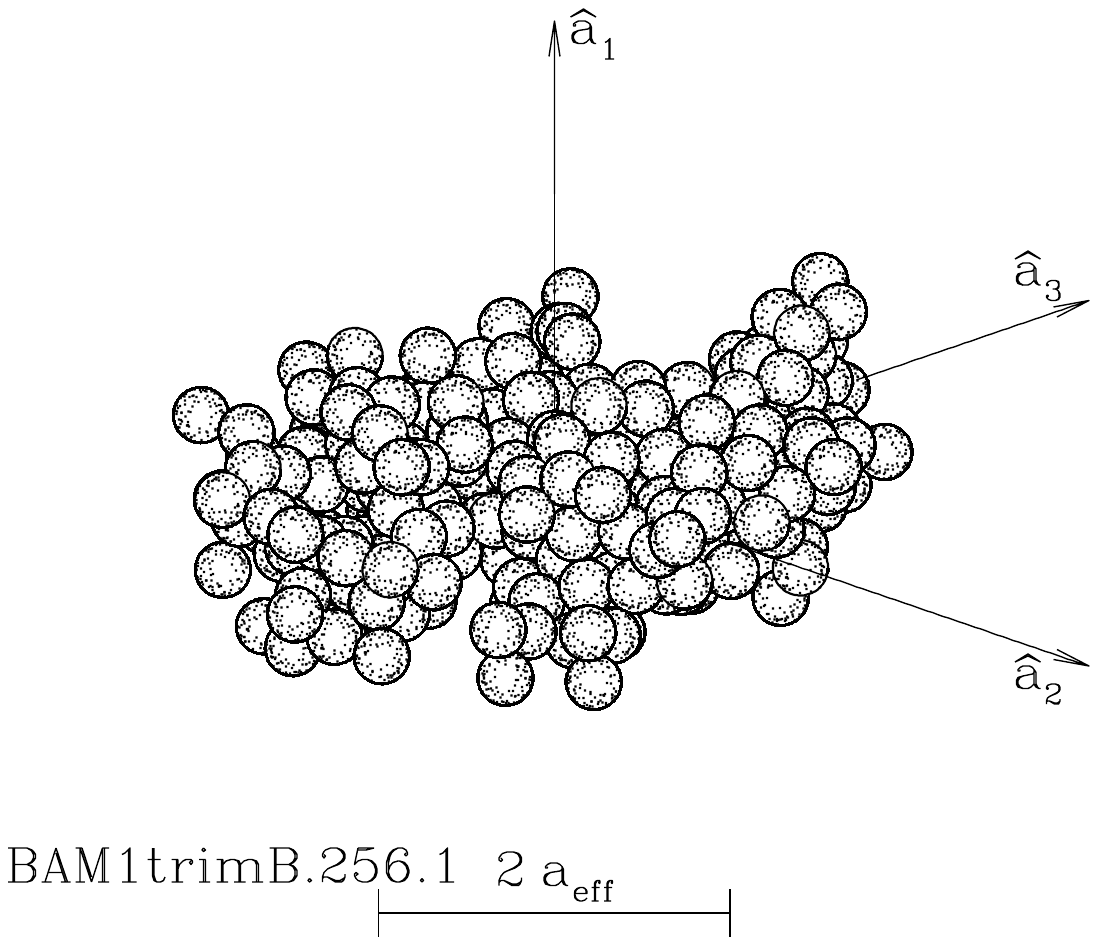}\\

\includegraphics[angle=0,height=\figheight,
                 clip=true,trim={\triml} {\trimb} {\trimr} \trimt]
{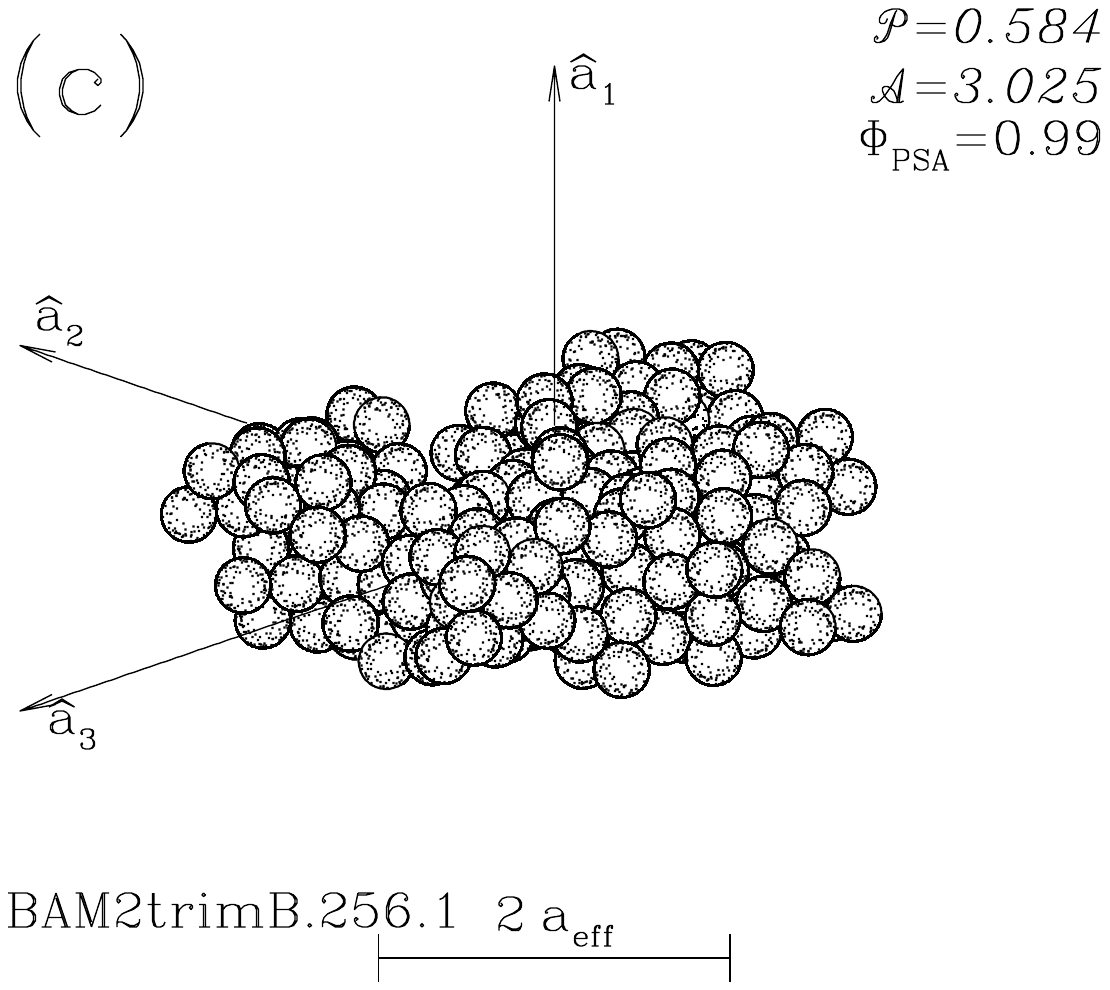}
\hspace*{\negspace}
\includegraphics[angle=0,height=\figheight,
                 clip=true,trim={\triml} {\trimb} {\trimr} \trimt]
{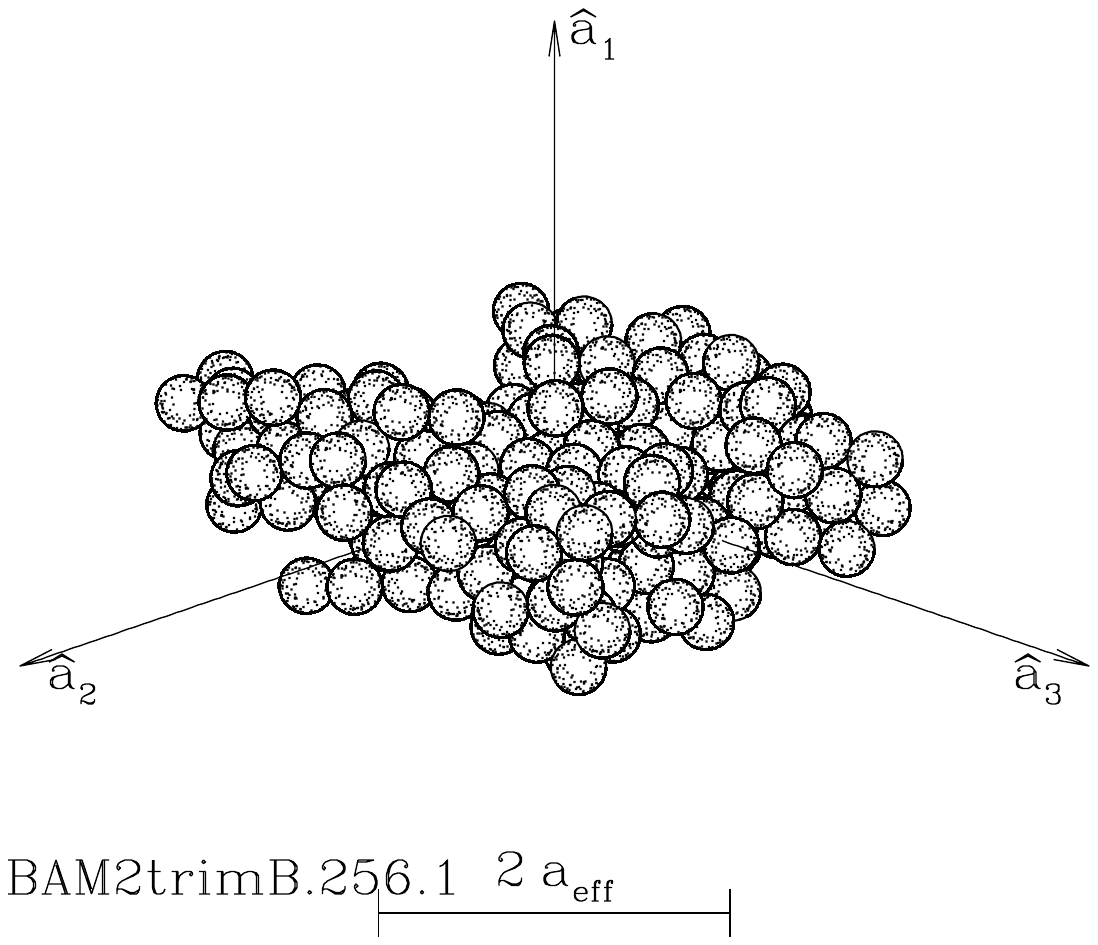}
\hspace*{\negspace}
\includegraphics[angle=0,height=\figheight,
                 clip=true,trim={\triml} {\trimb} {\trimr} \trimt]
{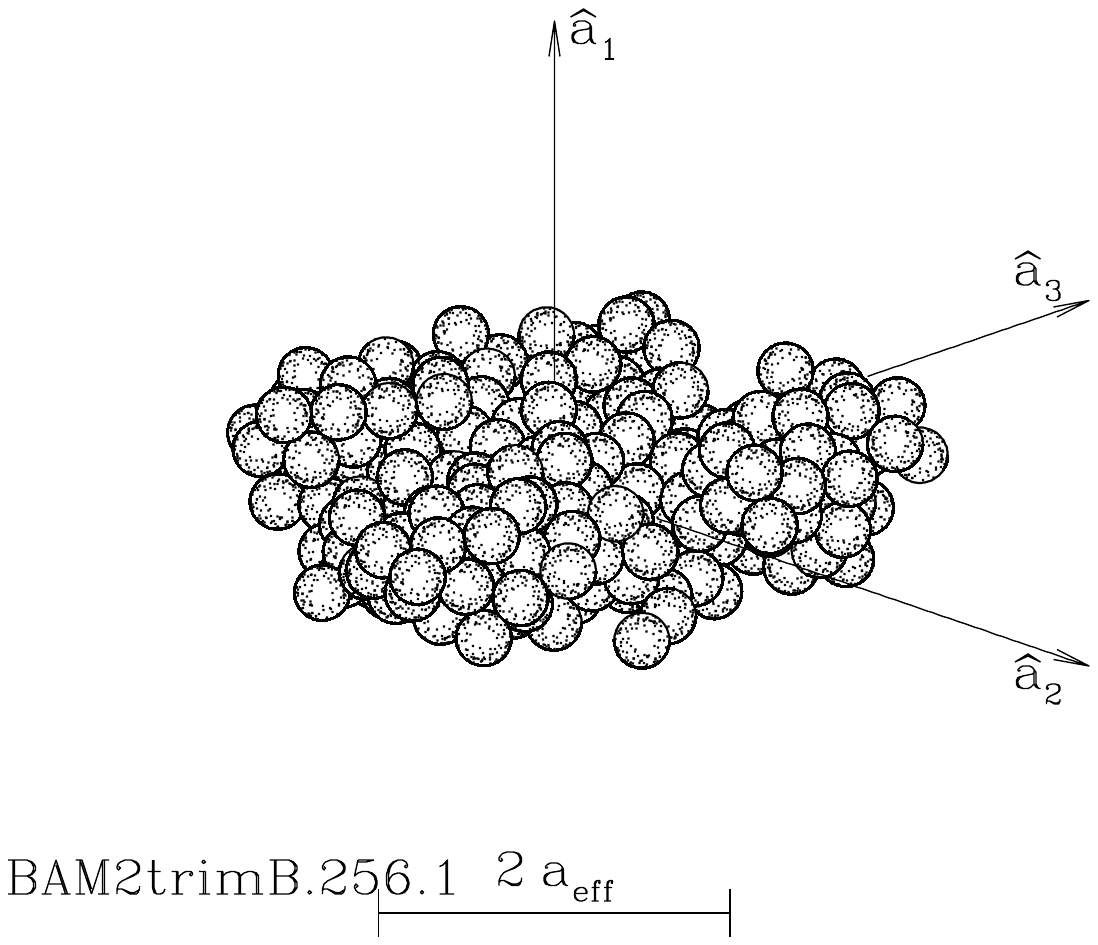}
\caption{\label{fig:shapes_trimB}\footnotesize Same as Figure
  \ref{fig:shapes_babam1bam2}, but for aggregates created by the
  \emph{trimB} procedure: (a) BAtrimB.256.1, (b) BAM1trimA.256.1, and
  (c) BAM2trimB.256.1.$^{\ref{fn:website}}$}
\end{center}
\end{figure}
\begin{figure}
\begin{center}
\includegraphics[angle=0,height=\figheight,
                 clip=true,trim={\triml} {\trimb} {\trimr} \trimt]
{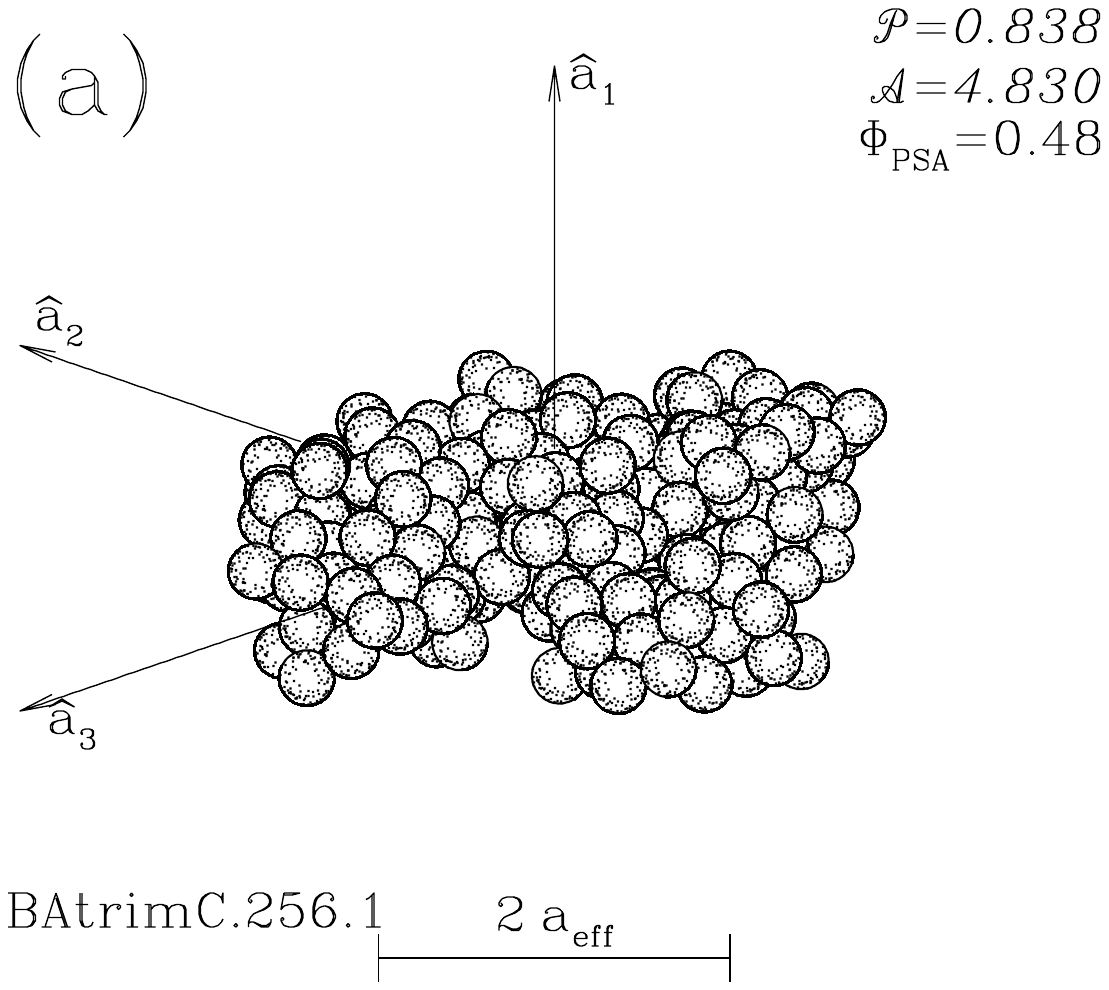}
\hspace*{\negspace}
\includegraphics[angle=0,height=\figheight,
                 clip=true,trim={\triml} {\trimb} {\trimr} \trimt]
{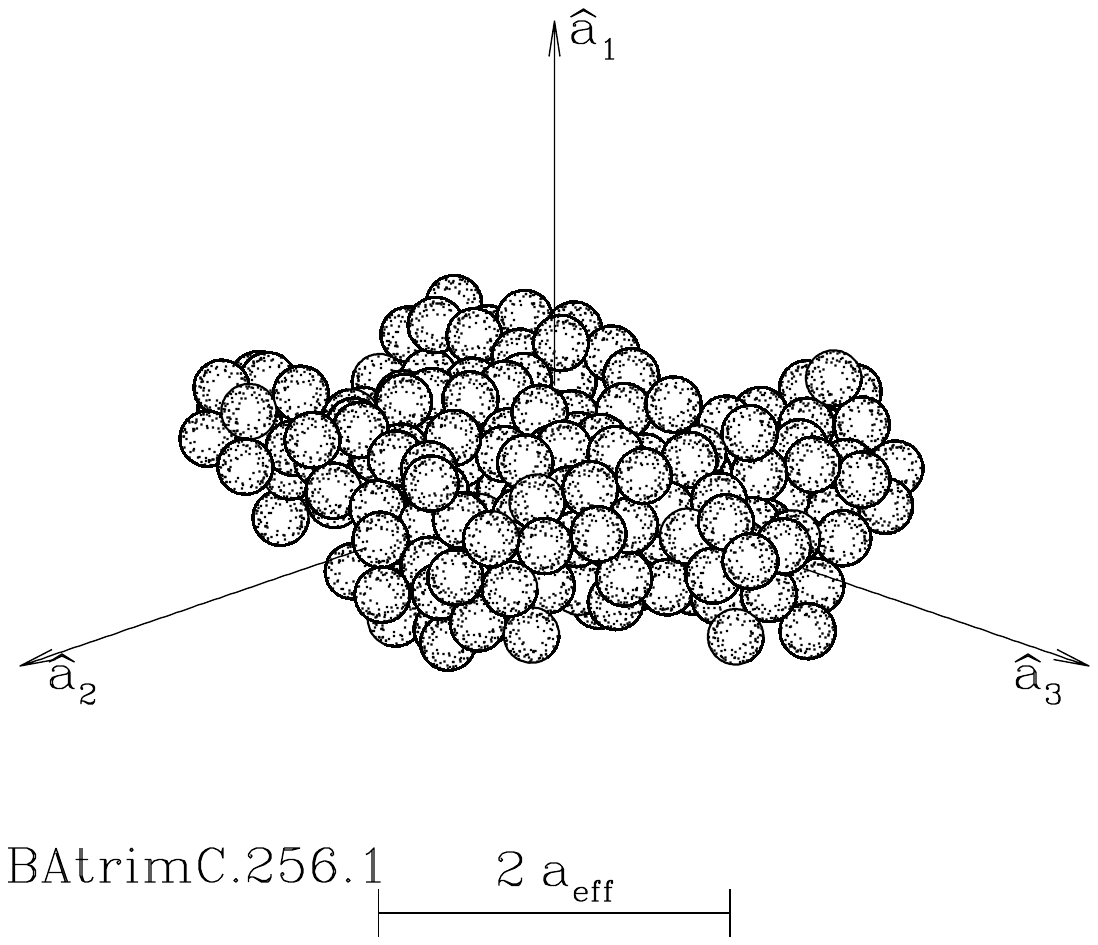}
\hspace*{\negspace}
\includegraphics[angle=0,height=\figheight,
                 clip=true,trim={\triml} {\trimb} {\trimr} \trimt]
{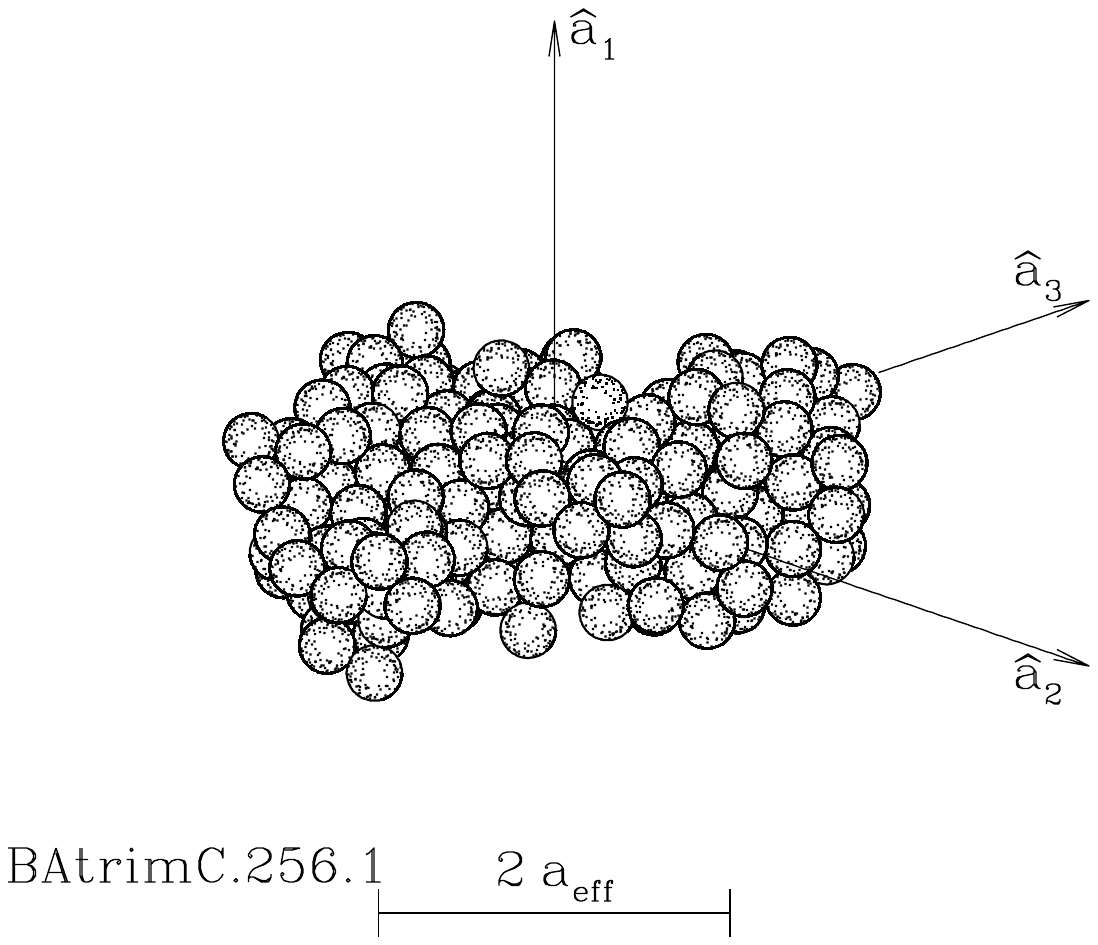}\\

\includegraphics[angle=0,height=\figheight,
                 clip=true,trim={\triml} {\trimb} {\trimr} \trimt]
{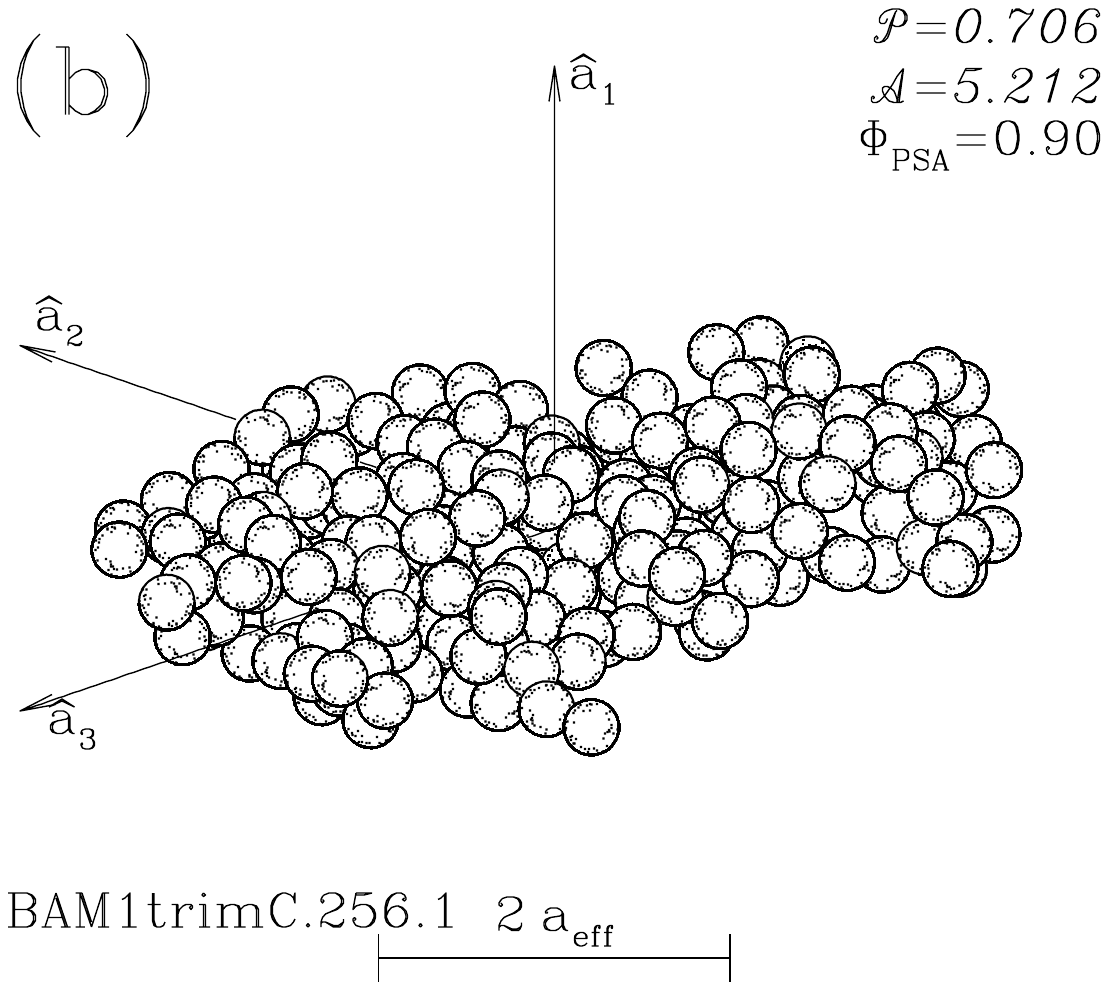}
\hspace*{\negspace}
\includegraphics[angle=0,height=\figheight,
                 clip=true,trim={\triml} {\trimb} {\trimr} \trimt]
{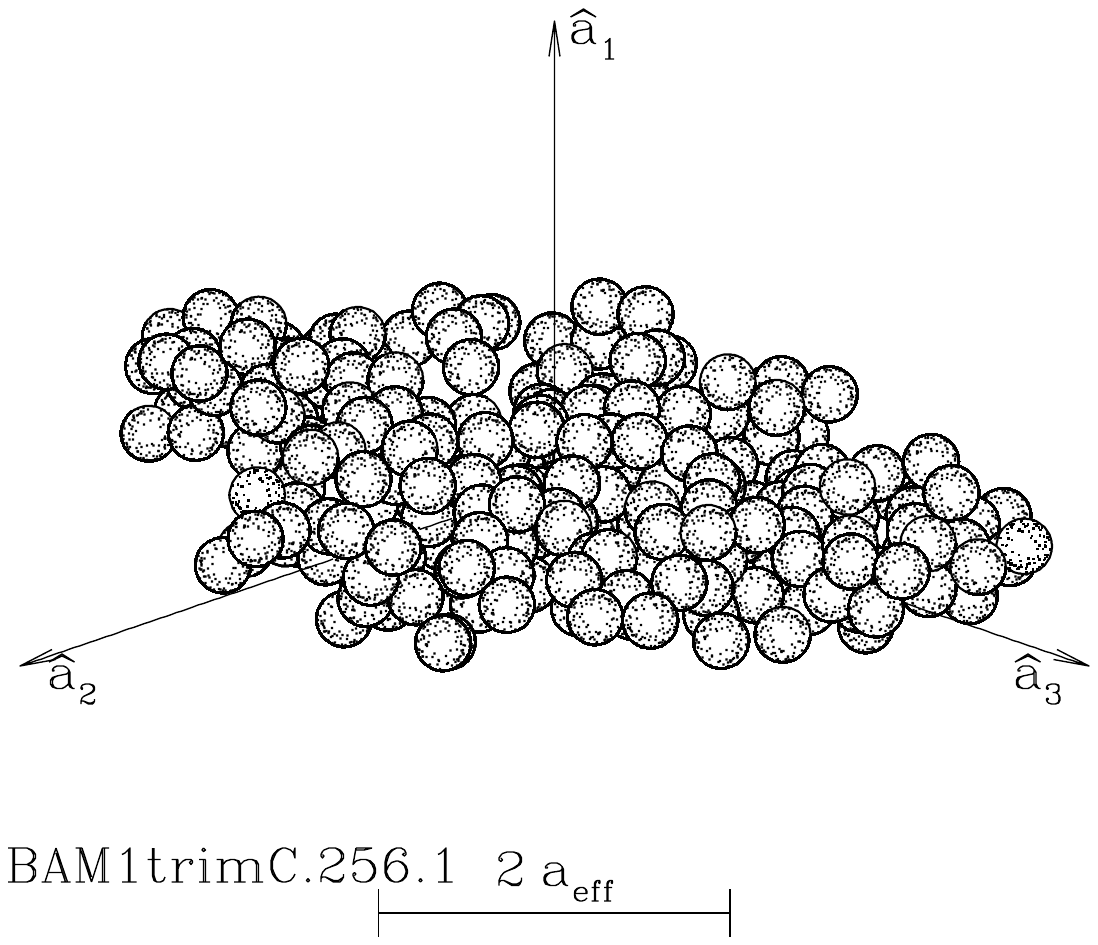}
\hspace*{\negspace}
\includegraphics[angle=0,height=\figheight,
                 clip=true,trim={\triml} {\trimb} {\trimr} \trimt]
{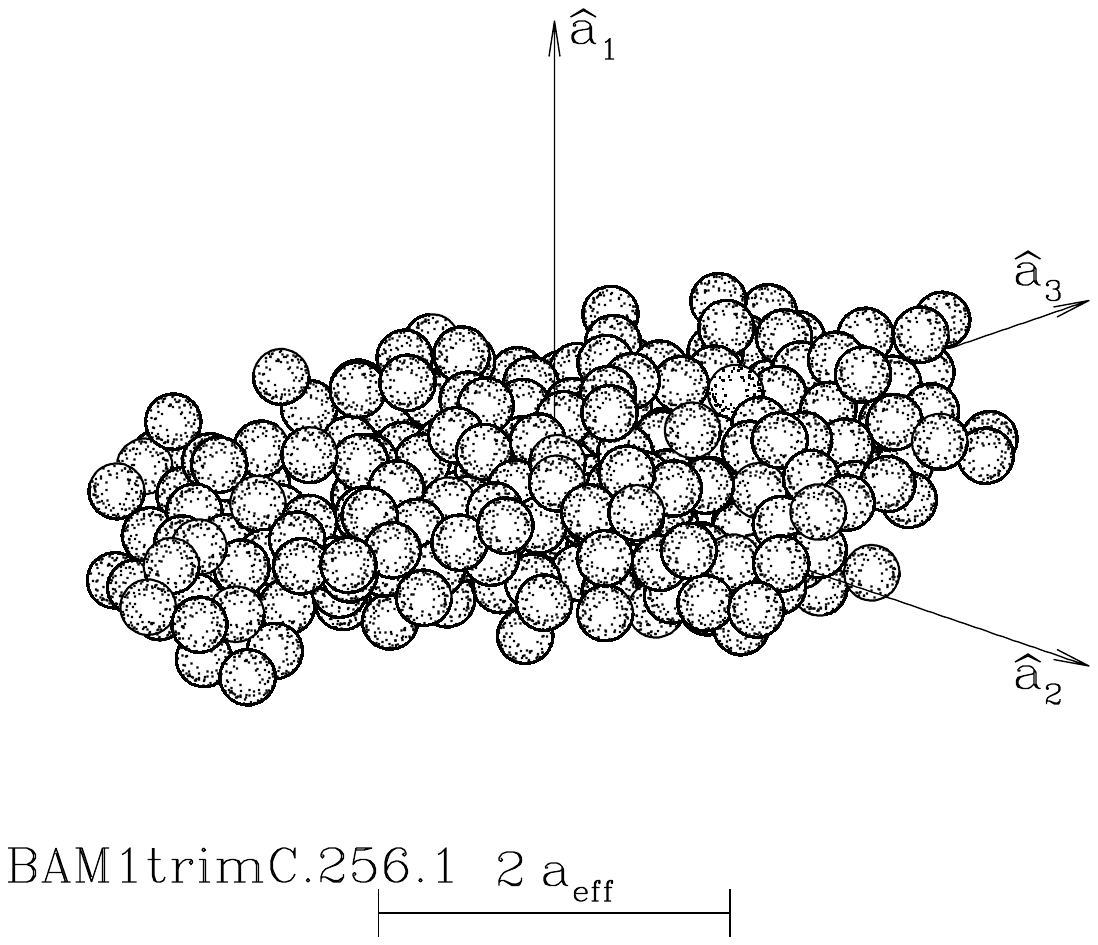}\\

\includegraphics[angle=0,height=\figheight,
                 clip=true,trim={\triml} {\trimb} {\trimr} \trimt]
{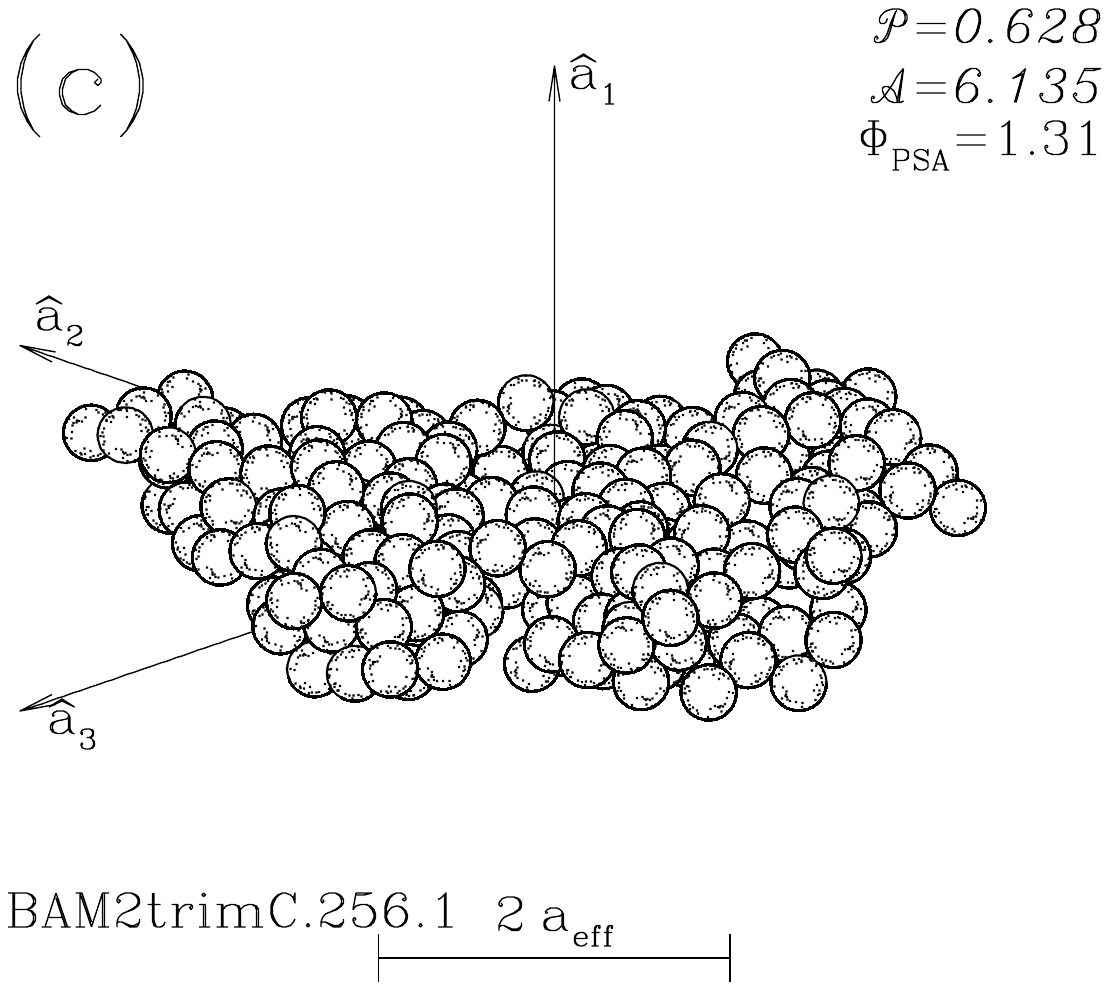}
\hspace*{\negspace}
\includegraphics[angle=0,height=\figheight,
                 clip=true,trim={\triml} {\trimb} {\trimr} \trimt]
{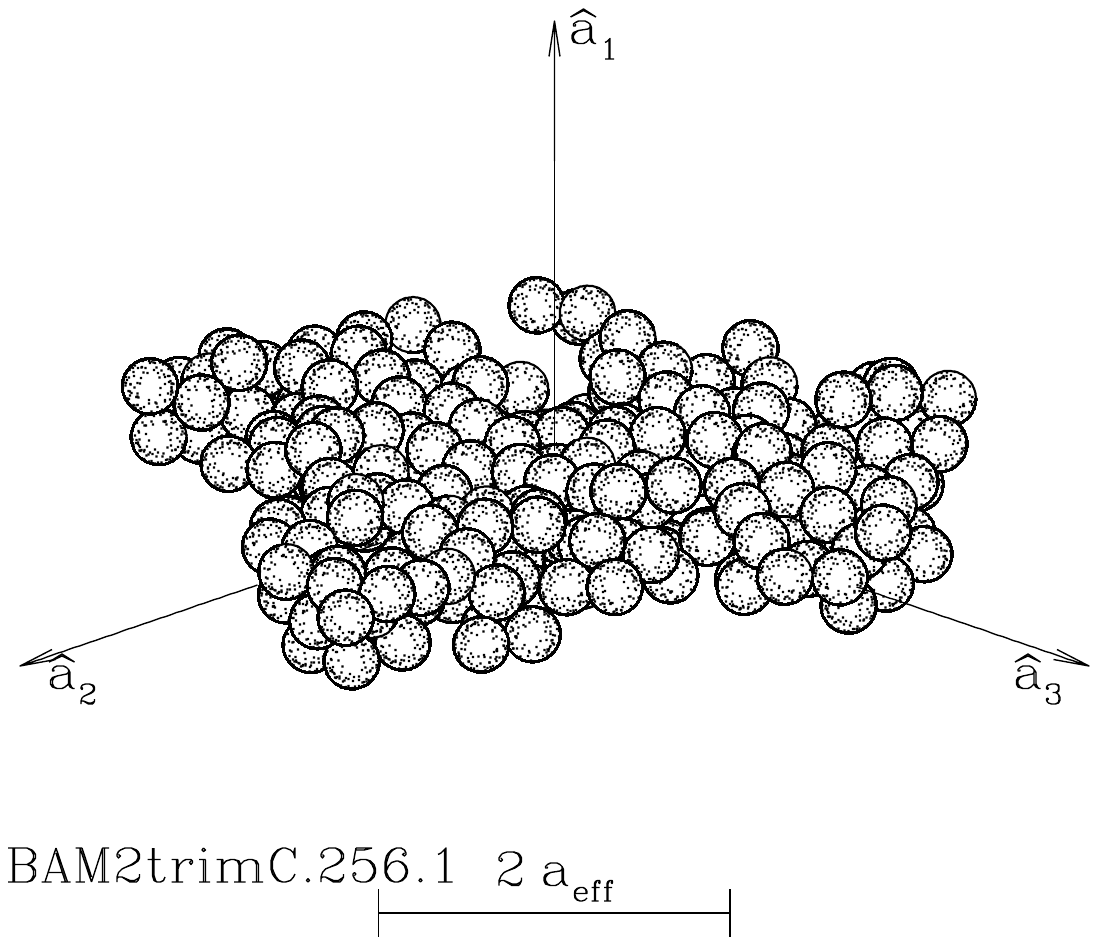}
\hspace*{\negspace}
\includegraphics[angle=0,height=\figheight,
                 clip=true,trim={\triml} {\trimb} {\trimr} \trimt]
{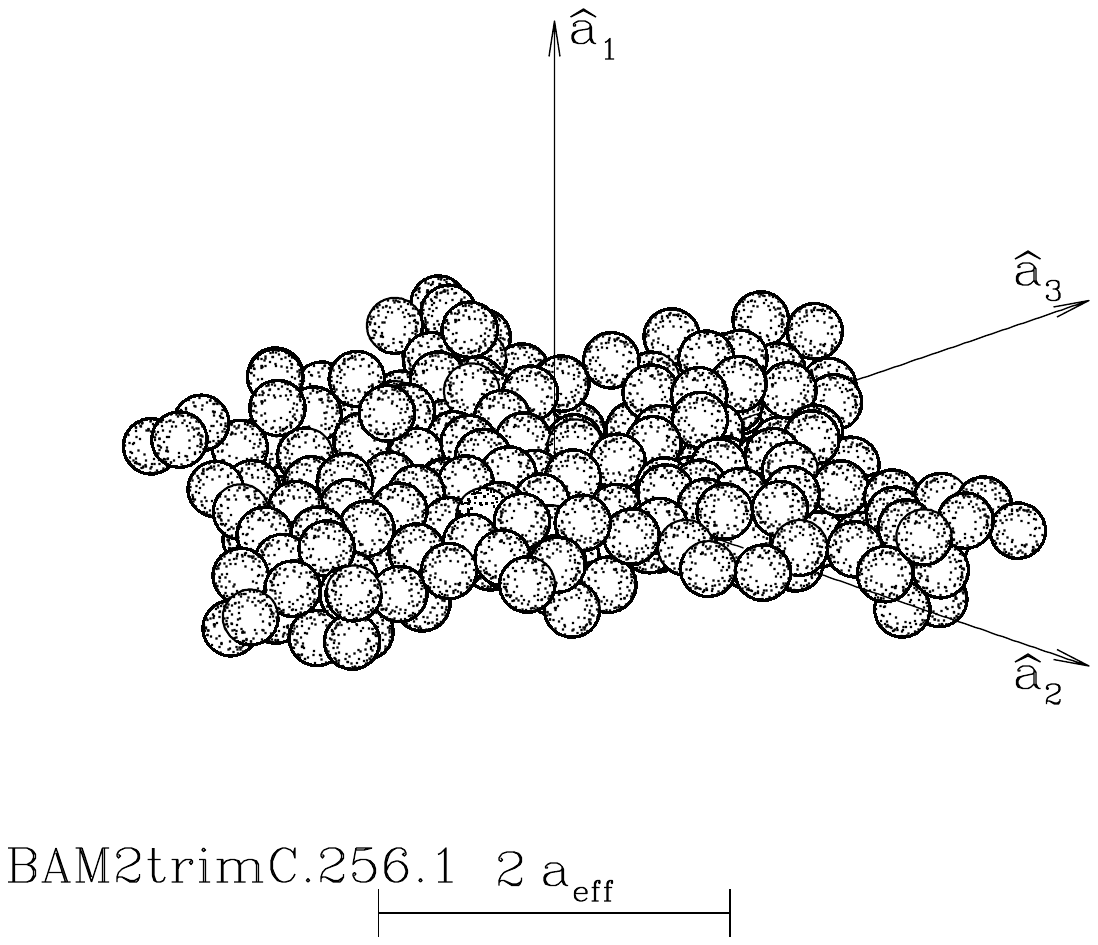}
\caption{\label{fig:shapes_trimC}\footnotesize Same as Figure
  \ref{fig:shapes_babam1bam2}, but for aggregates created by the
  \emph{trimC} procedure: (a) BAtrimC.256.1, (b) BAM1trimC.256.1, and (c)
  BAM2trimC.256.1.$^{\ref{fn:website}}$}
\end{center}
\end{figure}
\newcommand{\co}{\,:\,}
\begin{center}
\begin{table}
\caption{\label{tab:geom}
         Geometric Parameters for the Considered Shapes}
{\footnotesize
\begin{tabular}{r l l c c c c c c}
\hline
$N$ & shape         & axial ratios &$\alpha_1$&$\alpha_2$&$\alpha_3$
                                                              &$\Asymm$\,$^a$
                                                                      &$\Stretch$\,$^b$
                                                                              &$\poromacro$\,$^c$ \\
\hline
2 & bisphere        & 1\co1\co2      & 2.2049 & 2.2049 & 0.6300 & 1.871 & 1.871 & 0.184 \\
3 & threesphere     & 1\co1\co3      & 3.6857 & 3.6857 & 0.4808 & 2.769 & 2.769 & 0.208 \\
3 & trisphere       & 1\co1.87\co2   & 2.0833 & 1.2820 & 1.2820 & 2.082 & 0.785 & 0.308 \\
\hline
256& BA.256.1       &1\co1.10\co1.22 & 4.5421 & 3.8713 & 3.1928 & 1.342 & 1.017 & 0.860 \\
256& BA.256.2       &1\co1.20\co1.40 & 4.5146 & 3.9351 & 2.9464 & 1.381 & 1.079 & 0.852 \\
256& BA.256.3       &1\co1.18\co1.28 & 4.5474 & 3.7800 & 3.1125 & 1.387 & 1.010 & 0.855 \\
256& BA.256.4       &1\co1.10\co1.37 & 3.8261 & 3.5056 & 3.1018 & 1.173 & 1.018 & 0.842 \\
\hline
256& BAM1.256.1     &1\co1.09\co1.43 & 2.8462 & 2.5011 & 1.8513 & 1.375 & 1.090 & 0.706 \\
256& BAM1.256.2     &1\co1.21\co1.42 & 3.2580 & 2.7559 & 1.9919 & 1.479 & 1.082 & 0.741 \\
256& BAM1.256.3     &1\co1.04\co1.60 & 2.7464 & 2.6270 & 1.7413 & 1.301 & 1.201 & 0.698 \\
256& BAM1.256.4     &1\co1.05\co1.32 & 3.0355 & 2.7150 & 2.1538 & 1.287 & 1.062 & 0.752 \\
\hline
256& BAM2.256.1     &1\co1.28\co1.30 & 2.0827 & 1.6846 & 1.6078 & 1.312 & 0.921 & 0.563 \\
256& BAM2.256.2     &1\co1.18\co1.43 & 2.3293 & 1.9550 & 1.4589 & 1.466 & 1.061 & 0.578 \\
256& BAM2.256.3     &1\co1.00\co1.37 & 2.2886 & 2.1853 & 1.3588 & 1.350 & 1.239 & 0.582 \\
256& BAM2.256.4     &1\co1.27\co1.31 & 2.2880 & 1.9949 & 1.6011 & 1.323 & 1.042 & 0.612 \\
\hline
256& BAtrimA.256.1  &1\co1.10\co2.09 & 5.5978 & 5.2975 & 2.0947 & 1.766 & 1.547 & 0.837 \\
256& BAtrimA.256.2  &1\co1.00\co2.15 & 5.9604 & 5.6640 & 1.9524 & 1.897 & 1.660 & 0.833 \\
256& BAtrimA.256.3  &1\co1.19\co2.61 & 6.5335 & 6.0392 & 1.8181 & 2.222 & 1.752 & 0.826 \\
256& BAtrimA.256.4  &1\co1.23\co2.15 & 5.0958 & 4.9174 & 1.8739 & 1.734 & 1.591 & 0.812 \\
\hline
256& BAM1trimA.256.1&1\co1.15\co2.57 & 4.0083 & 3.7684 & 1.2284 & 2.014 & 1.698 & 0.676 \\
256& BAM1trimA.256.2&1\co1.13\co2.38 & 4.3939 & 4.1387 & 1.2761 & 2.075 & 1.748 & 0.703 \\
256& BAM1trimA.256.3&1\co1.10\co2.40 & 3.9182 & 3.7365 & 1.2240 & 1.939 & 1.706 & 0.674 \\
256& BAM1trimA.256.4&1\co1.31\co2.23 & 4.1224 & 3.5831 & 1.4893 & 2.083 & 1.446 & 0.711 \\
\hline
256& BAM2trimA.256.1&1\co1.08\co2.15 & 2.7909 & 2.6722 & 1.1466 & 1.648 & 1.494 & 0.578 \\
256& BAM2trimA.256.2&1\co1.27\co2.65 & 3.4278 & 3.2202 & 0.9193 & 2.195 & 1.814 & 0.533 \\
256& BAM2trimA.256.3&1\co1.06\co2.23 & 3.6061 & 3.5617 & 0.8524 & 2.113 & 2.032 & 0.533 \\
256& BAM2trimA.256.4&1\co1.11\co2.31 & 3.0484 & 2.8989 & 1.0067 & 1.886 & 1.655 & 0.548 \\
\hline
256& BAtrimB.256.1  &1\co3.06\co3.29 & 6.9806 & 4.4450 & 3.3357 & 2.954 & 0.921 & 0.838 \\
256& BAtrimB.256.2  &1\co3.09\co3.52 & 7.6158 & 4.7914 & 3.4839 & 3.398 & 0.930 & 0.836 \\
203& BAtrimB.256.3  &1\co2.50\co2.70 & 6.5451 & 4.3900 & 3.0701 & 2.675 & 0.979 & 0.837 \\
256& BAtrimB.256.4  &1\co2.82\co3.47 & 7.2562 & 4.6455 & 3.2323 & 3.417 & 0.959 & 0.822 \\  
\hline
256& BAM1trimB.256.1&1\co2.67\co3.40 & 4.6552 & 3.3062 & 1.9012 & 2.904 & 1.111 & 0.697 \\
256& BAM1trimB.256.2&1\co2.62\co3.52 & 5.4627 & 3.7929 & 2.1502 & 3.372 & 1.107 & 0.723 \\
256& BAM1trimB.256.3&1\co2.81\co3.69 & 4.9165 & 3.4406 & 1.9504 & 3.219 & 1.111 & 0.690 \\
256& BAM1trimB.256.4&1\co2.95\co3.39 & 5.4628 & 3.6533 & 2.2820 & 3.401 & 1.035 & 0.725 \\
\hline
256& BAM2trimB.256.1&1\co2.82\co2.87 & 3.7554 & 2.1944 & 1.9714 & 3.025 & 0.807 & 0.584 \\
256& BAM2trimB.256.2&1\co3.42\co3.66 & 4.1837 & 2.6593 & 1.8385 & 3.709 & 0.955 & 0.558 \\
256& BAM2trimB.256.3&1\co2.66\co3.57 & 4.5707 & 3.4745 & 1.4238 & 3.735 & 1.362 & 0.577 \\
256& BAM2trimB.256.4&1\co2.73\co3.18 & 3.9970 & 2.6027 & 1.7863 & 3.197 & 0.974 & 0.591 \\
\hline
256& BAtrimC.256.1&1\co4.05\co4.99 & 9.0308 & 5.4698 & 4.0287 & 4.830 & 1.109 & 0.838 \\
256& BAtrimC.256.2&1\co4.13\co5.10 & 7.4399 & 4.6722 & 3.3762 & 3.497 & 0.932 & 0.825 \\
\hline
256& BAM1trimC.256.1&1\co3.68\co5.34 & 7.2802 & 5.3448 & 2.2033 & 5.212 & 1.335 & 0.706 \\
256& BAM1trimC.256.2&1\co4.08\co6.49 & 9.0775 & 6.9573 & 2.3396 & 6.432 & 1.510 & 0.727 \\
\hline
256& BAM2trimC.256.1&1\co5.51\co5.53 & 6.4871 & 3.4908 & 3.1687 & 6.135 & 0.770 & 0.628 \\
256& BAM2trimC.256.2&1\co5.72\co7.08 & 6.9519 & 4.2215 & 2.6719 & 6.994 & 1.026 & 0.606 \\
\hline
\multicolumn{8}{l}{$^a$ Asymmetry parameter (Eq.\ \ref{eq:Anew}).}\\
\multicolumn{8}{l}{$^b$ Stretch parameter \citep[Eq.\ 3 in][]{Draine_2024a}.}\\
\multicolumn{8}{l}{$^c$ Macroporosity parameter (Eq.\ \ref{eq:poromacro}).}
\end{tabular}
}
\end{table}
\end{center}

\subsection{$N=256$ Trimmed Aggregates}

In addition to the BA, BAM1, and BAM2 random aggregates described
above, we also study aggregates that have been trimmed to make them
more flattened or elongated.  We first construct random aggregates
following the BA, BAM1 and BAM2 prescriptions, and find the principal
axes $\bahat_1$, $\bahat_2$, $\bahat_3$.  We then systematically
remove spheres in a manner designed to make the final structure either
more flattened or more elongated.  Let $\br_j$ be the center of sphere
$j$, with the initial centroid at $\br=0$.

We consider three types of trimmed aggregates of $M$ spheres:

\begin{itemize}

\item {\bf\emph{trimA}:} Start with a $N=2M$ random aggregate (BA or
  BAM1 or BAM2).  Remove the sphere with the largest value of
  $\br_j\cdot\bahat_1$, next remove the sphere with the most negative
  value of $\br_j\cdot\bahat_1$, next remove the sphere with the
  largest value of $\br_j\cdot\bahat_2$, then remove the sphere with
  the most negative value of $\br_j\cdot\bahat_2$.  Repeat and
  continue until $M$ spheres remain.  This process tends to produce
  elongated structures, extended in the $\bahat_3$ direction -- see
  Figure \ref{fig:shapes_trimA}.

\item {\bf\emph{trimB}:} Start with a $N=2M$ random aggregate (BA or
  BAM1 or BAM2).  Remove the sphere with the largest value of
  $\br_j\cdot\bahat_1$, then remove the sphere with the most negative
  value of $\br_j\cdot\bahat_1$.  Repeat and continue until $M$
  spheres remain.  This process tends to produce flattened structures,
  extended in the $\bahat_2-\bahat_3$ plane -- see Figure
  \ref{fig:shapes_trimB}.

\item {\bf\emph{trimC}:} Same as \emph{trimB}, but starting with a
  $N=4M$ random aggregate.  Because 75\% of the original spheres have
  been removed, the \emph{trimC} aggregates tend to be more flattened
  than the \emph{trimB} aggregates -- see Figure
  \ref{fig:shapes_trimC}.

\end{itemize}

For each of these procedures: after each removal, check to see if any
of the remaining spheres or groups of spheres have become disconnected
from the main body; if so, remove them.  This procedure can sometimes
result in a trimmed aggregate with fewer than the intended final
number of $M$ spheres, although this occurred for only one\footnote{
Target BAtrimB.256.3 consists of $N=203$ spheres.}
of the examples encountered in the present study.\footnote{
The \emph{trimA}, \emph{trimB}, and \emph{trimC} aggregates studied
here can be found at
\url{www.astro.princeton.edu/~draine/agglom.html}.}

\medskip

For each aggregate we find the $\alpha_j$, $\poromacro$, and $\Asymm$:
see Table \ref{tab:geom}. We also measure the spatial extent of the
aggregate along the $\bahat_1,\bahat_2,\bahat_3$ directions, resulting
in the ``axial ratios'' listed in Table \ref{tab:geom}.

\medskip

The BAM2trimB and BAM2trimC aggregates are the most asymmetric shapes
considered here (see Figures \ref{fig:shapes_trimB} and
\ref{fig:shapes_trimC}), but their asymmetries do not seem especially
extreme when compared to the IDPs in Figure \ref{fig:IDP}.

\section{\label{sec:DDA}
         Scattering and Absorption}
\subsection{Discrete Dipole Approximation}

Cross sections for scattering and absorption are calculated using the
DDA, for wavelengths $\lambda$ from the FUV ($\lambda=0.1\micron$) to
the FIR ($\lambda=100\micron$).  We use the public-domain code {\tt
  DDSCAT}\,\footnote{
{\tt DDSCAT} 7.3.3,
available at \url{www.ddscat.org}\label{fn:ddscat}}
\citep{Draine+Flatau_1994}.  For the solid material we use the
Astrodust dielectric function obtained by \citet{Draine+Hensley_2021a}
for 5:7:7 oblate spheroids with $\poromicro=0.2$ (the ``oblate''
dielectric function in Figure 2 of Paper I).  Our goal is to explore
the effects of shape for a fixed dielectric function for the solid material.
 
Calculations are carried out for $151$ wavelengths from $0.1\micron$
to $100\micron$, uniformly spaced in $\log\lambda$.  For each shape,
we consider 6 values of $\aeff$ between $0.05\micron$ to
$0.30\micron$; for some shapes we extend the calculations to larger
values of $\aeff$ as required to have the linear polarization peaking
near $\lambda=0.55\micron$ as in the interstellar medium.  As in Paper
I, for each shape, size $\aeff$, orientation, and wavelength
$\lambda$, we calculate cross sections for two orthogonal linear
polarizations using three different numbers $N_d$ of dipoles, and then
extrapolate cross sections to $N_d\rightarrow\infty$ using Equation
(8) of Paper I.  See Appendix \ref{app:comp} for further details.

\subsection{Orientational Averages}
\subsubsection{Extinction}

For each shape and size $\aeff$, and wavelength $\lambda$, we
calculate the extinction cross section, $C_{\rm ext}\equiv C_{\rm
  abs}+C_{\rm sca}$, averaged over random orientations:
\beq
C_{\rm ext,ran}(\lambda)\equiv
\big\langle C_{\rm ext}(\lambda)\big\rangle_{\rm ran}
~~~.
\eeq
Orientational averaging is discussed in Appendix D of Paper I.  In brief, let
$\Theta$ be the angle between the direction of propagation $\bkhat$ and an
axis $\bahat$ fixed in the grain.  Let $\beta$ be an angle for rotation
of the grain around axis $\bahat$.  For extinction, we average over both
incident polarization states.

For the bisphere and threesphere we let $\bahat$ be the axis of
rotational symmetry.  These shapes have reflection symmetry ($\beta$
is irrelevant).  We sample uniformly in $\cos\Theta\in[0,1]$ (11
values of $\cos\Theta$).

For the trisphere we take $\bahat=\bahat_1$, the axis of 3-fold
rotational symmetry.  The trisphere also has reflection symmetry.  We
sample uniformly in $\cos\Theta\in[0,1]$ (11 values of $\cos\Theta$) and
$\beta\in(0,\pi/3)$ (3 values of $\beta$), for a total of 33
orientations.

The $N=256$ random aggregates have no symmetries.  Letting
$\bahat=\bahat_1$, we sample uniformly in $\cos\Theta\in[-1,1]$ (11
values of $\Theta$), and uniformly in $\beta\in(0,2\pi)$ (12 values of
$\beta$), for a total of 132 orientations.

\subsubsection{Polarization}

``Perfect spinning alignment'' (PSA) is
the optimal configuration for spinning grains to produce polarization:
the grain is assumed to be spinning around $\bahat_1$, the principal
axis of largest moment of inertia, with $\bahat_1$ parallel to the
local magnetic field $\bB_0$, and radiation propagating perpendicular
to $\bahat_1$.  For linear polarization states
$\bE\parallel\bahat_1$, $\bE\perp\bahat_1$,
\beq \label{eq:CpolPSA}
C_{\rm pol,PSA}(\lambda) \equiv \frac{1}{2}
\Big\langle C_{\rm ext}(\lambda,\bE\perp\bahat_1)-
             C_{\rm ext}(\lambda,\bE\parallel\bahat_1)
\Big\rangle_{\rm PSA}
~~~.
\eeq

For the bisphere and threesphere ($\bahat_1\perp\bahat$) we
interpolate among the 11 values of $\cos\Theta \in [0,1]$ to average
over $\Theta$.

For the trisphere ($\bahat_1=\bahat$), we set $\Theta=\pi/2$ and
average over $\beta$ (3 values $\in[0,\pi/3$]).

For the random aggregates ($\bahat_1=\bahat$), we set $\Theta=\pi/2$,
and average over rotations around $\bahat_1$ (12 values of
$\beta\in(0,2\pi)$).

\medskip

\section{\label{sec:Q}
         Results: Extinction and Polarization by Aggregates}

\begin{figure}
\begin{center}
\includegraphics[angle=0,width=\fwidthb,
                 clip=true,trim=0.5cm 0.5cm 0.5cm 0.5cm]
{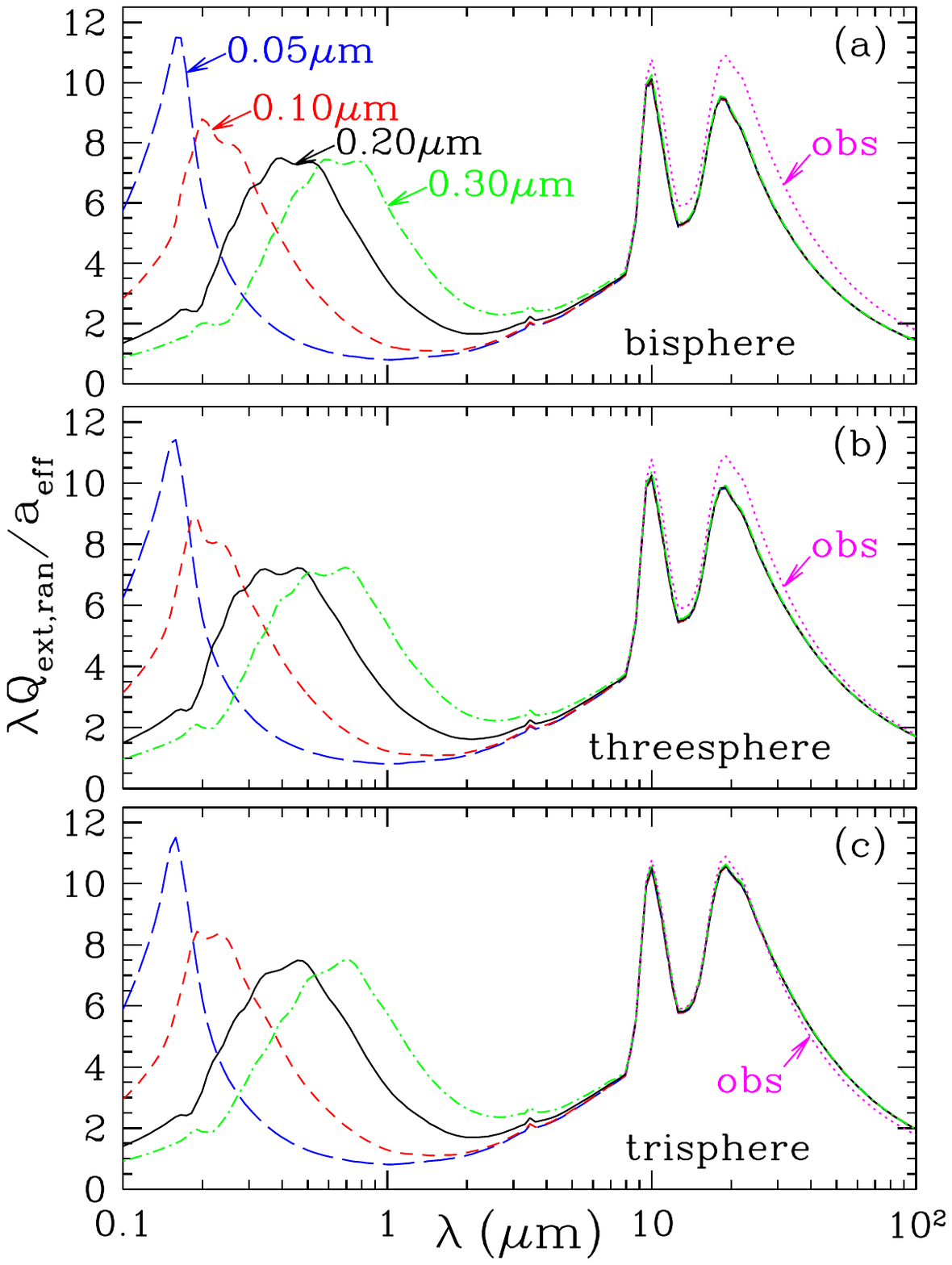}
\includegraphics[angle=0,width=\fwidthb,
                 clip=true,trim=0.5cm 0.5cm 0.5cm 0.5cm]
{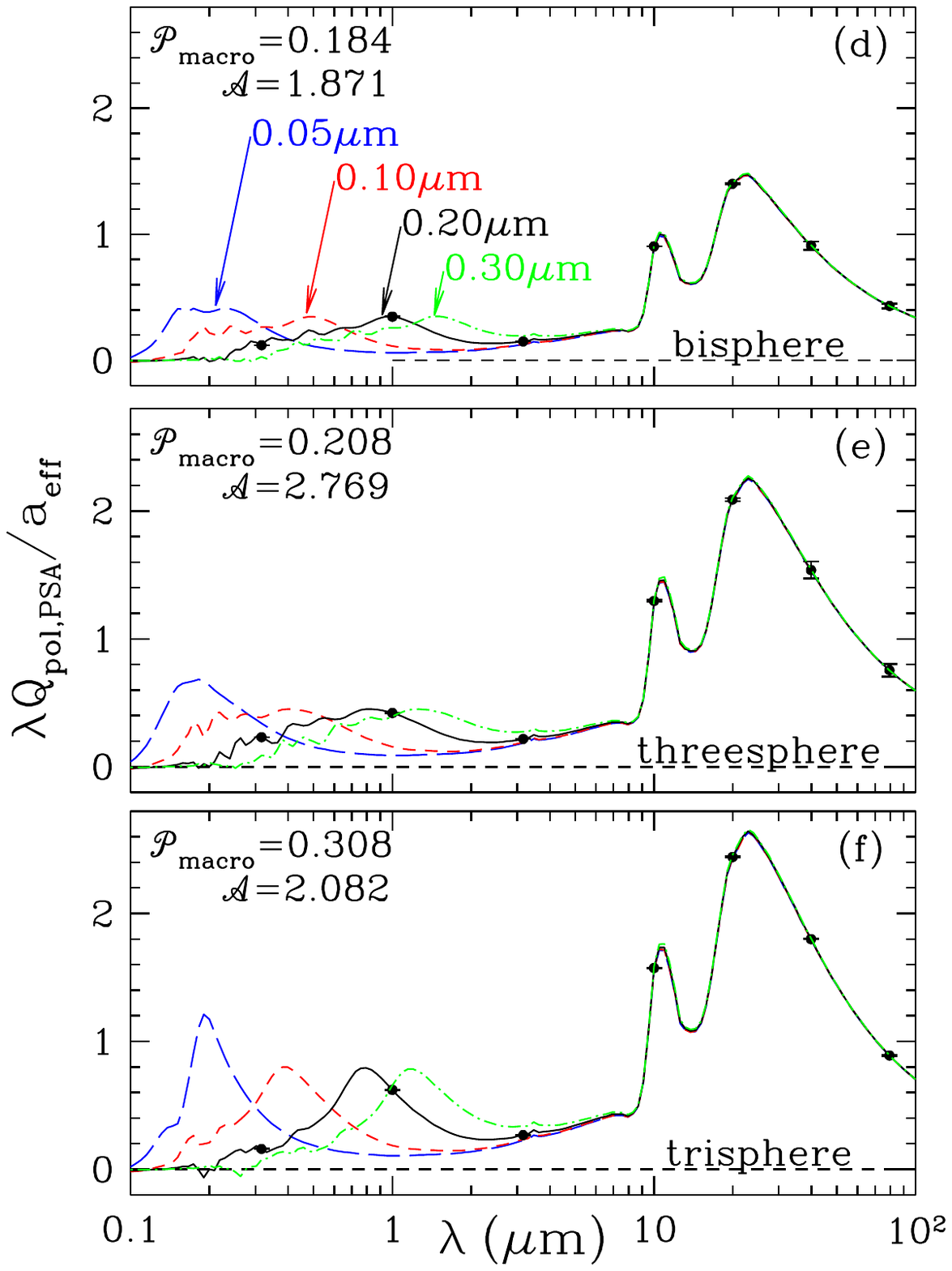}
\caption{\label{fig:Q_bsts}\footnotesize \emph{a-c}: $\lambda Q_{\rm
    ext,ran}/\aeff$ for bisphere, threesphere, and trisphere
  geometries (see Figure \ref{fig:shapes_23}) for
  $\aeff=0.05,0.1,0.2,0.3\micron$.  For $\lambda>8\micron$, the dotted
  curves show the ``observed'' value of $\lambda \Qabs/\aeff$ (see
  text).  \emph{d-f}: $\lambda \QpolPSA/\aeff$ for the same
  geometries.  For $\aeff=0.2\micron$, uncertainty intervals are shown
  at selected $\lambda$.}
\end{center}
\end{figure}
\begin{figure}
\begin{center}
\includegraphics[angle=0,width=\fwidthb,
                 clip=true,trim=0.5cm 0.5cm 0.5cm 0.5cm]
{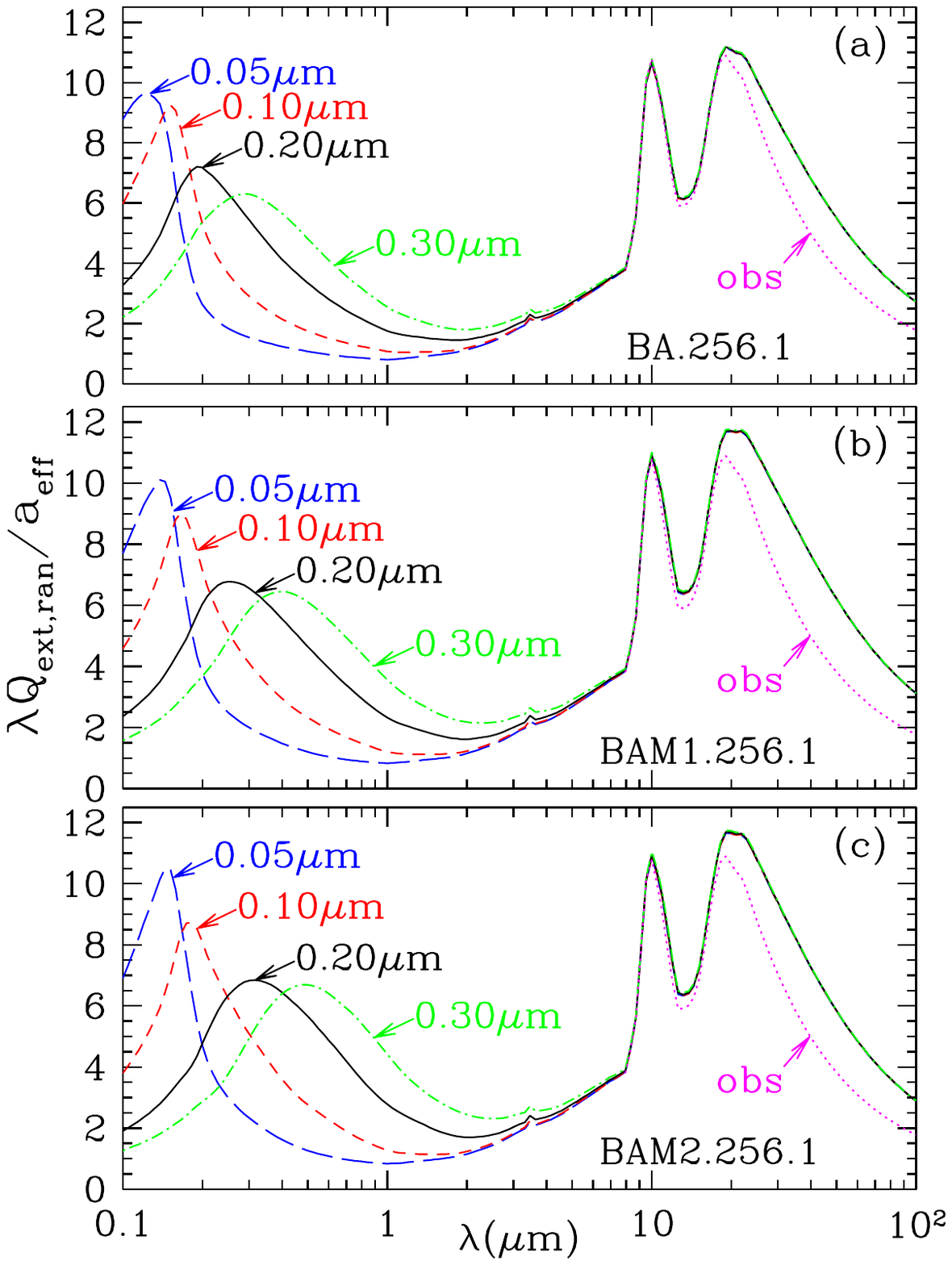}
\includegraphics[angle=0,width=\fwidthb,
                 clip=true,trim=0.5cm 0.5cm 0.5cm 0.5cm]
{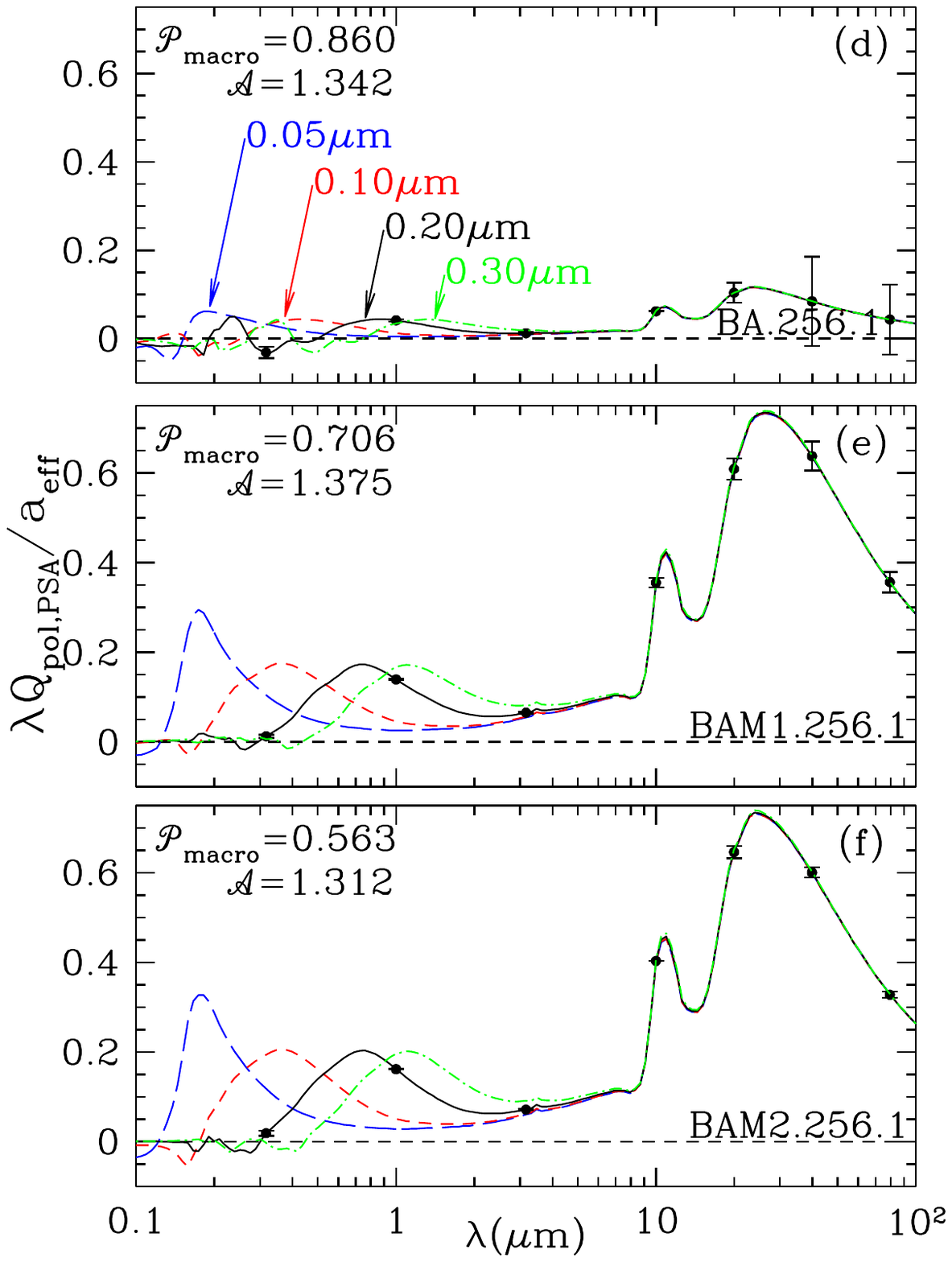}
\caption{\label{fig:Q_babam1bam2}\footnotesize Same as Figure
  \ref{fig:Q_bsts}, but for examples of $N=256$ BA, BAM1, and BAM2
  aggregates (see Figure
  \ref{fig:shapes_babam1bam2}).$^{\ref{fn:website}}$  Panels
  \emph{d,e,f}: Note change of scale relative to Figure
  \ref{fig:Q_bsts} -- these random aggregates are ineffective
  polarizers.}
\end{center}
\end{figure}
\begin{figure}
\begin{center}
\includegraphics[angle=0,width=\fwidthb,
                 clip=true,trim=0.5cm 0.5cm 0.5cm 0.5cm]
{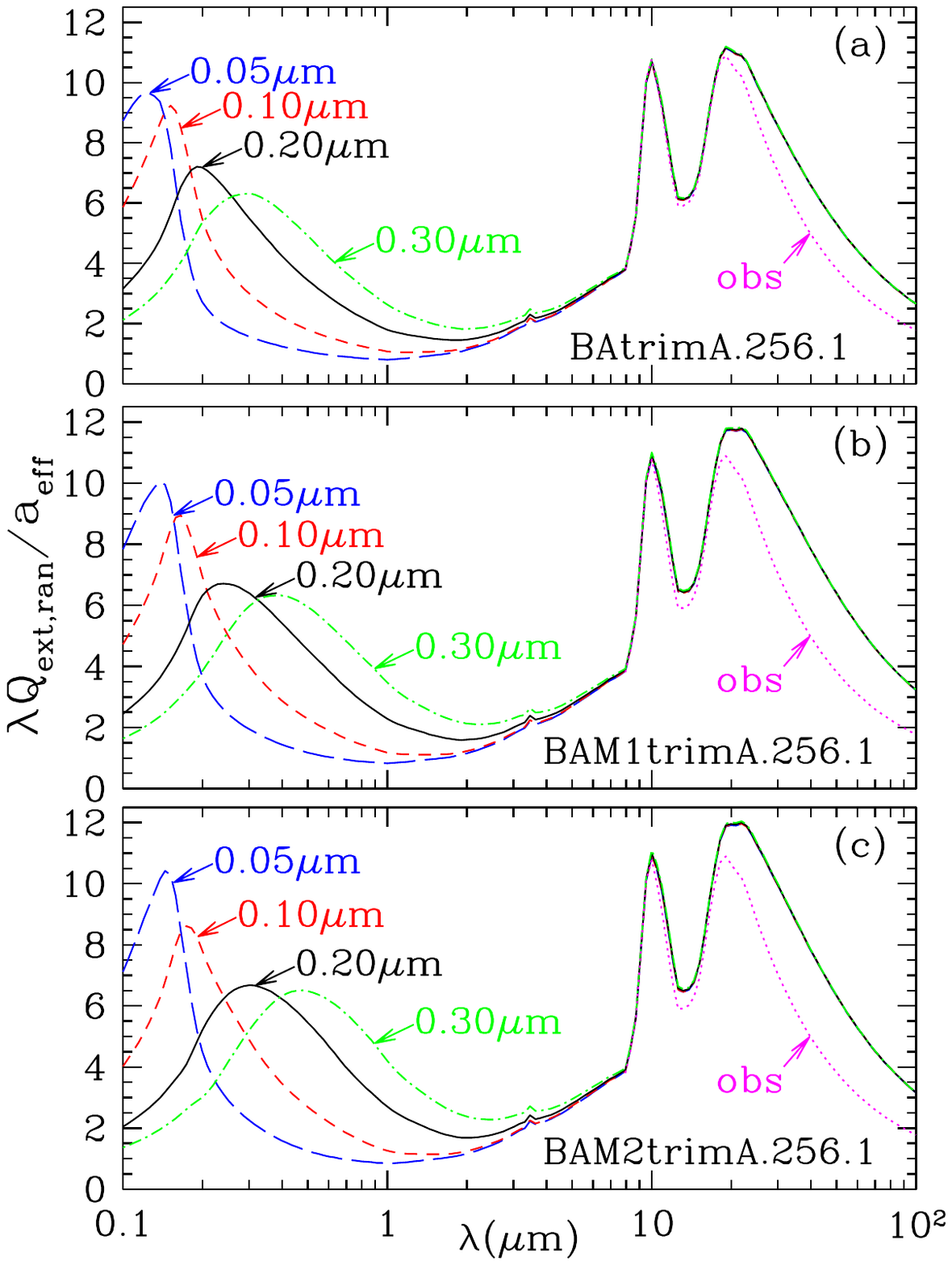}
\includegraphics[angle=0,width=\fwidthb,
                 clip=true,trim=0.5cm 0.5cm 0.5cm 0.5cm]
{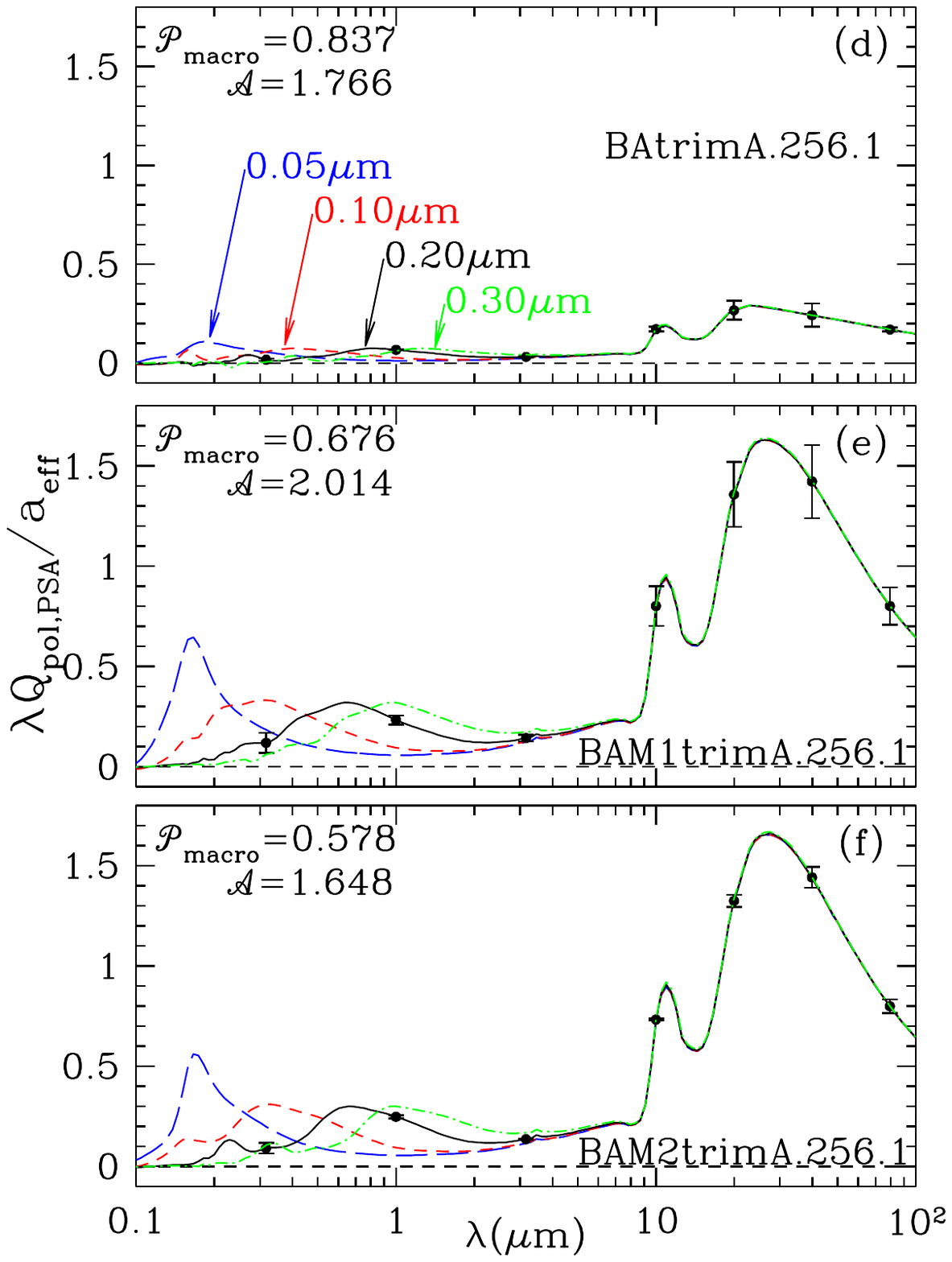}
\caption{\label{fig:Q_trimA} Same as Figure \ref{fig:Q_bsts}, but for
  ``\emph{trimA}'' aggregates (see Figure
  \ref{fig:shapes_trimA}).$^{\ref{fn:website}}$  Panels \emph{d,e,f}:
  Note change of scale relative to Figure \ref{fig:Q_babam1bam2}.}
\end{center}
\end{figure}
\begin{figure}
\begin{center}
\includegraphics[angle=0,width=\fwidthb,
                 clip=true,trim=0.5cm 0.5cm 0.5cm 0.5cm]
{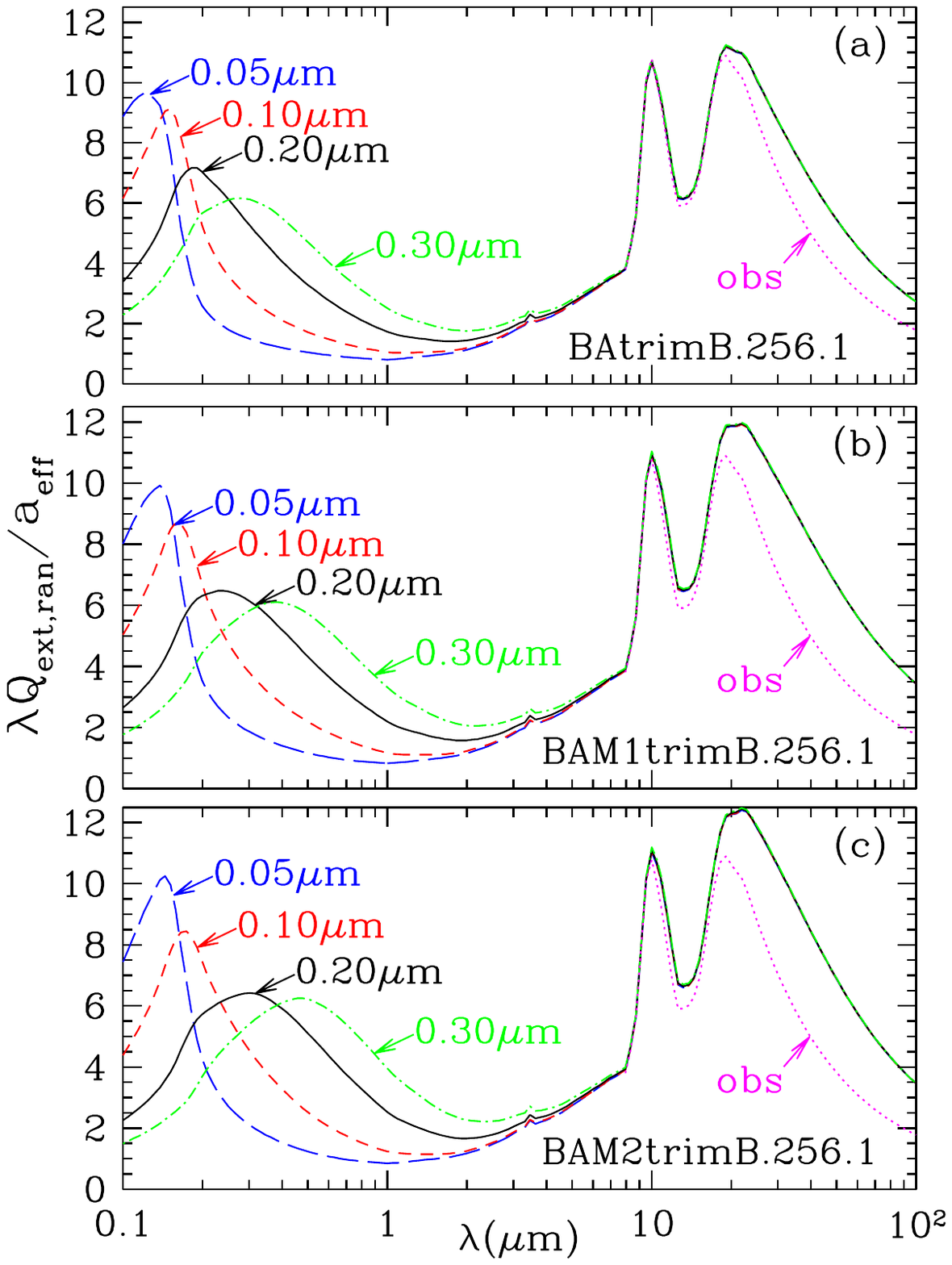}
\includegraphics[angle=0,width=\fwidthb,
                 clip=true,trim=0.5cm 0.5cm 0.5cm 0.5cm]
{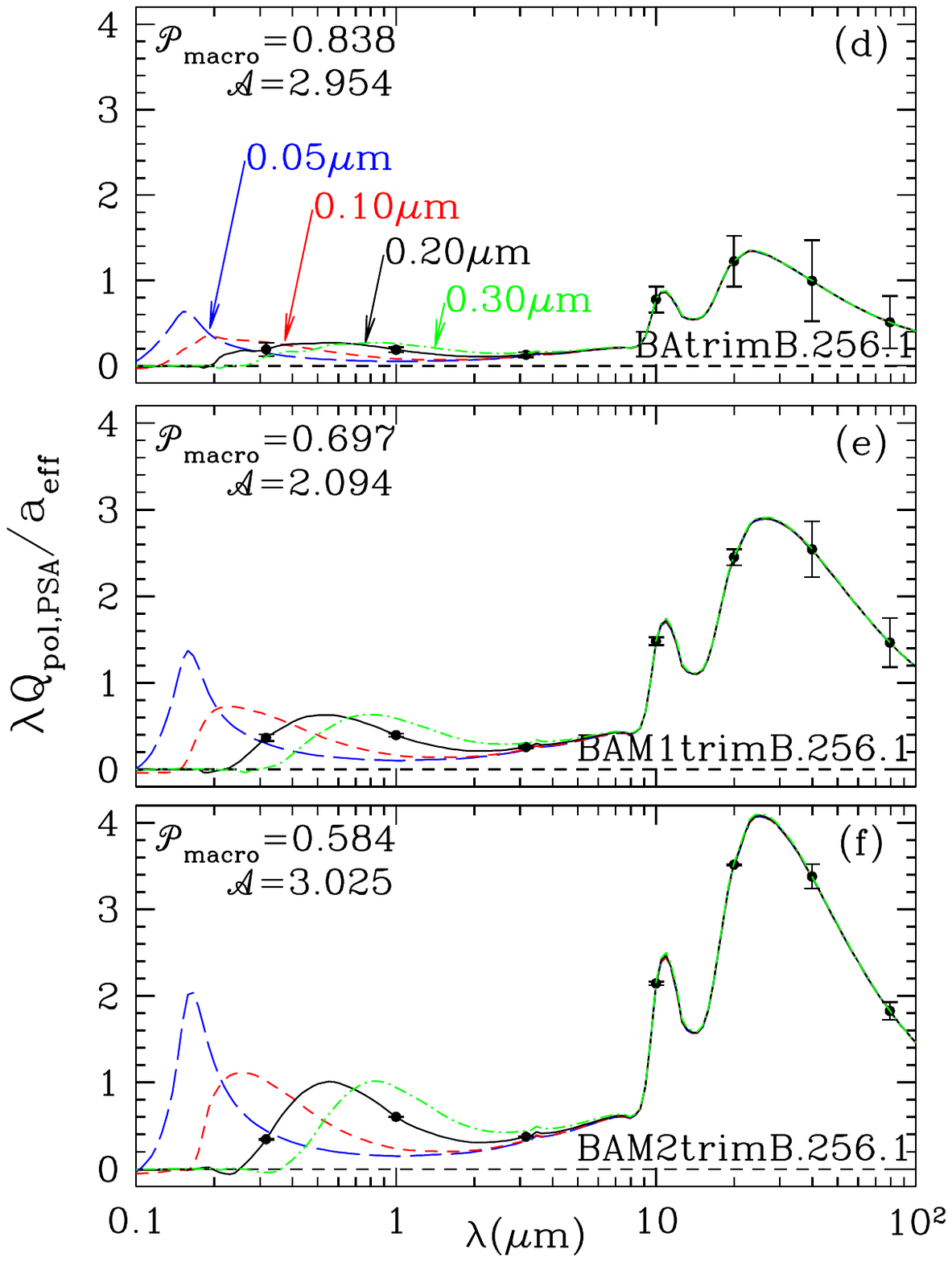}
\caption{\label{fig:Q_trimB}Same as Figure \ref{fig:Q_bsts}, but for
  ``\emph{trimB}'' aggregates (see Figure
  \ref{fig:shapes_trimB}).$^{\ref{fn:website}}$  Panels \emph{d,e,f}:
    Note change of scale relative to Figure \ref{fig:Q_babam1bam2}.}
\end{center}
\end{figure}
\begin{figure}
\begin{center}
\includegraphics[angle=0,width=\fwidthb,
                 clip=true,trim=0.5cm 0.5cm 0.5cm 0.5cm]
{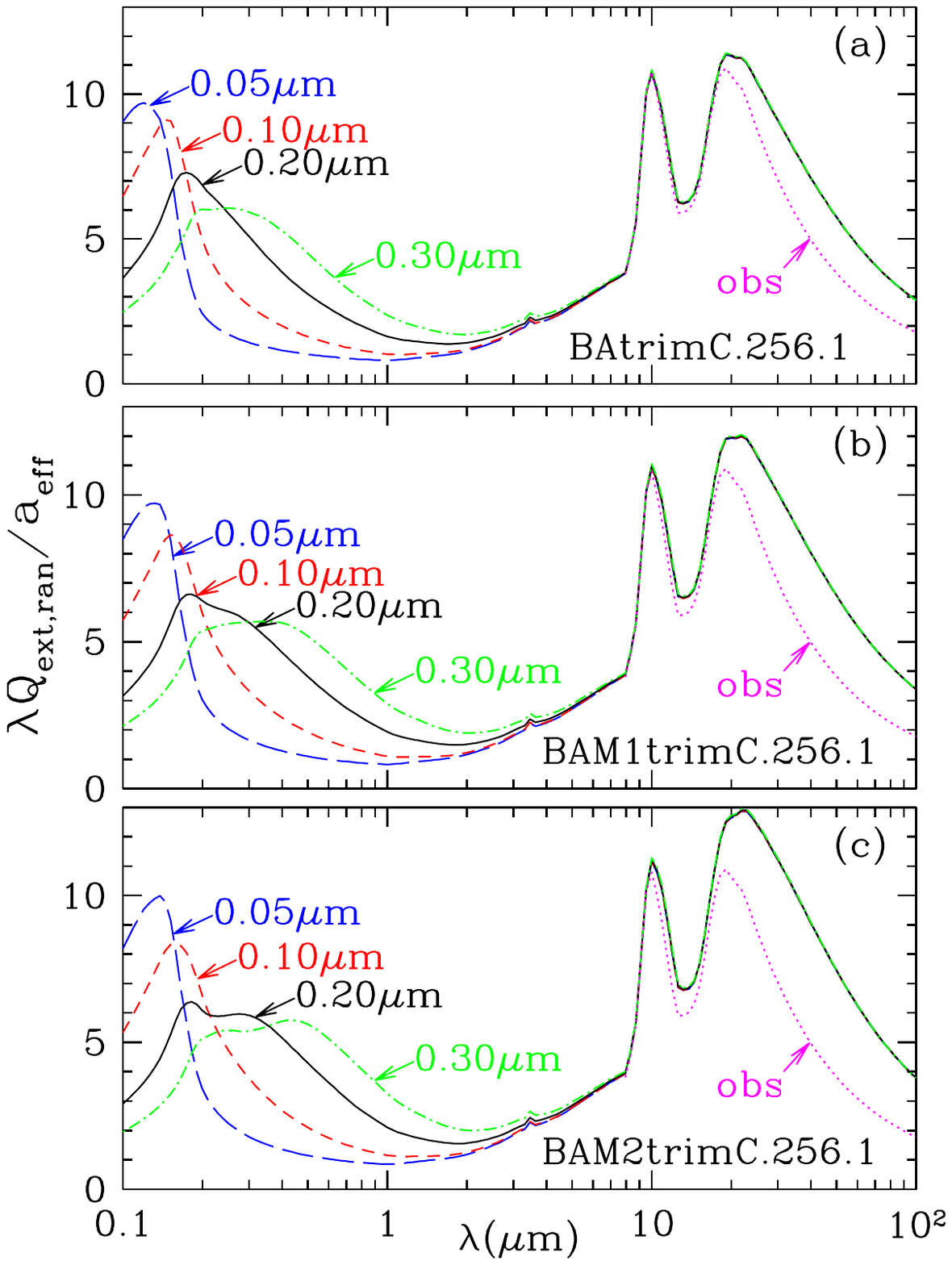}
\includegraphics[angle=0,width=\fwidthb,
                 clip=true,trim=0.5cm 0.5cm 0.5cm 0.5cm]
{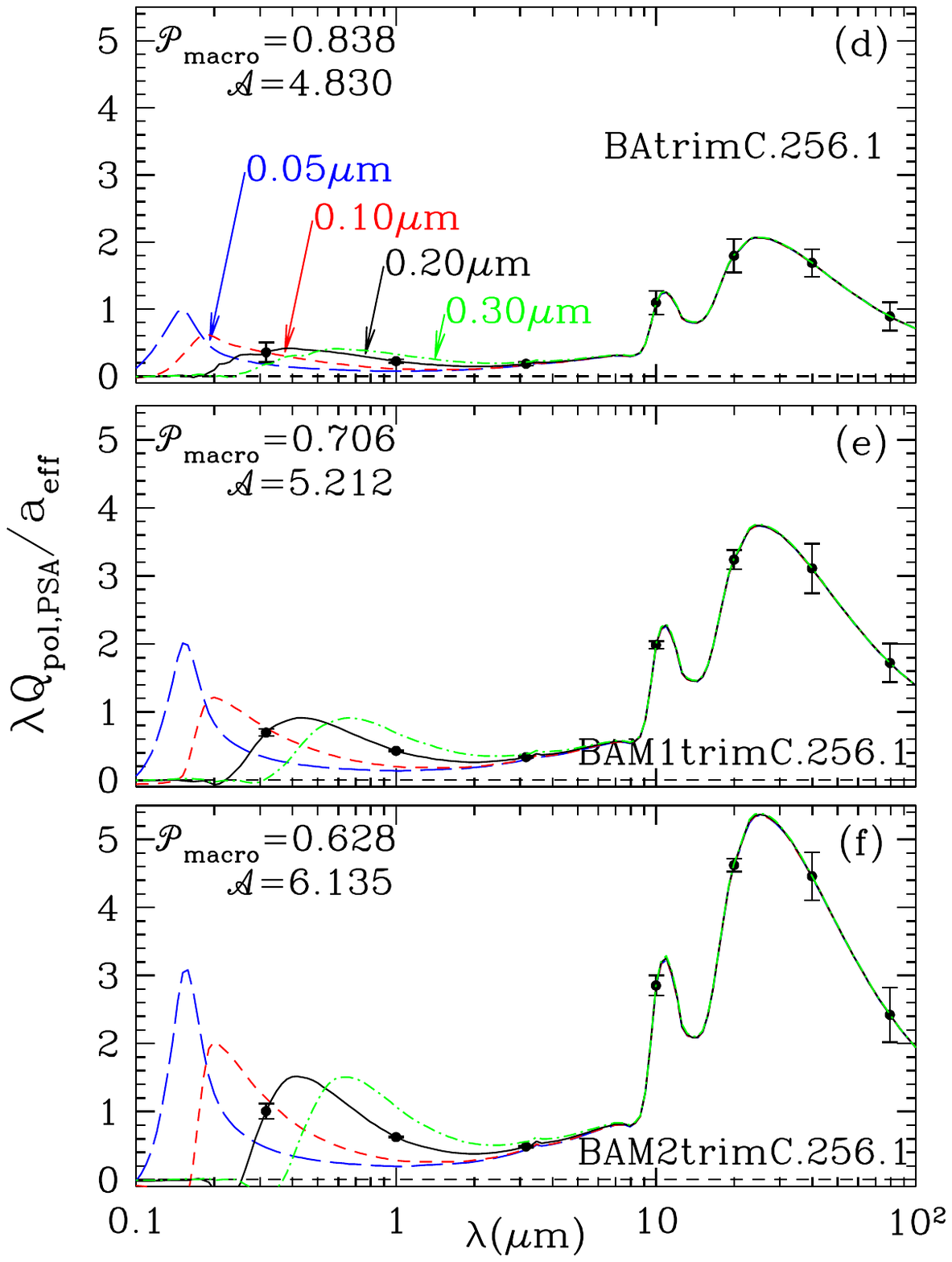}
\caption{\label{fig:Q_trimC}Same as Figure \ref{fig:Q_bsts}, but for
  ``\emph{trimC}'' aggregates (see Figure
  \ref{fig:shapes_trimC}).$^{\ref{fn:website}}$  Panels \emph{d,e,f}:
  Note change of scale relative to Figure \ref{fig:Q_babam1bam2}.}
\end{center}
\end{figure}

\subsection{Extinction}

We define the usual dimensionless efficiency factors:
\beq
Q_x({\rm shape},\lambda) \equiv \frac{C_x({\rm shape},\lambda)}{\pi\aeff^2}
~~~,
\eeq
where $C_x$ is a cross section.  Cross sections per unit
solid volume are proportional to $Q_x/\aeff$:
\beq
\frac{C_x({\rm shape},\lambda)}{\Vsolid} 
= \frac{3}{4}\,\frac{Q_x({\rm shape},\lambda)}{\aeff}
~~~.
\eeq
Figures \ref{fig:Q_bsts}--\ref{fig:Q_trimC} show the dimensionless
quantities $\lambda\Qextran/\aeff$ for extinction by randomly-oriented
particles, and $\lambda\QpolPSA/\aeff$ for particles in ``perfect
spinning alignment'' (PSA), for one example of each of the shape
classes studied here, from the FUV ($\lambda=0.1\micron$) to the FIR
($\lambda=100\micron$).  For each shape, results are shown for four
sizes $\aeff$.

The extinction cross sections per unit volume have broadly similar
behavior.  For $\aeff\approx 0.2\micron$ and $\lambda\ltsim1\micron$,
where scattering is important, $\lambda \Qextran/\aeff$ has a broad
peak, with peak value $\lambda \Qextran/\aeff \approx 7$ near
\beq
\lambda_{{\rm max},\lambda \Qext} \approx 
2.5\left(1-\poromacro\right)^{1/3}\aeff
~~~,
\eeq
e.g., for $\aeff=0.2\micron$, $\lambda_{{\rm max},\lambda\Qext}\approx
0.4\micron$ for the bisphere ($\poromacro = 0.18$, Figure
\ref{fig:Q_bsts}a), and $0.2\micron$ for BA.256.1 ($\poromacro\approx
0.85$, Figure \ref{fig:babam1bam2}a).

For $\lambda\gtsim 5\micron$, the grains have $\aeff\ll\lambda$,
scattering is small compared to absorption ($C_{\rm
  ext}\approx\Cabs$), and $\lambda \Qext/\aeff=(4/3)\lambda C_{\rm
  ext}/\Vsolid$ depends only on the shape and dielectric function (and
therefore $\lambda$), but not on $\aeff$.  For the ``astrodust''
dielectric function, with strong absorption in the silicate features
near 10$\micron$ and 18$\micron$, all shapes have $\lambda
\Qextran/\aeff$ peaking at $\sim$$10\micron$ and $\sim$$20\micron$,
with peak values ranging from $\sim$$10$ for the $N=2$ and $N=3$
aggregates (Figure \ref{fig:Q_bsts}), to $\sim$$12$ for the $N=256$
aggregates (Figures \ref{fig:Q_babam1bam2}-\ref{fig:Q_trimC}).

Figures \ref{fig:Q_bsts}-\ref{fig:Q_trimC} also show the observed
value of $\lambda\Qabs/\aeff=(4\lambda/3)\Cabs/\Vsolid$ for
$\lambda>8\micron$ \citep{Hensley+Draine_2021}.  

For a fixed dielectric function, the opacity depends on the grain
shape.  The dielectric function used in all calculations in the
present paper was devised so that randomly-oriented 1.4:1 oblate
spheroids would reproduce the observed IR-submm opacity
\citep{Draine+Hensley_2021c}.  Figure \ref{fig:Q_bsts} shows that the
trisphere geometry provides nearly the same opacity as the 1.4:1
oblate spheroid (i.e., the curve labelled ``obs''); the bisphere
provides somewhat \emph{less} absorption.

The present study uses the same dielectric function for all shapes.
The high porosity of the BA, BAM1, and BAM2 aggregates results in
enhanced absorption: at $\lambda=100\micron$, the BA, BAM1, BAM2
examples provide more opacity than the 1.4:1 oblate spheroid, by
factors ranging from 1.8--2.3 (see Figure \ref{fig:Q_babam1bam2}).  A
fully self-consistent approach would require that for each shape we
derive a new dielectric function that would reproduce the observed
absorption at long wavelengths.  For spheroidal shapes,
\citet{Draine+Hensley_2021c} developed an iterative technique
employing the analytic result for absorption by spheroids in the
Rayleigh limit.  Unfortunately, analytic results for these complex
shapes are unavailable.  The numerically difficult problem of devising
a self-consistent dielectric function for these complex shapes is
deferred to future work.

\subsection{Polarization}

The polarization cross sections vary greatly from one aggregate to
another.  The original BA, BAM1, and BAM2 aggregates (see Figure
\ref{fig:Q_babam1bam2}) are ineffective polarizers.  The polarization
cross section per volume varies considerably from one random
realization to another, but with a general trend of decreasing
$\QpolPSA/\aeff$ as the porosity increases from
$\poromacro\approx0.58$ to $\poromacro\approx0.85$.

Of the 45 nonconvex shapes studied here, the BAM2trimB and BAM2trimC
geometries, with $\poromacro\approx 0.58$ and aspect ratios in the
range 2.8-7 (see Table \ref{tab:geom}) have the highest polarization
cross section per volume, both in the optical and in the infrared (see
Figure \ref{fig:Q_trimB}c): for the BAM2trimC.256.1 aggregate with
$\aeff=0.20\micron$, $\lambda \QpolPSA/\aeff$ peaks at $\sim$$6$ in
the optical, and $\sim$5 near $25\micron$.  The strength of the
polarization will be discussed further in Section \ref{sec:poleffint}
below.

\medskip

\section{\label{sec:lambdap}
         Starlight Polarization: Effective Wavelength $\lambdap$
         and Profile Width $\sigmap$}

\citet{Draine+Hensley_2021c} showed that interstellar grains must have
certain integral properties to be compatible with the observed
polarization of starlight.  For a grain with a given shape and size
$\aeff$, the effective wavelength for polarization of starlight is
defined to be
\beq
\lambdap({\rm shape},\aeff) \equiv 
\exp\left[
\frac{\int_{\lambda_1}^{\lambda_2}\ln\lambda~\QpolPSA(\lambda) ~d\ln\lambda}
     {\int_{\lambda_1}^{\lambda_2}\QpolPSA(\lambda) ~d\ln\lambda}
\right]
~~~,
\eeq
with $\lambda_1=0.15\micron$ and $\lambda_2=2.5\micron$ to cover the
wavelength range over which starlight polarization is well-observed.

\begin{figure}
\begin{center}
\includegraphics[angle=0,width=\fwidth,
                 clip=true,trim=0.5cm 5.0cm 0.5cm 2.5cm]
{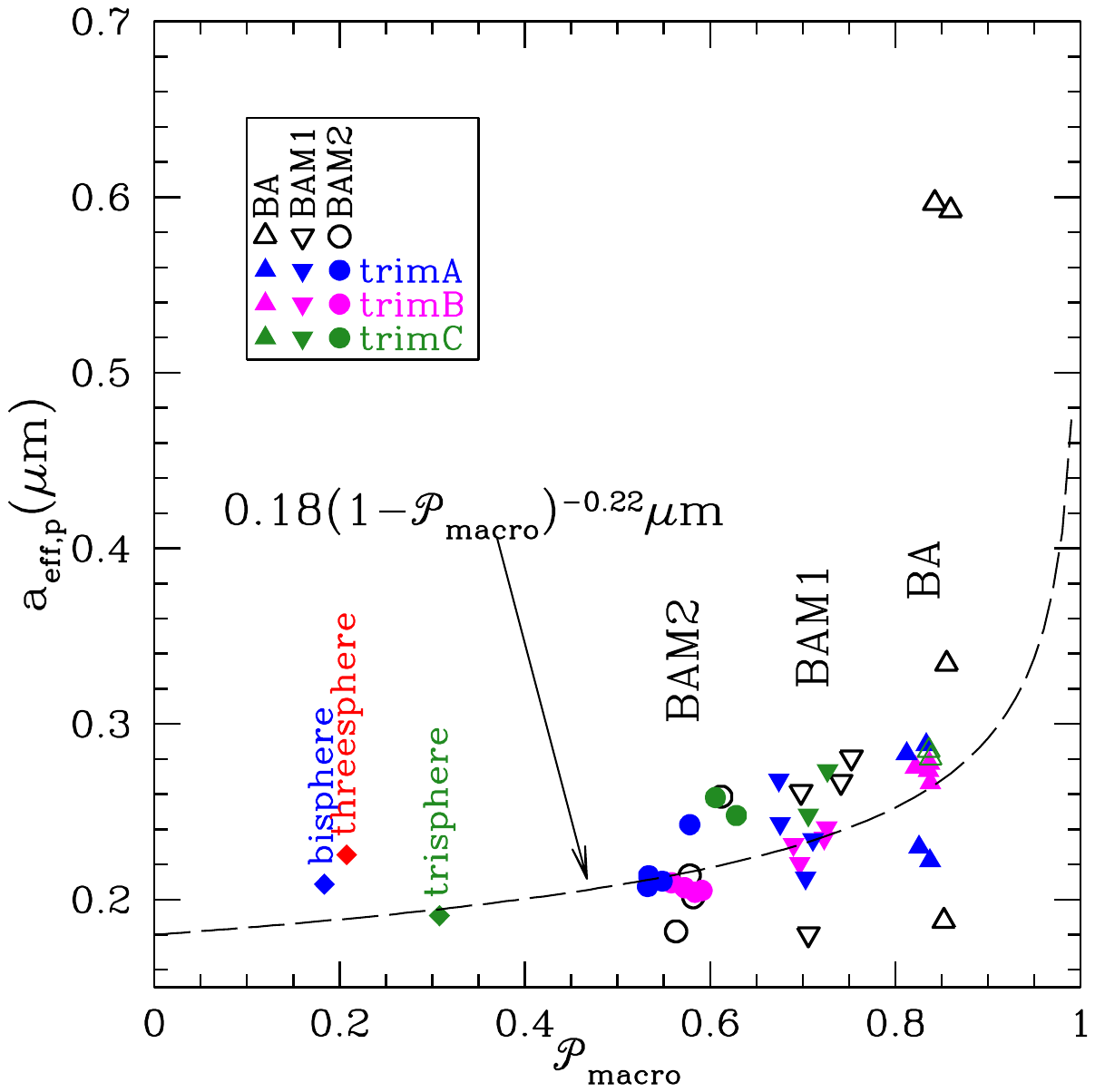}
\caption{\label{fig:aeffp}\footnotesize
  Effective radius $\aeffp$ such that aggregate has $\lambda_p=0.567\micron$,
  plotted against macroporosity $\poromacro$.  
  Broken line: Eq.\ (\ref{eq:aeffp_vs_poro}).}
\end{center}
\end{figure}

To be consistent with the average observed polarization of starlight
(peaking at $\lambda_{\rm max}\approx 0.55\micron$), dust grains
should have sizes $\aeff$ near a characteristic size $\aeffp$ for
which $\lambdap\approx0.567\micron$ (see Paper I).  Using
$\lambdap({\rm shape},\aeff)$ calculated for selected sizes $\aeff$,
we interpolate to estimate $\aeffp({\rm shape})$; the results are
shown in Figure \ref{fig:aeffp} and Table \ref{tab:results}.  The
aligned grains responsible for starlight polarization have a
distribution of sizes, but the distribution should be peaked near
$\aeffp$ in order for the starlight polarization to peak near
$\lambda_{\rm max}\approx 0.55\micron$.

The convex shapes in Paper I \citep{Draine_2024a} had $\aeffp\approx
0.18\micron$.  For the 45 non-convex shapes in the present study,
Figure \ref{fig:aeffp} shows that $\aeffp$ tends to increase with
increasing $\poromacro$.  With the exception of the BA aggregates
(which are very poor polarizers), the other shapes are approximately
consistent with the empirical relation
\beq \label{eq:aeffp_vs_poro}
\aeffp\approx 0.18\micron (1-\poromacro)^{-0.22}
~~~,
\eeq
although individual shapes in Figure \ref{fig:aeffp} scatter above or
below Equation (\ref{eq:aeffp_vs_poro}) by $\pm15\%$ or so.  The
characteristic mass of the polarizing grains is
\beq
M_p = \frac{4\pi}{3}\rhosolid\aeffp^3 \approx
7\xtimes10^{-14}(1-\poromacro)^{-0.66}\gm 
~~~,
\eeq
where we have assumed $\rhosolid\approx2.74\gm\cm^{-3}$ for astrodust
material with $\poromicro\approx0.2$ \citep{Draine+Hensley_2021b}.
The characteristic linear extent of the polarizing grains is
\beq \label{eq:D_p}
D_{\rm p} \approx (1-\poromacro)^{-1/3}\times(2\aeffp) \approx 0.36
(1-\poromacro)^{-0.55}\micron
~~~,
\eeq

In the ISM, the observed dependence of starlight polarization $p_{\rm
  obs}(\lambda)$ on wavelength arises from a mixture of grain sizes.
To reproduce the observed wavelength dependence $p_{\rm
  obs}(\lambda)$, individual grains must have polarization profiles
that are \emph{narrower} than $p_{\rm obs}(\lambda)$.  The width of
the polarization profile for a single grain size and shape is measured
by the dimensionless quantity
\beq \label{eq:sigmap}
\sigmap({\rm shape},\aeff) \equiv
\left[ 
\frac{\int_{\lambda_1}^{\lambda_2} \left[\ln(\lambda/\lambda_p)\right]^2
|\QpolPSA({\rm shape},\aeff,\lambda)| \,d\ln\lambda}
{\int_{\lambda_1}^{\lambda_2}|\QpolPSA({\rm shape},\aeff,\lambda)|
\,d\ln\lambda}
\right]^{1/2}
~~~.
\eeq
\citep{Draine+Hensley_2021c}.  In Paper I, it was argued that
individual grains should have $\sigmap<0.6$ in order to be able to
reproduce the observed polarization profile.  Table \ref{tab:results}
lists $\sigmap({\rm shape},\aeffp)$ for each shape.

\begin{table}
\vspace*{-1.0cm}
\begin{center}
\caption{\label{tab:results}Polarization Parameters for Nonconvex Shapes$^a$}
{\footnotesize
\begin{tabular}{c c c c c}
\hline
shape           & $\aeffp (\mu{\rm m})$$^b$
                        & $\PhiPSA(\aeffp)$$^c$
                                          & $\pfirmax$$^d$ 
                                                      & $\sigmap(\aeffp)$$^e$ \\
\hline
bisphere        & 0.209 & $0.373\pm0.001$ & $0.176\pm0.006$ & $0.603\pm0.001$ \\
threesphere     & 0.226 & $0.538\pm0.001$ & $0.265\pm0.011$ & $0.599\pm0.003$ \\
trisphere       & 0.191 & $0.686\pm0.002$ & $0.274\pm0.001$ & $0.575\pm0.001$ \\
\hline
BA.256.1        & 0.592 & $0.003\pm0.002$ & $0.008\pm0.014$ & $0.637\pm0.067$ \\
BA.256.2        & 0.188 & $0.052\pm0.007$ & $0.005\pm0.026$ & $0.693\pm0.016$ \\
BA.256.3        & 0.334 & $0.030\pm0.001$ & $-0.005\pm0.047$& $0.781\pm0.010$ \\
BA.256.4        & 0.600 & $0.006\pm0.023$ & $0.002\pm0.014$ & $0.465\pm0.182$ \\
\hline
BAM1.256.1      & 0.180 & $0.162\pm0.005$ & $0.065\pm0.005$ & $0.549\pm0.007$ \\  
BAM1.256.2      & 0.267 & $0.232\pm0.021$ & $0.098\pm0.007$ & $0.585\pm0.005$ \\
BAM1.256.3      & 0.261 & $0.095\pm0.021$ & $0.049\pm0.020$ & $0.694\pm0.032$ \\
BAM1.256.4      & 0.281 & $0.131\pm0.017$ & $0.045\pm0.006$ & $0.597\pm0.020$ \\
\hline
BAM2.256.1      & 0.182 & $0.182\pm0.001$ & $0.066\pm0.002$ & $0.534\pm0.012$\\
BAM2.256.2      & 0.214 & $0.356\pm0.013$ & $0.133\pm0.004$ & $0.560\pm0.003$\\
BAM2.256.3      & 0.201 & $0.191\pm0.009$ & $0.069\pm0.015$ & $0.593\pm0.011$ \\
BAM2.256.4      & 0.259 & $0.154\pm0.009$ & $0.074\pm0.005$ & $0.640\pm0.002$ \\
\hline
BAtrimA.256.1   & 0.222 & $0.074\pm0.006$ & $0.038\pm0.008$ & $0.682\pm0.030$ \\
BAtrimA.256.2   & 0.288 & $0.144\pm0.011$ & $0.055\pm0.013$ & $0.643\pm0.012$ \\
BAtrimA.256.3   & 0.230 & $0.139\pm0.001$ & $0.056\pm0.008$ & $0.615\pm0.008$ \\
BAtrimA.256.4   & 0.283 & $0.125\pm0.048$ & $0.031\pm0.010$ & $0.711\pm0.011$ \\
\hline
BAM1trimA.256.1 & 0.242 & $0.358\pm0.048$ & $0.146\pm0.017$ & $0.597\pm0.065$ \\
BAM1trimA.256.2 & 0.212 & $0.298\pm0.012$ & $0.139\pm0.010$ & $0.542\pm0.006$ \\
BAM1trimA.256.3 & 0.268 & $0.291\pm0.085$ & $0.136\pm0.030$ & $0.635\pm0.025$ \\
BAM1trimA.256.4 & 0.233 & $0.415\pm0.029$ & $0.176\pm0.011$ & $0.529\pm0.003$ \\
\hline
BAM2trimA.256.1 & 0.231 & $0.327\pm0.003$ & $0.149\pm0.006$ & $0.618\pm0.003$ \\
BAM2trimA.256.2 & 0.214 & $0.579\pm0.001$ & $0.222\pm0.010$ & $0.548\pm0.016$ \\
BAM2trimA.256.3 & 0.207 & $0.453\pm0.011$ & $0.191\pm0.004$ & $0.502\pm0.072$ \\
BAM2trimA.256.4 & 0.211 & $0.350\pm0.048$ & $0.159\pm0.018$ & $0.634\pm0.014$ \\
\hline
BAtrimB.256.1   & 0.267 & $0.337\pm0.076$ & $0.112\pm0.048$ & $0.545\pm0.018$ \\
BAtrimB.256.2   & 0.274 & $0.447\pm0.002$ & $0.171\pm0.013$ & $0.527\pm0.004$ \\
BAtrimB.256.3   & 0.278 & $0.285\pm0.024$ & $0.109\pm0.008$ & $0.569\pm0.016$ \\
BAtrimB.256.4   & 0.270 & $0.434\pm0.012$ & $0.163\pm0.012$ & $0.520\pm0.007$ \\
\hline
{    BAM1trimB.256.1} & 0.221 & {    0.681$\pm$0.034} & $0.262\pm0.033$ & $0.487\pm0.008$ \\
{\bf BAM1trimB.256.2} & 0.235 & {\bf 0.756$\pm$0.053} & $0.244\pm0.017$ & $0.492\pm0.012$ \\
{\bf BAM1trimB.256.3} & 0.231 & {\bf 0.811$\pm$0.020} & $0.241\pm0.012$ & $0.493\pm0.004$ \\
{    BAM1trimB.256.4} & 0.241 & {    0.744$\pm$0.058} & $0.261\pm0.001$ & $0.508\pm0.020$ \\
\hline
{\bf BAM2trimB.256.1} & 0.204 & {\bf 0.991$\pm$0.005} & $0.225\pm0.003$ & $0.471\pm0.001$ \\
{\bf BAM2trimB.256.2} & 0.210 & {\bf 1.220$\pm$0.018} & $0.209\pm0.004$ & $0.467\pm0.008$ \\
{\bf BAM2trimB.256.3} & 0.207 & {\bf 1.066$\pm$0.024} & $0.224\pm0.004$ & $0.471\pm0.003$ \\
{\bf BAM2trimB.256.4} & 0.205 & {\bf 0.960$\pm$0.018} & $0.230\pm0.001$ & $0.474\pm0.012$ \\
\hline
BAtrimC.256.1         & 0.285 & {    0.484$\pm$0.081} & $0.186\pm0.016$ & $0.501\pm0.013$ \\
BAtrimC.256.2         & 0.281 & {    0.527$\pm$0.018} & $0.210\pm0.004$ & $0.499\pm0.008$ \\
\hline
{\bf BAM1trimC.256.1} & 0.248 & {\bf 0.901$\pm$0.036} & $0.244\pm0.004$ & $0.478\pm0.002$ \\
{\bf BAM1trimC.256.2} & 0.273 & {\bf 0.905$\pm$0.004} & $0.245\pm0.007$ & $0.488\pm0.015$ \\
\hline
{\bf BAM2trimC.256.1} & 0.248 & {\bf 1.311$\pm$0.054} & $0.206\pm0.020$ & $0.462\pm0.002$ \\
{\bf BAM2trimC.256.2} & 0.258 & {\bf 1.475$\pm$0.007} & $0.197\pm0.004$ & $0.469\pm0.005$ \\
\hline
\multicolumn{5}{l}{$a$ Cases with $\PhiPSA>0.7$ and 
$\pfirmax\in[0.20,0.26]$ are shown in bold.}\\
\multicolumn{5}{l}{$b$ $\aeff$ such that $\lambdap=0.567\micron$ (see text).}\\
\multicolumn{5}{l}{$c$ Starlight polarization efficiency integral: Equation (\ref{eq:PhiPSA}).}\\
\multicolumn{5}{l}{$d$ Maximum FIR-submm polarization fraction: Equation (8) from \citet{Draine_2024a}.}\\
\multicolumn{5}{l}{$e$ Polarization profile width parameter: Equation (\ref{eq:sigmap}).}
\end{tabular}
}
\end{center}
\end{table}

\medskip

\section{\label{sec:poleffint}
         Starlight Polarization Efficiency Integral}

If a candidate grain shape is to account for the strength of the
observed starlight polarization, it must be a relatively efficient
polarizer.  \citet{Draine+Hensley_2021c} defined the
dimensionless starlight polarization efficiency integral for grains in
perfect spinning alignment:
\beq \label{eq:PhiPSA}
\PhiPSA(\aeff,{\rm shape}) \equiv 
\int_{\lambda_1}^{\lambda_2} 
\frac{\CpolPSA({\rm shape},\aeff,\lambda)}{\Vsolid} \,d\lambda
~~~.
\eeq
Based on estimates for the solid volume per H nucleon, and
observations of starlight polarization per unit H column density,
\citet{Draine+Hensley_2021c} showed that the grains responsible for
the polarization must have $\PhiPSA \gtsim 0.7$.  For the astrodust
dielectric function for $\poromicro=0.2$, the requirement
$\PhiPSA>0.7$ was satisfied by oblate spheroids with axial ratio $\geq
1.4$, or prolate spheroids with axial ratio $\geq 2$
\citep{Draine+Hensley_2021c}.

\begin{figure}
\begin{center}
\includegraphics[angle=0,width=\fwidth,
                 clip=true,trim=0.5cm 5.0cm 0.5cm 2.5cm]
{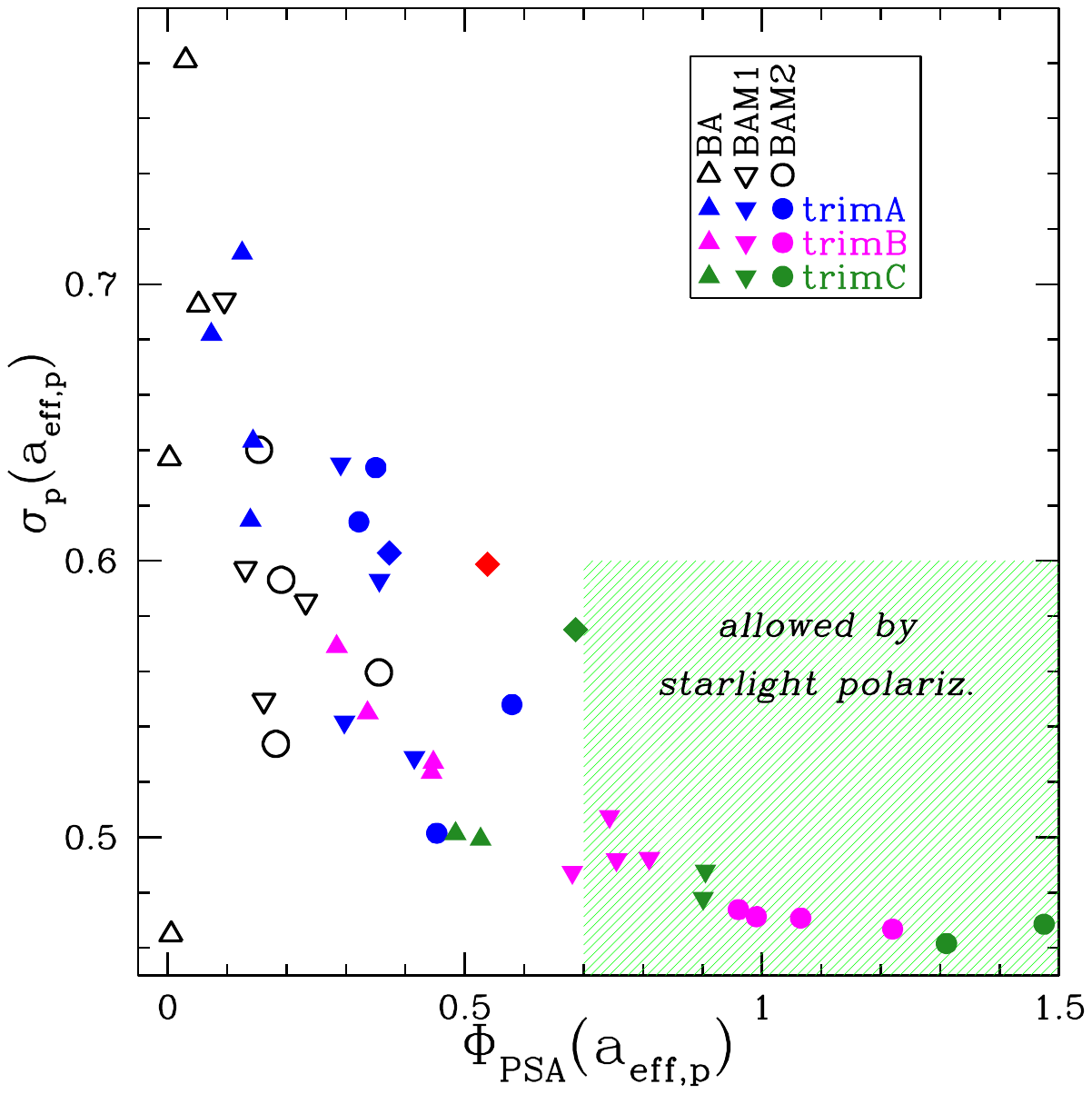}
\caption{\label{fig:sigma_vs_phi}\footnotesize Polarization parameters
  $\sigmap$ and $\PhiPSA$ for 45 aggregates, each with size
  $\aeff=\aeffp$ such that $\lambdap=0.567\micron$.  The allowed
  region $\PhiPSA>0.7$, $\sigmap<0.6$ (see Paper I) is shown in green.
  Only the trisphere, BAM1trimB, BAM1trimC, BAM2trimB, and BAM2trimC
  geometries are within or close to the allowed region -- other shapes
  considered here have $\PhiPSA$ too small to account for the observed
  polarization of starlight.}
\end{center}
\end{figure}

Table \ref{tab:results} gives $\PhiPSA(\aeffp)$ for the 45 shapes
studied here, and Figure \ref{fig:sigma_vs_phi} shows
$\PhiPSA(\aeffp)$ and $\sigmap(\aeffp)$.   

The original BA, BAM1, and BAM2 geometries, with asymmetry parameters
$\Asymm < 1.48$, are not sufficiently asymmetric to account for the
observed polarization.  Macroporosities $\poromacro > 0.5$ are only
viable if the aggregate is more flattened or elongated than the
results of the BA, BAM1, or BAM2 random aggregation processes.  Ten of
the ``trimmed'' BAM1 and BAM2 aggregates (with $\poromacro$ as large
as 0.72) \emph{are} sufficiently flattened or elongated to be able to
satisfy the requirement $\PhiPSA>0.7$ (see Figure
\ref{fig:sigma_vs_phi}).

\begin{figure}
\begin{center}
\includegraphics[angle=0,width=\fwidthc,
                 clip=true,trim=0.5cm 5.0cm 0.5cm 2.5cm]
{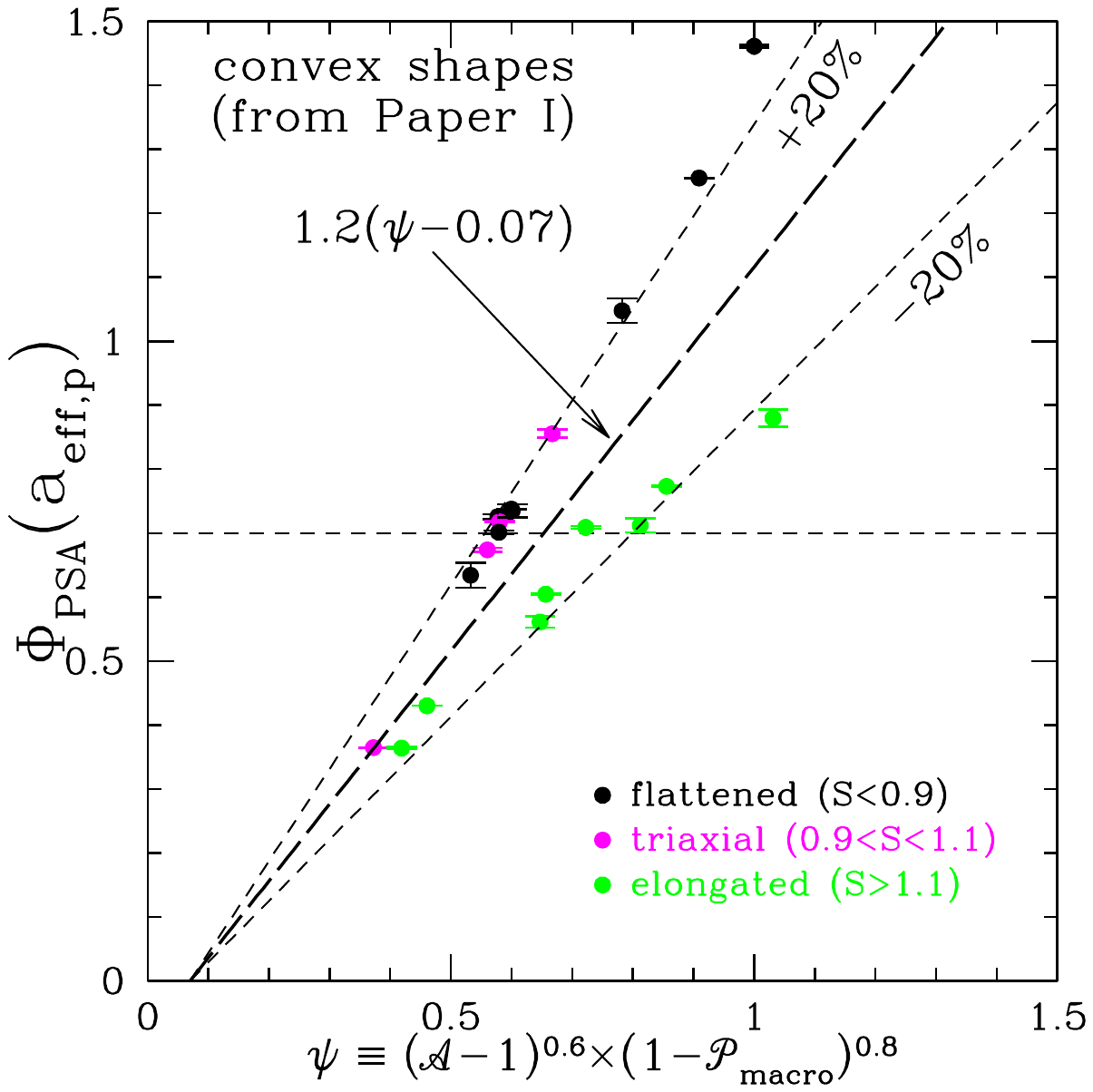}
\includegraphics[angle=0,width=\fwidthc,
                 clip=true,trim=0.5cm 5.0cm 0.5cm 2.5cm]
{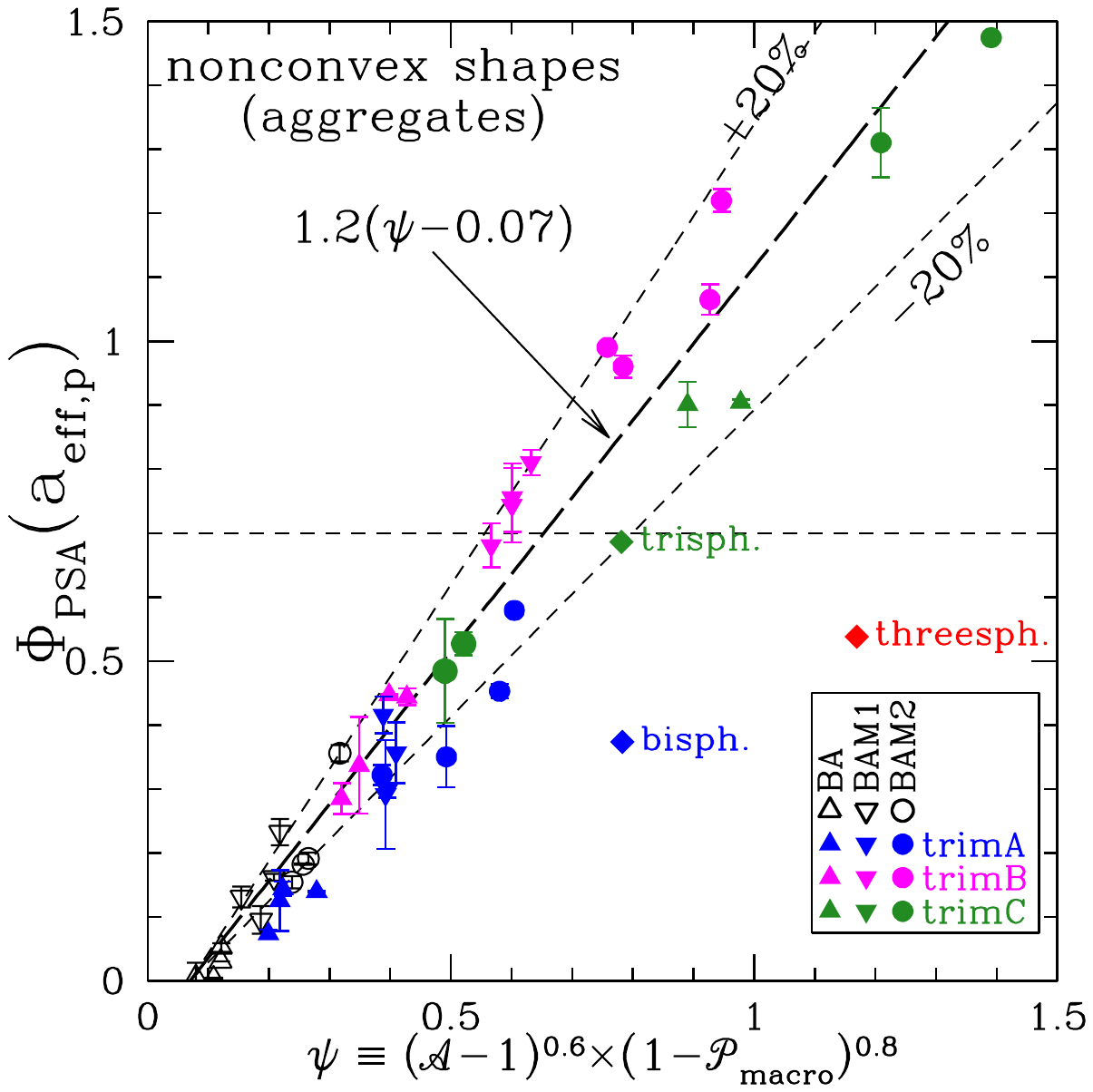}
\caption{\label{fig:phi_vs_A}\footnotesize Starlight polarization
  efficiency factor $\PhiPSA$
  vs.\ $\psi\equiv(\Asymm-1)^{0.8}(1-\poromacro)^{0.6}$ for (a) convex
  shapes from \citet{Draine_2024a} (b) nonconvex aggregates in present
  study.}
\end{center}
\end{figure}

Using the DDA to obtain $\PhiPSA$ for a given shape and size is
computationally demanding.  It would be valuable to be able to
estimate $\PhiPSA(\aeffp)$ from $\poromacro$ and a simple measure of
grain geometric asymmetry.  Figure \ref{fig:phi_vs_A} shows that the
calculated values of $\Phi({\rm shape},\aeffp)$ are approximately
consistent with a simple empirical relation:
\beqa \label{eq:Phiapprox}
\PhiPSA({\rm shape},\aeffp) &~\approx~& 1.2(\psi-0.07)
\\
\psi&\equiv&
(\Asymm-1)^{0.6}(1-\poromacro)^{0.8}
\eeqa
for the 20 convex shapes (with $0\leq \poromacro < 0.12$) in Paper I,
and the 45 nonconvex aggregates (with $0.18<\poromacro<0.84$) studied
here.  Evidently the single parameter $\psi$ is a good predictor of a
given shape's ability to polarize starlight.  For shapes with
$\PhiPSA(\aeffp)\gtsim 0.3$, Eq.\ (\ref{eq:Phiapprox}) predicts
$\PhiPSA$ to within $\sim$$\pm20$\%, although there are a few
conspicuous deviations [e.g., the bisphere and threesphere shapes fall
  a factor $\sim$2 below Equation (\ref{eq:Phiapprox})].

The requirement $\PhiPSA(\aeffp)\gtsim 0.7$ places a lower bound on
the asymmetry parameter
\beq \label{eq:Asymm_min}
\Asymm ~\gtsim~ 1 + \frac{0.49}{(1-\poromacro)^{4/3}}
~~~.
\eeq
According to Equation (\ref{eq:Asymm_min}), macroporosity
$\poromacro>0.9$ would require very extreme shapes with $\Asymm\gtsim
12$ (e.g., oblate spheroids with axial ratio exceeding $\sim$12:1, or
prolate spheroids with axial ratio exceeding $\sim$16:1) to achieve
$\PhiPSA\gtsim 0.7$ as required by observations.  Such extreme shapes
seem implausible.

\medskip
\section{\label{sec:infrared}
         Polarization in the Infrared}

\subsection{Far-Infrared}

For a given grain shape, the polarization fraction of the FIR emission
depends on the degree of alignment of the grains.  For the degree of
alignment that reproduces the observed maximum amount of starlight
polarization per unit reddening, one can predict the maximum amount of
FIR polarization, $\pfirmax$.  Because the fractional polarization is
essentially independent of $\lambda$ for $\lambda\gtsim 50\micron$
(see Fig.\ 11 of Paper I), we can compare the calculated maximum
polarization fraction $[p_{\rm em}(100\micron)]_{\rm max}$ with
$[p_{\rm em}(850\micron)]_{\rm max}=0.220_{-0.014}^{+0.035}$
determined by \citet{Planck_2018_XII}.

\begin{figure}
\begin{center}
\includegraphics[angle=0,width=\fwidth,
                 clip=true,trim=0.5cm 5.0cm 0.5cm 2.5cm]
{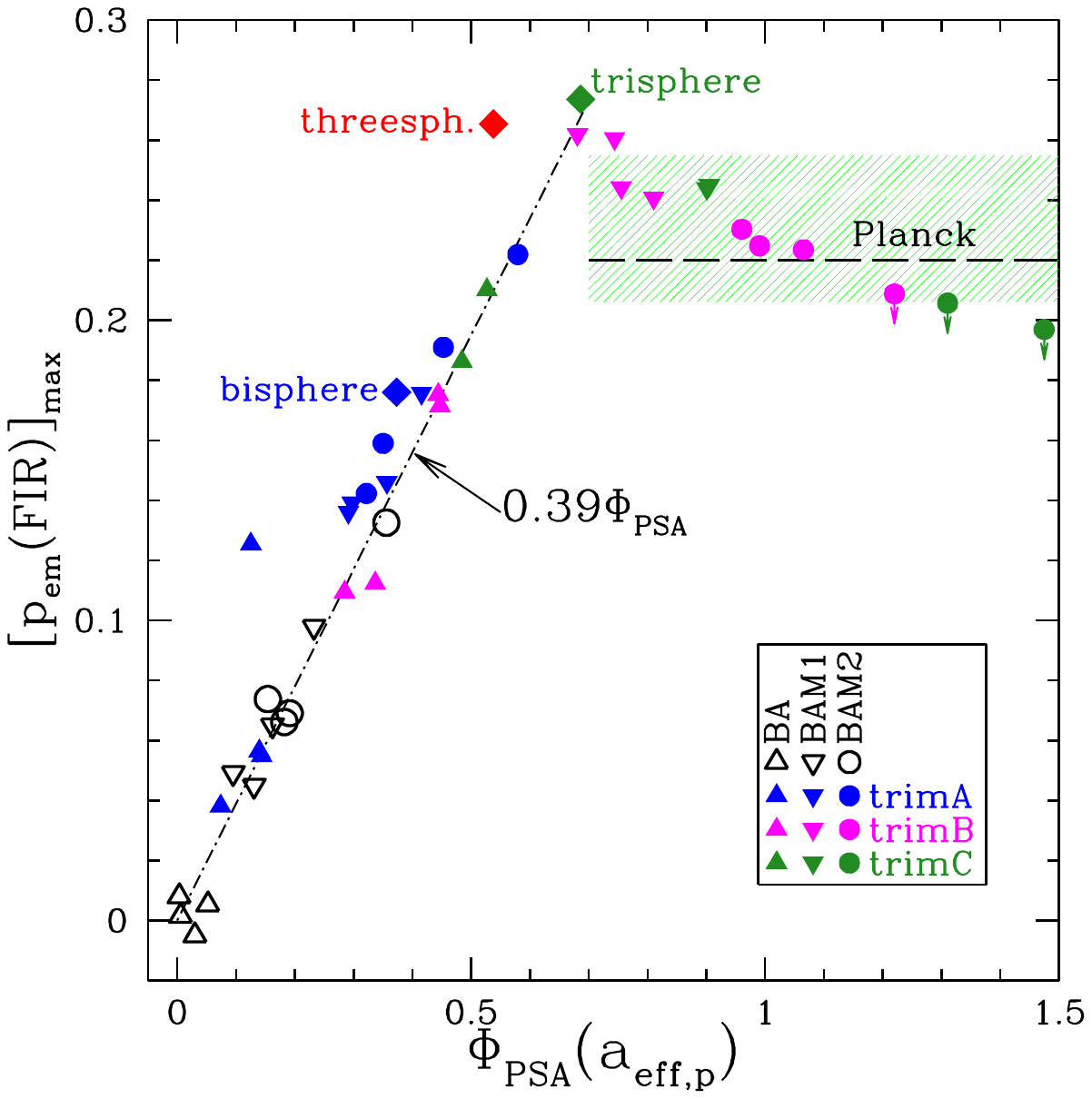}
\caption{\label{fig:firpol}\footnotesize Polarization parameters
  $\pfirmax$ and $\PhiPSA(\aeffp)$ for the 45 aggregates.  The region
  allowed by {\it Planck} polarization is shown in green.  7 shapes
  fall within the allowed region.}
\end{center}
\end{figure}

Table \ref{tab:results} lists $\pfirmax$ evaluated using Equation (28)
from Paper I.\footnote{
For shapes with $\PhiPSA(\aeffp)<0.7$ (35 of the 45 shapes in the
present study), $\pfirmax$ is evaluated
assuming that a fraction $\falign=0.7$ of the grain mass is in perfect
spinning aligment, and the remaining 30\% of the mass is in small
grains that are randomly oriented.  For shapes with
$\PhiPSA(\aeffp)>0.7$, $\pfirmax$ is
calculated assuming $\falign=0.49/\PhiPSA(\aeffp)$.}
Figure \ref{fig:firpol} shows $\pfirmax$
versus $\PhiPSA(\aeffp)$ for the 45 nonconvex shapes studied here.
For $\PhiPSA(\aeffp)< 0.7$, $\pfirmax$ varies
approximately linearly with $\PhiPSA(\aeffp)$ (dot-dash line in Figure
\ref{fig:firpol}).

The calculated values of $\pfirmax$ are very small for the BA, BAM1,
and BAM2 random aggregates, particularly the highest-porosity BA
aggregates, but these shapes are already ruled out by their inability
to provide enough starlight polarization.  Figure
\ref{fig:sigma_vs_phi} showed that 10 of the trimmed aggregates were
consistent with starlight polarization, including the requirement
$\sigma_p<0.6$; Figure \ref{fig:firpol} shows that 7 of these are also
consistent with the {\it Planck} polarization constraints.

It is also notable in Figure \ref{fig:firpol} that the 13 shapes with
$\PhiPSA>0.65$ have $\pfirmax$ tending to decline with increasing
$\PhiPSA$: as the shapes become more asymmetric, the starlight
polarization efficiency $\PhiPSA$ increases more rapidly than
$\QpolPSA/\Qran$, resulting in a drop in $\pfirmax$.  The same trend
was previously seen for prolate spheroids, oblate spheroids, and
square prisms \citep{Draine_2024a}.  For all studied shapes (including
the convex shapes in Paper I), we have no examples of shapes with
$\PhiPSA>1.3$ that have $\pfirmax$ large enough to be consistent with
the \emph{Planck} constraint.

It is important to note that the present results were all calculated
using a single dielectric function.  As discussed above, for the
$M=256$ aggregates, this dielectric function gives a FIR opacity
exceeding the observed value (see Figures
\ref{fig:Q_babam1bam2}-\ref{fig:Q_trimC}).  A dielectric function with
reduced FIR absorption (as required for these aggregates to be
consistent with the observed FIR-submm opacity) would give less total
emission and less polarized emission in the FIR.  The reduction in
both Re($\epsilon$) and Im($\epsilon$) will result in a modest reduction in
the polarization fraction in the FIR.\footnote{
Figure 11 of Paper I shows the fractional polarization for 2:1 oblate
spheroids calculated with two different dielectric functions.  The
change in dielectric function resulted in a drop in $\pfirmax$ from
$0.171$ to $0.166$ -- a fractional reduction of $\sim$3\% in
$\pfirmax$.}
Allowing for this reduction in $\pfirmax$, we conclude that the three
shapes with $\PhiPSA>1.2$ fall below the allowed range of $\pfirmax$:
only seven of the 45 shapes considered here have $\PhiPSA>0.7$ and
$\pfirmax$ consistent with \emph{Planck}.

\medskip

\subsection{10$\mu${\rm m} Polarization}

\begin{figure}
\begin{center}
\includegraphics[angle=0,width=\fwidth,
                 clip=true,trim=0.5cm 5.0cm 0.5cm 2.5cm]
{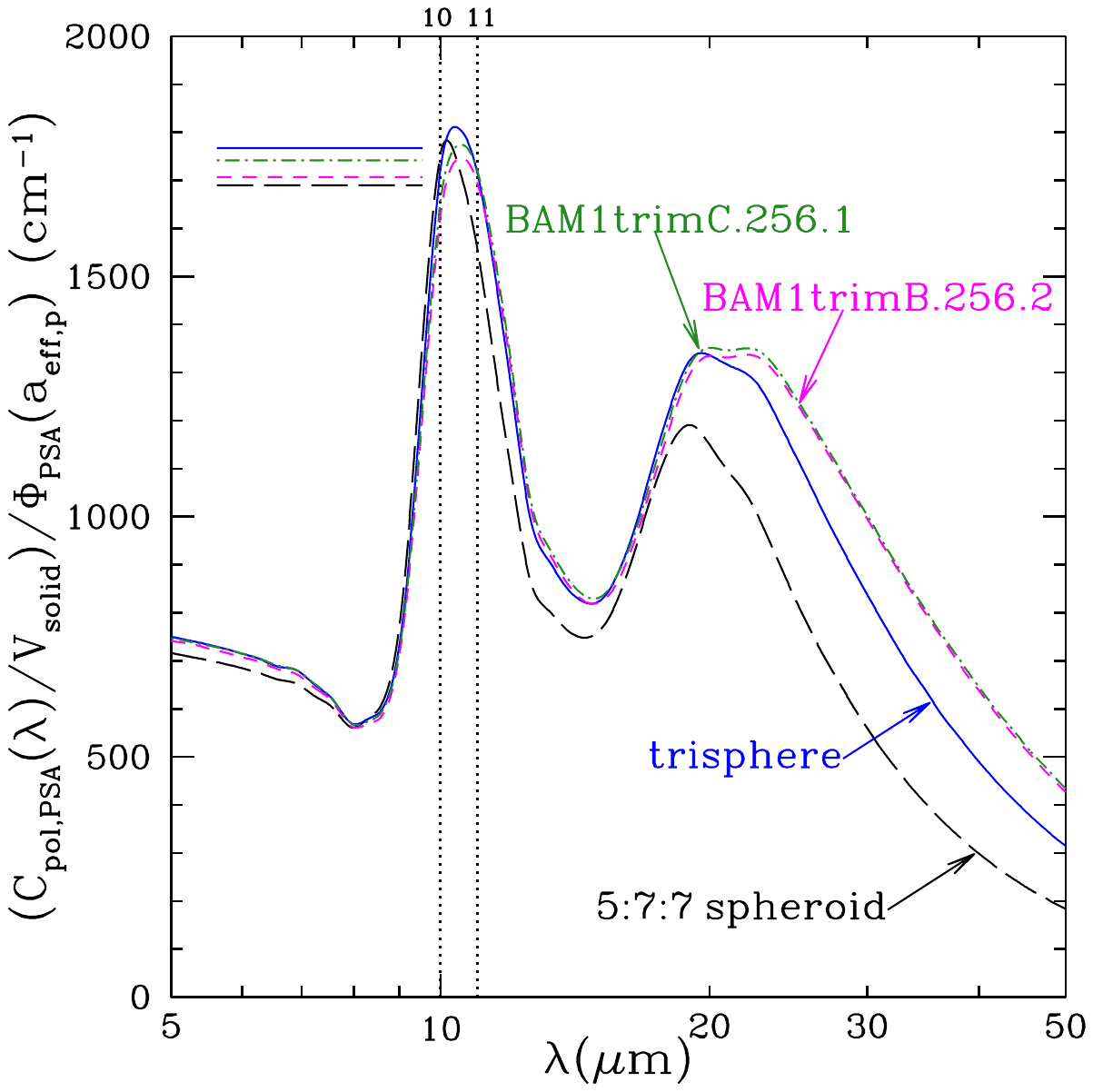}
\caption{\label{fig:silpol}\footnotesize Polarization in the 10 and
  18$\mu$m silicate features relative to starlight polarization for
  three aggregates: the trisphere, BAM2trimB.256.1, and
  BAM2trimA.256.2.  Result for 1.4:1 oblate spheroid is shown for
  comparison.  Horizontal lines indicate values of
  $(\overline{\CpolPSA}(10-11\micron)/\Vsolid)/\PhiPSA(\aeffp)$ (see
  Table \ref{tab:10um}).}
\end{center}
\end{figure}

In the astrodust model \citep{Hensley+Draine_2023}, the grains
providing starlight polarization are composed of a mixture of
materials, including the amorphous silicates that produce infrared
extinction.  In this model, starlight polarization in the optical must
be accompanied by polarization in the 10$\micron$ silicate feature.
We calculate the ratio of the $10\micron$ polarization cross section
per unit volume to the starlight polarization efficiency integral
$\PhiPSA(\aeffp)$ following the treatment in Paper I.

The untrimmed BA, BAM1, and BAM2 random aggregates are not of
interest: for these shapes, $\PhiPSA$ is so small that such grains are
incapable of contributing appreciably to the observed starlight
polarization.  In Figure \ref{fig:silpol} we show
$\CpolPSA(\lambda)/(\Vsolid\PhiPSA)$ for the trisphere geometry and
examples of BAM1trimB and BAM1trimC aggregates that are compatible
with starlight polarization, as well as a 5:7:7 oblate spheroid
\citep{Draine_2024a}.  Table \ref{tab:10um} lists results for the 7
aggregates that are compatible with starlight polarization and {\it
  Planck} polarization (see Figure \ref{fig:firpol}).  The $10\micron$
polarization profile is nearly unaffected by shape or porosity.

Averaging the
polarization from 10$\micron$ to $11\micron$, all seven shapes fall in
the relatively narrow range
%
\beq
\frac{\Delta\CpolPSA({\rm sil.})}{\Vsolid\,\PhiPSA} 
\equiv
\frac{\overline{\CpolPSA}(10\!-\!11\micron)-\overline{\CpolPSA}(8\!-\!9\micron)}
     {\Vsolid\,\PhiPSA} =  1114\pm 29 \cm^{-1}
~~~.
\eeq
This predicts polarization in the $10\micron$ feature, relative to the
optical polarization $p_{\rm max}$
\citep{Draine+Hensley_2021c,Draine_2024a}:
\beqa \label{eq:polsil/pmax}
\left(\frac{\Delta p_{\rm sil}}{p_{\rm max}}\right)
&=&
1.23\xtimes10^{-4}\cm \times 
\frac{\Delta\CpolPSA({\rm sil.})}{\Vsolid\,\PhiPSA}
\\ \nonumber 
&=& 0.137 \pm 0.004 ~~~.
\eeqa
This will be further discussed in Section \ref{subsec:10um}.
\begin{table}
\begin{center}
\caption{\label{tab:10um}10 Micron Polarization}
{\footnotesize
\begin{tabular}{c c c c c}
\hline shape & $\poromacro$ & $\PhiPSA$ & $\left[\frac{\overline{\CpolPSA}(10-11\,\mu{\rm
      m})}{\Vsolid~\PhiPSA(\aeffp)}\right]$ &
$\left[\frac{\Delta\CpolPSA({\rm
      sil.})}{\Vsolid~\PhiPSA(\aeffp)}\right]$ \\
  & & & $(\cm^{-1})$ & $(\cm^{-1})$ \\
\hline
5:7:7 spheroid$^a$&0.000 & $0.726\pm0.004$ & $1682 \pm 1$ & $1075 \pm 3$ \\
\hline
BAM2trimB.256.1 & 0.584 & $0.991\pm0.005$ &$1707 \pm 1$  & $1103 \pm 2$ \\
%
%
BAM2trimB.256.3 & 0.577 & $1.067\pm0.022$ &$1709 \pm 6$  & $1109 \pm 7$ \\
BAM2trimB.256.4 & 0.591 & $0.961\pm0.018$ & $1702 \pm 16$ & $1101 \pm 16$ \\
%
%
%
BAM1trimB.256.2 & 0.723 & $0.756\pm0.054$ &$1721 \pm 9$  & $1109 \pm 11$ \\
BAM1trimB.256.3 & 0.690 & $0.811\pm0.020$ &$1741 \pm 9$  & $1129 \pm 6$ \\
BAM1trimC.256.1 & 0.706 & $0.935\pm0.026$ &$1742 \pm 10$ & $1129 \pm 9$ \\
BAM1trimC.256.2 & 0.727 & $0.905\pm0.004$ & $1749 \pm 11$ & $1131 \pm 12$ \\
\hline
range for allowed aggregates           && & $1733\pm37$   & $1114\pm29$ \\
\hline
\multicolumn{4}{l}{$a$ from Paper I \citep{Draine_2024a}.}\\
\end{tabular}
}
\end{center}
\end{table}

\medskip

\medskip
\section{\label{sec:discuss}
         Discussion}


\subsection{$N=2$ and $N=3$ Aggregates}

Grain-grain collisions at speeds resulting in sticking are expected to
take place in the ISM.  \citet{Jura_1980} pointed out that coagulation
appears to be \emph{required} to explain the observed grain growth on
the sightline to $\rho$\,Ophiuchi.

Coagulation of approximately equal-size grains might result in
particles resembling the symmetric bisphere or trisphere geometries
considered here.  Prior to the present study, the symmetric bisphere
geometry seemed likely (at least to the author) to be an efficient
polarizer, at both optical and FIR wavelengths.  It was anticipated
that at optical wavelengths the same near-field wave effects that
result in effective polarization by 2:1:1 prolate spheroids would make
bispheres similarly effective.

Therefore, it was surprising to find that symmetric bispheres are
significantly \emph{less} effective for starlight polarization than
2:1:1 spheroids or 2:1:1 cylinders (see Figure
\ref{fig:sigma_vs_phi}), with $\PhiPSA(\aeffp)\approx 0.37$ -- only
$\sim$50\% of $\PhiPSA\approx 0.7$ found in Paper I for 2:1:1
spheroids and 2:1:1 cylinders.  The low polarizing efficiency of
bispheres implies that grains with bisphere morphology \emph{cannot}
contribute significantly to either the observed extinction curve or
the observed starlight polarization.

Similarly, the $N=3$ aggregates were expected to be effective
polarizers.  The trisphere and threesphere structures both have
starlight polarization efficiency factors $\PhiPSA$ that are larger
than the bisphere, but the threesphere geometry ($\PhiPSA\approx0.54$)
still falls below the lower limit $\PhiPSA>0.7$ required to reproduce
the observed polarization of starlight.  The trisphere geometry, with
$\PhiPSA\approx0.69$, is marginally compatible with starlight
polarization.

A population of $N=2$ and $N=3$ aggregates would presumably include a
substantial fraction of bispheres, and $N=3$ shapes (e.g., the
threesphere) that are less effective polarizers than the trisphere.
Such a mixture would have $\langle\PhiPSA\rangle$ well below the lower
limit 0.7~.  We conclude that the observed starlight polarization
cannot be accounted for by $N=2$ and $N=3$ aggregates of more-or-less
spherical monomers.

\medskip

\subsection{FIR and Submm Polarization}

The same grains that account for polarization of starlight must also
reproduce the observed linear polarization of thermal emission from
dust.  At each point in the grain the local electric field is the sum
of the incident wave plus radiation from other parts of the grain.  In
the optical, where starlight polarization peaks, the grain size
$\aeff$ is comparable to the wavelength $\lambda$, radiation arrives
at different parts of the grain with different phase shifts, and
interference effects are important.  However, phase shifts are
negligible at FIR wavelengths $\lambda\gg\aeff$.  Thus, it was
anticipated that the ratio of FIR polarization cross section to
optical polarization cross section might be sensitive to both grain
shape and porosity, so that measurement of polarization at both FIR
and optical wavelengths might allow us to constrain the porosity.

In Paper I, which was limited to convex shapes with $\poromacro<0.12$,
it was found that all shapes which were compatible with the
requirements $\PhiPSA>0.7$ and $\sigmap\ltsim0.6$ predicted maximum
fractional polarization $\pfirmax$ in the range 0.17-0.25, close to or
within the {\it Planck} constraint $0.220_{-0.014}^{+0.035}$.

The present study has extended the calculations to irregular shapes
with porosities $\poromacro$ up to 0.85\,.  Seven of the BAM1trimB and
BAM2trimB aggregates with $\PhiPSA>0.7$ have $\pfirmax$ within the
range allowed by \emph{Planck}.  We find that above the minimum value
$\PhiPSA=0.7$ there appears to be a systematic trend for $\pfirmax$ to
decrease for increasing $\PhiPSA$: the three shapes with the larges
$\PhiPSA$ have $\pfirmax$ below the range allowed by \emph{Planck}.  A
similar trend was previously seen for prolate spheroids and oblate
spheroids \citep[see Figure 12b in][]{Draine_2024a}.  We conclude that
if interstellar grains are porous aggregates with sufficient
flattening or elongation to account for the polarization, they will
have $0.7<\PhiPSA\ltsim 1.2$, implying a relatively high fractional
alignment $\falign\approx 0.7/\PhiPSA \gtsim 0.6$ for the $\aeff\gtsim
0.05\micron$ grains that account for more than 50\% of the grain mass.

\medskip

\subsection{\label{subsec:10um} $10\micron$ Polarization}

The astrodust model -- or any model which posits that one grain type
dominates both starlight polarization and far-infrared emission --
predicts that the silicate material in interstellar grains will
produce polarization of starlight in the $10\micron$ silicate feature,
with $\Delta p({\rm sil}) \approx 0.137 p_{\rm max}$, where $p_{\rm
  max}$ is the polarization maximum (near $0.55\micron$):
see Equation (\ref{eq:polsil/pmax}).  The only star where the optical
polarization and the $10\micron$ polarization have both been measured
is the blue hypergiant Cyg\,OB2-12, where the optical polarization has
$p_{\rm max}=0.0967\pm 0.0010$ \citep{Whittet+Martin+Hough+etal_1992}
and the 10$\micron$ feature has excess polarization $\Delta p_{\rm
  sil}\approx0.0070\pm0.0035$ \citep{Telesco+Varosi+Wright+etal_2022}.
For Cyg\,OB2-12, the observed $\Delta p_{\rm sil}/p_{\rm
  max}=0.072\pm0.036$ is $\sim$$1.8\sigma$ lower than the value
$0.137$ predicted by Eq. (\ref{eq:polsil/pmax}).

The theoretical prediction (\ref{eq:polsil/pmax}) was found to be
approximately independent of shape, for both the convex shapes in
Paper I and the complex aggregate shapes studied here.  Therefore, the
discrepancy between the observed $\Delta p_{\rm sil}/p_{\rm max}$ and
the prediction (Equation \ref{eq:polsil/pmax}) cannot be attributed to
uncertainties concerning the detailed shape of the grains providing
the starlight polarization.

If the \citet{Telesco+Varosi+Wright+etal_2022} result for $\Delta
p_{\rm sil}/p_{\rm max}$ is confirmed, any grain model (e.g., the
astrodust model) that posits a single grain component dominating both
starlight polarization and far-infrared emission would appear to be
ruled out.  However, \citet{Draine_2024a} has argued that Cyg\,OB2-12
may have significant \emph{intrinsic} (apparently time-variable)
polarization, complicating determination of the \emph{interstellar}
contribution to the polarization.  More polarimetry of Cyg OB2-12 in
the optical and mid-IR is needed to confirm the determination of
$\Delta p_{\rm sil}/p_{\rm max}$ for interstellar dust.  Studies of
other sightlines are needed, to determine if this result is
representative of the diffuse ISM.

\medskip

\subsection{Determination of Grain Porosities Using X-Rays}

Small-angle scattering of X-rays by interstellar grains
\citep{Overbeck_1965} can provide an independent constraint on
porosity.  Suppose that dust is located on the path to a compact X-ray
source, with $d_s$ and $d_X$ the distance from us to the dust and
X-ray source, respectively.  The halo angle $\theta_h \approx
(1-\xi)\theta_s$, where $\theta_s$ is the scattering angle, and
$\xi\equiv d_s/d_X < 1$.

At $E\gtsim 1\keV$ the scattering cross section $C_{\rm sca}\propto
\aeff^6$, and the scattering is dominated by the larger grains.  The
halo has a uniform surface brightness ``core'', and extended wings
with surface brightness $\propto \theta_s^{-4}$ \citep{Draine_2003c}.
The transition from core to wing takes place near the angle
$\theta_{s,50}\approx \lambda/D_X$ containing 50\% of the scattered
power, where $D_X$ is the characteristic diameter of the grains
dominating the X-ray scattering.

For compact grains, the grain size distribution has $D_X\approx
0.6\micron$, hence $\theta_{s,50}\approx 360\arcsec(\keV/h\nu)$
\citep{Draine_2003c}.  If the grains are porous, we expect $D_X$ to
scale as $D_{\rm p}$ (see Eq.\ \ref{eq:D_p}), i.e., $D_X\propto
(1-\poromacro)^{-0.55}$.  Thus we estimate
\beq
\theta_{s,50} \approx 360\arcsec 
\left(\frac{\keV}{h\nu}\right)
\left(1-\poromacro\right)^{0.55}
~~~.
\eeq
If the distribution of dust along the line-of-sight is known, then measurement of the dust-scattered halo can be used to constrain the
porosities of interstellar grains.

The best measured X-ray halo is toward GX\,13+1.
\citet{Smith+Edgar+Shafer_2002} and \citet{Smith_2008} found the
observed $2\arcsec < \theta_h < 900\arcsec$ X-ray halo to be generally
consistent with models with compact grains, for uniform dust density
along the path to the source, although the instrumental PSF dominates
the observed halo profile for $\theta_h\ltsim 5\arcsec$.
\citet{Smith+Edgar+Shafer_2002} argued that highly porous grains
(e.g., $\poromacro=0.8$) were ruled out because they would produce
insufficient X-ray scattering, but did not allow for the increased
size of the grains in such models as well as the nonuniform density
within a grain with macroporosity.

\citet{Heng+Draine_2009} calculated X-ray scattering by BA, BAM1,
and BAM2 aggregates.  For equal-mass grains, increased porosity
(BAM2$\rightarrow$BAM1$\rightarrow$BA) resulted in reduced
$\theta_{s,50}$, as expected.  Although computational limitations
precluded use of realistic size distributions,
\citet{Heng+Draine_2009} concluded that the observed X-ray halo around
GX\,13+1 was not consistent with random aggregates, while calling for
more detailed modeling to confirm this.  

With improved computational methods \citep{Hoffman+Draine_2016},
calculations of X-ray scattering for aggregates with size
distributions that are compatible with interstellar reddening and
polarization should now be feasible, enabling interstellar grain
models with porous grains to be tested.

\medskip

\subsection{Implications for Grain Evolution}

Grain growth by coagulation is expected to be important in the diffuse
ISM.  If we suppose that all of the grain mass is in spheres of
mass-equivalent radius $\aeff$ and density $\rho=\rho_{\rm
  solid}(1-\poromacro)$, then in a region with H nucleon density
$\nH$, the time for a grain to collide with an equal mass of dust
(i.e., another grain) is
\beqa \label{eq:taucoll}
\tau_{\rm coll} &~\approx~& 
\frac{1}{n_{\rm gr}4\pi\aeff^2 (1-\poromacro)^{-2/3} v_{\rm rel}}
\\
&=&
1.2\xtimes10^7\yr 
\left(\frac{\aeff}{0.2\micron}\right)
\left(\frac{30\cm^{-3}}{\nH}\right)
\left(\frac{\kms}{v_{\rm rel}}\right)
\left(\frac{0.007}{Z_{\rm d}}\right)
\left(1-\poromacro\right)^{2/3}
,
\eeqa
where $v_{\rm rel}$ is the characteristic grain-grain velocity, and
$Z_{\rm d}$ is the mass fraction in dust.  A more careful treatment
would allow for a distribution of grain sizes, but (\ref{eq:taucoll})
remains \replaced{a reasonable}{an informative} estimate.

\added{We consider a scenario where the large grains are aggregates of
  smaller particles of varied composition.}  In the diffuse ISM,
grains are typically charged, and MHD turbulence is expected to result
in $v_{\rm rel}\approx 1\kms$ for grains with $\aeff\gtsim 0.1\micron$
\citep{Yan+Lazarian+Draine_2004}, with larger grains moving faster
than smaller grains.  For $v_{\rm rel}\gtsim 0.4\kms$, $\tau_{\rm
  coll}$ is comparable to or shorter than the timescale $1/\nH
R_{\HH}$ for conversion of H to H$_2$ by grain surface catalysis
\citep[for the empirical rate coefficient
  $R_{\HH}\approx3\xtimes10^{-17}\cm^3\s^{-1}$;][]{Jura_1975}.  Thus,
the typical grain in molecular gas will have collided with several
times its own mass of dust grains while the gas was converting from
H\,I to H$_2$.

Collisions between fast-moving large grains seem likely to result in
shattering, particularly if one or both are aggregates;
\citet{Yan+Lazarian+Draine_2004} suggest that this may account for the
sharp cutoff in the grain size distribution at $\sim$$0.3\micron$,
\added{as well as replenishing the population of small grains}. Grain
growth may proceed by larger grains ``sweeping up'' smaller grains.
When a very small grain ($a\ltsim 0.03\micron$) hits a much larger
grain ($a\gtsim 0.1\micron$) with an impact speed of $\sim$$0.5\kms$,
it seems possible that some impacts may result in the impactor
sticking to or embedding itself within the larger grain, resulting in
grain growth.

\deleted{Even the BAM2 growth scheme produces aggregates with
  $\poromacro \approx 0.58$ \citep{Shen+Draine+Johnson_2008}.
  Therefore, coagulation seems likely to produce relatively porous
  grains, unless grain growth by coagulation is accompanied by
  processes that can act to reduce the grain porosity either during
  the collision event, or over the time between grain-grain
  collisions.}

Coagulation acting alone will produce high-porosity aggregates
resembling the BA, BAM1, or BAM2 structures, which we have seen above
to be very weak polarizers, incompatible with the observed
polarization of starlight and polarized far-infrared emission.  There
must be processes acting in the interstellar medium to increase the
polarizing abilities of the aggregates, by reducing the porosity
and/or increasing the asymmetry of the aggregates.
 
\medskip

\subsubsection{Densification by Crushing?}

A porous grain undergoing a collision with another grain may undergo
some degree of ``crushing'', lowering the porosity.  For aggregates of
monodisperse SiO$_2$ spheres with initial porosities
$\poromacro\approx 0.85$, \citet{Blum+Schraepler_2004} found that the
porosity could be reduced to $\poromacro\approx 0.66$ if subjected to
a static pressure $p\approx 10^4 {\rm \,dyne}\cm^{-2}$.  For initial
porosities $\poromacro\approx 0.8$, simulations of aggregate-aggregate
collisions with collision velocity $0.005\kms$ resulted in partial
fragmentation, but with a substantial fraction of the material
remaining in a structure with lower porosity $\poromacro\approx 0.6$
\citep{Gunkelmann+Ringl+Urbassek_2016,Planes+Millan+Urbassek+Bringa_2021}.
It seems plausible that continued aggregate-aggregate collisions with
velocities in a suitable range might result in much of the mass
residing in relatively compact aggregates, perhaps with porosities
$\poromacro\ltsim 0.6$.  However, based on our studies of BAM2
aggregates, even porosities $\poromacro \approx 0.6$ appear to be
inconsistent with polarization observations, unless the structures are
significantly more asymmetric than the BAM2 aggregates studied here.

\medskip

\subsubsection{Photolytic Densification?}

In the diffuse ISM, the FUV intensity is such that a single electron
in a chemical bond will be photoexcited to a higher energy orbital on
a timescale of $\sim$$10^{11}\sec$.  Thus, on the
$\sim$$3\xtimes10^8\yr$ lifetime of a grain
\citep[e.g.][]{Barlow_1978, Draine+Salpeter_1979b,
  Zhukovska+Dobbs+Jenkins+Klessen_2016,
  Hu+Zhukovska+Somerville+Naab_2019}, a single bonding electron will
be photoexcited to a new electronic state 
$\sim$$10^5$ times.  If the new electronic state happens to be
repulsive, an atom or radical may be displaced.  It is conceivable
that this process might result in a systematic tendency to reduce the
porosity of the grain, with atoms or radicals gradually relocating to
higher-coordination sites where they are more permanently bonded.
Note that FUV photons can penetrate $\sim$$0.1\micron$ into the grain,
allowing this process to act even in the interior of a submicron
aggregate.

Laboratory experiments on materials of technological interest for
production of thin film transistors have demonstrated that some
compounds can be ``annealed'' at room temperature if exposed to FUV
radiation \citep[e.g.,][]{Kim+Heo+Kim+etal_2012,Park+Kang+Kim_2020}.
However, there do not appear to have been any laboratory studies of
high-porosity materials to see whether densification occurs.  Lab
studies of FUV irradiation of silica aerogels would be valuable.

Cosmic rays are also important for exciting grain material,
particularly the heavy ions (O$^{+8}$, Ne$^{+10}$, Fe$^{+26}$)
\citep[see, e.g.,][]{Leger+Jura+Omont_1985}.  A low energy or high-$Z$
cosmic ray passing through the grain leaves behind a hot channel which
permits atomic rearrangement as the ``thermal spike'' cools by
diffusion.  The atomic rearrangement will tend to ``anneal'' the
material, putting atoms into higher-coordination sites, reducing the
spatial extent of the grain, and decreasing the porosity.

\medskip

\subsection{Processes that May Increase Grain Asymmetry}

Growth by addition of single-size monomers produces aggregates
resembling the BA, BAM1, or BAM2 structures, if the monomers stick at
or close to the point of first contact.  

However, grains in diffuse clouds with $\aeff\gtsim 0.1\micron$ are
expected to be spinning suprathermally, as the result of systematic
torques resulting from photoelectron emission, $\HH$ formation, and
starlight \citep{Purcell_1975, Draine+Weingartner_1996}; starlight
torques are expected to dominate.  \replaced{Conceivably, rapid rotation might
lead to migration of weakly bound physisorbed atoms, molecules, or
nanoparticles to points near the ``equator'', but an
$\aeff\approx0.2\micron$ grain in the diffuse ISM is expected to have
equatorial rotation speeds $v_{\rm rot} \ltsim 0.01\kms$
\citep{Draine+Weingartner_1996}, which seems too small to
significantly affect the shape of the growing
aggregate.}{In the diffuse ISM, $\aeff\approx0.2\micron$ grains are expected to have rotation speeds $v_{\rm rot} \ltsim 0.01\kms$
\citep{Draine+Weingartner_1996}, which seems too small to disrupt the
grain,\footnote{
In a spinning solid sphere, the greatest stresses are near the center.
The limiting equatorial speed is \citep{Draine+Salpeter_1979a}
$$
v_{\rm rot,max} \approx c_s \times 
\left(\frac{S_{\rm yield}}{Y}\right)^{1/2} ~~,
$$
where $Y$ is Young's modulus, $S_{\rm yield}$ is the yield stress,
and $c_s\approx(Y/\rho)^{1/2}\gtsim 1 \kms$ is the sound speed in the
solid .  For typical solids, $S_{\rm yield}\gtsim0.01 Y$, hence
$v_{\rm rot,max} \gtsim 0.1 c_s \approx 0.1\kms$}
but could conceivably affect the shape of the growing aggregate by
favoring migration of weakly bound physisorbed atoms, molecules, or
nanoparticles to locations near the ``equator''.}  Higher rotation
speeds occur for grains passing close to stars \citep[see,
  e.g.,][]{Silsbee+Draine_2016, Hoang_2019}, but such events are too
rare to affect the overall grain population.

Significant asymmetry could result from coagulation of two
already-large aggregates of comparable size, forming a bisphere-like
geometry.  However, we have seen that the bisphere produced by joining
together two $\poromacro=0$ spheres has $\PhiPSA=0.37$, well below the
$\PhiPSA>0.7$ required to explain the observed starlight polarization.
Unless the two colliding aggregates fuse into a shape that is a
significantly better polarizer than the bisphere, the final structure
will have a low $\PhiPSA$.

If interstellar grains are high-porosity aggregates, there must be
some mechanism to produce extreme asymmetries, e.g.  $\Asymm\gtsim
3.4$ if $\poromacro\approx 0.7$ (see Equation \ref{eq:Asymm_min}).
Suitable shapes can be obtained by systematically ``trimming''
initially random aggregates (e.g., the \emph{trimB}, and \emph{trimC}
aggregates in Figures \ref{fig:shapes_trimB}-\ref{fig:shapes_trimC})
but it is not clear what interstellar processes would impart such
asymmetric shapes to porous aggregates.  Atomic sputtering or
grain-grain collisions can ``trim'' weakly-bound monomers from the
surface, but the systematic effects on shape are unclear.  The IDPs in
Figure \ref{fig:IDP} are quite asymmetric, but the processes that
shaped them remain unknown.

\medskip

\begin{figure}
\begin{center}
\includegraphics[angle=0,width=\fwidth,
                 clip=true,trim=0.5cm 5.0cm 0.5cm 2.5cm]
{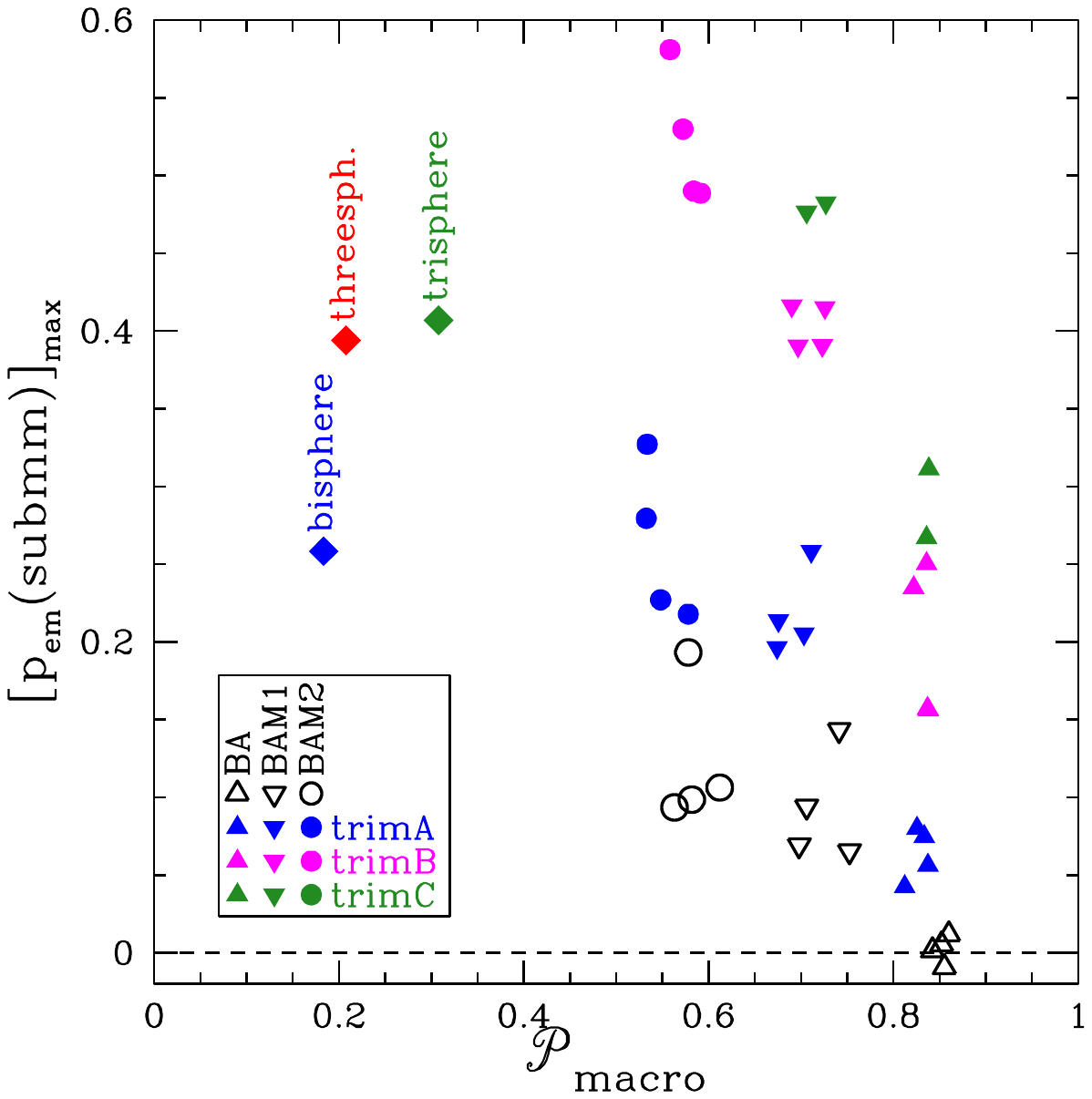}
\caption{\label{fig:submmpol}\footnotesize Maximum polarization of
  thermal emission for grains in the Rayleigh limit
  ($\aeff\ltsim\lambda/2\pi$) and perfect spinning alignment, versus
  macroporosity.  The plain BA, BAM1, and BAM2 aggregates (open
  symbols) have $[p_{\rm em}]_{\rm max} < 0.20$.  Polarizations
  exceeding 20\% require that aggregate grains be densified or
  reshaped (see text).}
\end{center}
\end{figure}

\added{
\subsection{Grains in High Density Regions}

Grain growth is seen in molecular clouds: changes in the extinction
curve are observed, characterized by increases of $R_V\equiv
A_V/E(B-V)$ from $\sim$3.1 in the diffuse ISM to $\sim$4.2 in the
$\rho$\,Oph molecular cloud \citep{Martin+Whittet_1990}, and values as
large as $R_V\approx 6$ in other clouds \citep{Whittet_2022}.
Increased grain size is also revealed by increased scattering in the
mid-infrared (``cloudshine'' and ``coreshine'') indicating growth of
grains to radii $\sim$$1.0\micron$ in the externally illuminated
regions of dense clouds \citep{Foster+Goodman_2006,
  Pagani+Steinacker+Bacmann+etal_2010,
  Andersen+Steinacker+Thi+etal_2013,
  Steinacker+Andersen+Thi+etal_2015}.

The observed grain growth must be due to coagulation, and is
presumably more advanced in denser regions.  For grains in the
Rayleigh limit $\aeff \ll \lambda$, Figure \ref{fig:submmpol} shows
the maximum possible polarization at mm and submm wavelengths, for
grains in perfect spinning alignment.  Plain BA, BAM1, or BAM2
coagulation results in porous aggregates that are poor polarizers.
Therefore, one might have expected lower fractional polarization of
submm emission in high column density regions.

BLASTPol observations of emission from dust in the Vela C molecular
cloud \citep{Fissel+Ade+Angile+etal_2016} found that the fractional
polarization decreases with increasing column density.
\citet{Fissel+Ade+Angile+etal_2016} concluded that even after allowing
for magnetic field disorder, the observations suggest that dust in
denser regions is less effective at producing polarization than dust
in diffuse regions.

However, \citet{LeGouellec+Maury+Guillet+etal_2020} analyzed ALMA
observations of 12 nearby low- and intermediate-mass Class 0
protostellar cores, and concluded that observed reductions in
fractional polarization may be primarily due to magnetic field
disorder, with no evidence of reductions in polarizing ability due to
reduced degree of grain alignment.  The low-mass Class 0 protostellar
core NGC\,1333 IRAS4A1 is a conspicuous example: at $870\micron$,
regions $\sim$400\,AU NE and $\sim$400\,AU SW of IRAS4A1 have
polarization fractions exceeding 20\%, with a similar polarization
fraction at $6.9\mm$ \citep{Ko+Liu+Lai+etal_2020}.  The dust at these
locations must have a submm polarizing efficiency comparable to dust
in the diffuse ISM.  If, as expected, the dust particles have grown by
coagulation, there must be processes acting to increase the flattening
or elongation of the aggregates, and/or reduce their porosity, even in
these dense regions.}

\medskip

\section{\label{sec:summary}
         Summary}

The principal results are as follows:
\begin{enumerate}

\item The grains producing the observed polarization of starlight
  cannot resemble bispheres: they are too inefficient as polarizers,
  and therefore cannot be common in the interstellar grain population.

\item The observed polarizing properties of interstellar dust cannot
  be accounted for by mixtures of $N=2$ and $N=3$ aggregates of
  more-or-less spherical monomers.

\item The empirical scaling relation (\ref{eq:Phiapprox}) can be used
  to estimate $\PhiPSA$ from the porosity $\poromacro$ and asymmetry
  parameter $\Asymm$.

\item If interstellar grains have high porosity, they must also be
  extremely flattened or elongated in order to reproduce polarization
  observations.  Porosity $\poromacro = 0.90$ requires $\Asymm \gtsim
  12$.  Unless there is a process to produce very extreme aspect
  ratios, porosities $\poromacro \gtsim 0.90$ are ruled out by
  starlight polarization.

\item Seven of the \emph{trimB} and \emph{trimC} aggregates studied
  are consistent with starlight polarization ($\PhiPSA>0.7$) and the
  submm polarization fraction observed by \emph{Planck}.
  All have $\poromacro < 0.73$, and $\PhiPSA< 1.2$.
  
\item All shapes that are able to reproduce the starlight polarization
  have similar $[\Delta\CpolPSA({\rm sil})/V]/\PhiPSA$, where
  $\Delta\CpolPSA({\rm sil})$ is the polarization cross section due to
  the $10\micron$ silicate feature.  The different shapes, therefore,
  predict similar ratios $\Delta p({\rm sil})/\pmax$, where $\pmax$ is
  the peak starlight polarization, and $\Delta p({\rm sil})$ is the
  polarization due to the silicate feature.

\item The discrepancy between the predicted
  \citep{Draine+Hensley_2021c} and observed
  \citep{Telesco+Varosi+Wright+etal_2022} 10$\micron$ polarization of
  Cyg\,OB2-12 is not resolved by varying the assumed grain shape.  If
  the measurement by \citet{Telesco+Varosi+Wright+etal_2022} is
  confirmed to characterize interstellar dust, the astrodust model
  (positing a single dominant grain type) will be disfavored.
  However, because Cyg OB2-12 itself appears to have time-variable
  intrinsic polarization, the ratio $\Delta p({\rm sil})/p_{\rm max}$
  should be measured on other sightlines.
  
\item X-ray scattering can be used to constrain grain porosities.  The
  observed X-ray scattering halo around GX\,13+1 may rule out models
  with $\poromacro>0.55$ \citep{Heng+Draine_2009}, but more extensive
  modeling is required to confirm this.

\item Random coagulation acting alone would result in high porosities
  and modest overall asymmetries that together are incompatible with
  the observed polarization of starlight and polarized FIR emission.
  Because coagulation is expected, the observed polarization requires
  processes that (1) increase the elongation or flattening and/or (2)
  reduce the porosity.  Crushing (during grain-grain collisions) can
  reduce the porosities of aggregates that are not disrupted.

\item The FUV radiation in diffuse clouds may lead to reduction in
  porosity through ``photolytic densification''.  Transient ionization
  and heating by cosmic rays may also contribute to densification.
  Even in the presence of grain growth by coagulation, photolytic
  densification may keep grain porosity low, so that interstellar
  polarization can be explained by axial ratios that are not extreme.

\item \added{The polarized emission observed in some Class 0 cores,
  with fractional polarization $p>20\%$, implies that grain growth in
  dense clouds must be accompanied by processes that either reduce the
  porosity or enhance the shape asymmetry.}

\end{enumerate}
\begin{acknowledgements}

I thank Don Brownlee, Henner Busemann, and Nicole Spring for
permission to use the images in Figure 1. I thank Massimo Bertini,
Larry Nittler and Rhonda Stroud for helpful discussions, and Robert
Lupton for availability of the SM package.  I thank the anonymous
referee for helpful suggestions that improved the manuscript.  I
gratefully acknowledge support from NFS grant AST-1908123, and from
the Institute for Advanced Study.

\end{acknowledgements}

\bibliography{/u/draine/work/libe/btdrefs}

\appendix
\section{\label{app:comp}Computational Details}
The computations were carried out using the DDA code {\tt DDSCAT
  7.3.3},\textsuperscript{\ref{fn:ddscat}} with error tolerance {\tt
  TOL} $=1\xtimes10^{-5}$.  The numbers of dipoles $N_1$, $N_2$, $N_3$
used for the different target realizations are listed in Table
\ref{tab:Ndip}.  In Paper I, it was feasible to carry out calculations
with much larger values of $N_1,N_2,N_3$ than for the BA, BAM1, and
BAM2 aggregates in the present study, because:
\begin{enumerate}
\item The symmetries of the convex targets considered in Paper I meant
  that fewer target orientations were needed than for the asymmetric
  BA, BAM1, and BAM2 targets (e.g., for the cylindrical targets in
  paper I, only 11 orientations were needed, vs.\ the $12\times11=132$
  orientations used for the BA, BAM1, and BAM2 targets).
\item The FFT methodology \citep{Goodman+Draine+Flatau_1991} employed
  in DDSCAT requires a rectangular ``computational volume'' that
  contains all occupied lattice sites.  For porous structures, the
  many unoccupied lattice sites within this computational volume
  impose a computational burden, which limits us to relatively modest
  numbers of occupied lattice sites $N_1$ in the largest calculations
  for the BA, BAM1, and BAM2 targets.
\end{enumerate}
Because the number of dipoles per spherical monomer is not
very large (e.g., $209917/256\approx 820$ for the BA.256.2 target),
the discreteness leads to limited fidelity in the calculated cross
sections for the $N_1,N_2,N_3$ values used.  As a result, the
extrapolation procedure has relatively large uncertainties,
particularly for the BA aggregates.

In all cases our uncertainty estimates are based on the difference
between extrapolation using $(N_1,N_2)$ and extrapolation using
$(N_2,N_3)$ and Equation (9) from \citet{Draine_2024a}.  We do not
include uncertainties resulting from the use of a finite number of
target orientations, or due to use of a finite number of wavelengths
in evaluation of the integral (\ref{eq:PhiPSA}) for $\PhiPSA$.

\begin{table}
\begin{center}
\caption{\label{tab:Ndip}Numbers of Dipoles Employed}
{\footnotesize
\begin{tabular}{c r r r}
\hline 
shape           & $N_1$~~ & $N_2$~~ & $N_3$~~ \\
\hline
bisphere        & 2209298 &  769872 & 277968 \\
threesphere     &  198608 &  102072 &  52264 \\
trisphere       & 2748360 & 1411420 & 421408 \\
\hline
BA.256.1        &  264622 &  135950 &  78757 \\
BA.256.2        &  209917 &  132484 &  76821 \\
BA.256.3        &  263017 &  165792 &  96341 \\
BA.256.4        &  275676 &  201131 & 141494 \\
\hline
BAM1.256.1      &  377960 &  115481 &  28326 \\
BAM1.256.2      &  281493 &  197919 & 114873 \\
BAM1.256.3      &  297503 &  172632 &  88663 \\
BAM1.256.4      &  268258 &  188641 & 109439 \\
\hline
BAM2.256.1      &  262890 &  135198 &  57378 \\
BAM2.256.2      &  319152 &  185369 &  95081 \\
BAM2.256.3      &  358900 &  208132 & 107021 \\
BAM2.256.4      &  249256 &  144616 &  74337 \\
\hline
BAtrimA.256.1   &  110338 &   56820 &  24101 \\
BAtrimA.256.2   &   61569 &   26179 &  15265 \\
BAtrimA.256.3   &   70907 &   41301 &  21273 \\
BAtrimA.256.4   &   50582 &   29359 &  15075 \\
\hline
BAM1trimA.256.1 &  231109 &   99109 &  57035 \\
BAM1trimA.256.2 &  216688 &  111419 &  45154 \\
BAM1trimA.256.3 &   58951 &   34288 &  17635 \\
BAM1trimA.256.4 &   50304 &   29192 &  15077 \\
\hline
BAM2trimA.256.1 &  137087 &   79709 &  41207 \\
BAM2trimA.256.2 &  153420 &   65103 &  37800 \\
BAM2trimA.256.3 &  106205 &   61793 &  31709 \\
BAM2trimA.256.4 &   49195 &   25421 &  13129 \\
\hline
BAtrimB.256.1   &  218532 &  126830 &  65153 \\
BAtrimB.256.2   &   91070 &   43515 &  22375 \\
BAtrimB.256.3   &   69764 &   35993 &  18595 \\
BAtrimB.256.4   &   55960 &   32387 &  16668 \\
\hline
BAM1trimB.256.1 &  148098 &   86079 &  44388 \\
BAM1trimB.256.2 &  147799 &   85718 &  44124 \\
BAM1trimB.256.3 &   96774 &   41132 &  23895 \\
BAM1trimB.256.4 &  108484 &   62895 &  32376 \\
\hline
BAM2trimB.256.1 &  231236 &  119352 &  61683 \\
BAM2trimB.256.2 &  178617 &  125619 &  72935 \\
BAM2trimB.256.3 &   55163 &   23504 &  12255 \\
BAM2trimB.256.4 &   51260 &   29865 &  15372 \\
\hline
BAtrimC.256.1   &   57004 &   33137 &  17189 \\
BAtrimC.256.2   &   60067 &   34836 &  17903 \\
\hline
BAM1trimC.256.1 &   70905 &   30045 &  17486 \\
BAM1trimC.256.2 &   87606 &   45078 &  19164 \\
\hline
BAM2trimC.256.1 &   77192 &   32730 &  18954 \\
BAM2trimC.256.2 &   75871 &   32110 &  18592 \\
\hline
\end{tabular}
}
\end{center}
\end{table}
\section{Supplementary Material}

Supplementary material, including images of all of the targets,
parameter files ({\tt ddscat.par}), as well as files specifying the
target geometries and dielectric function used in this study are
available at \website.

\end{document}